\documentclass[twocolumn]{aastex62}

\newcommand{\radm}{$\mathrm{rad\, m}^{-2}\,$}
\received{TBD}
\revised{TBD}
\accepted{TBD}
\submitjournal{ApJ}
\usepackage{textcomp}
\usepackage{rotating}
\usepackage{amsmath}
\newcommand{\GG}[1]{}
\shorttitle{A Wideband Polarization Study of Cygnus A}
\shortauthors{Sebokolodi et al.}

\begin{document}

\title{A Wideband Polarization Study of Cygnus A with the JVLA. I: The Observations and Data}

\correspondingauthor{Lerato Sebokolodi}
\email{mll.sebokolodi@gmail.com}

\author{M. Lerato L Sebokolodi}
\affiliation{Department of Physics and Electronics, Rhodes University, Grahamstown 6140, South Africa}
\affiliation{South African Radio Astronomy Observatory, SARAO, 2 Fir Street, Black River Park, Observatory, 7925, South Africa}
\affiliation{National Radio Astronomy Observatory, 1003 Lopezville Rd, Socorro, NM 87801}

\author{Rick Perley}
\affiliation{Department of Physics and Electronics, Rhodes University, Grahamstown 6140, South Africa}
\affiliation{National Radio Astronomy Observatory, 1003 Lopezville Rd, Socorro, NM 87801}

\author{Jean Eilek}
\affiliation{National Radio Astronomy Observatory, 1003 Lopezville Rd, Socorro, NM 87801}
\affiliation{New Mexico Tech, Socorro, NM 87801}

\author{Chris Carilli}
\affiliation{National Radio Astronomy Observatory, 1003 Lopezville Rd, Socorro, NM 87801}

\author{Oleg Smirnov}
\affiliation{Department of Physics and Electronics, Rhodes University, Grahamstown 6140, South Africa}
\affiliation{South African Radio Astronomy Observatory, SARAO, 2 Fir Street, Black River Park, Observatory, 7925, South Africa}

\author{Robert Laing}
 \affiliation{SKA Organization, Jodrell Bank Lower Withington, Macclesfield Cheshire SK11 9FT, United Kingdom}

\author{Eric Greisen}
\affiliation{National Radio Astronomy Observatory, 1003 Lopezville Rd, Socorro, NM 87801}

 \author{Michael Wise}
 \affiliation{SRON Netherlands Institute for Space Research, Sorbonnelaan 2, 3584 CA Utrecht, the Netherlands}



\begin{abstract}
We present results from deep, wideband, high spatial and spectral
resolution observations of the nearby luminous radio galaxy Cygnus A
with the Jansky Very Large Array. The high surface brightness of this
source enables detailed polarimetric imaging, providing images at
$0.75\arcsec$, spanning 2 - 18 GHz, and at 0.30$\arcsec$ (6 - 18 GHz). The fractional polarization from 2000 independent
lines of sight across the lobes decreases strongly with decreasing
frequency, with the eastern lobe depolarizing at higher frequencies
than the western lobe. The depolarization shows considerable
structure, varying from monotonic to strongly oscillatory.  The
fractional polarization in general increases with increasing
resolution at a given frequency, as expected. However, there are
numerous lines of sight with more complicated behavior.  We have
fitted the $0.3\arcsec$ images with a simple model incorporating
random, unresolved fluctuations in the cluster magnetic field to
determine the high resolution, high-frequency properties of the source
and the cluster. From these derived properties, we generate
predicted polarization images of the source at lower frequencies,
convolved to 0.75$\arcsec$. These predictions are remarkably
consistent with the observed emission.  The observations are
consistent with the lower-frequency depolarization being due to unresolved
fluctuations on scales $\gtrsim$ 300 - 700 pc in the magnetic field and/or electron density superposed
on a partially ordered field component. There is no
indication in our data of the location of the depolarizing screen or
the large-scale field, either, or both of which could be located
throughout the cluster, or in a boundary region between the lobes and the cluster.

\end{abstract}



\section{Introduction}\label{sec:Intro}
Cygnus A (3C 405) is the prototypical Fanaroff-Riley type II radio
galaxy \citep{1974FANAROFF}. It is one of the most luminous radio
galaxies known, and is exceptionally close
\citep[$z=0.056$;][]{1982SPINRAD}\footnote{Adopting the $\Lambda$CDM
  cosmology with $H_0=69.3$ km s$^{-1}$ Mpc$^{-1}$, $\Omega_m=0.288$,
  and $\Omega_{\Lambda}=0.712$ \citep{2013HINSHAW}, gives a distance
  of $227$ Mpc, and $1\arcsec \approx 1.1$ kpc.}  compared to galaxies
of similar radio luminosity \citep[see Fig. 1
  of][]{1996STOCKTON}. Cygnus A's high flux density ($\sim1000$ Jy at
$2$ GHz), combined with its relatively small angular size (maximum
projected extent of $\approx$ $120\arcsec$) means the source is
unusually bright, making it an outstanding target for high resolution
radio polarimetric imaging with synthesis telescopes like the Very
Large Array (VLA).

Cygnus A is located at the center of a dense cooling-core X-ray
emitting cluster \citep{1972GIACCONI,1979FABBIANO}. The cluster
has a core of radius $\sim 18$ kpc and density of $\sim
15 \times 10^{-26}$ g cm$^{-3}$ \citep{2002SMITH,2019HALBESMA}. The
gas temperature decreases from $\sim 9$ keV at $\sim 300$ kpc radius
to $\sim 3.5$ keV within the core \citep[see top right plot Fig. 4 in
][]{2018SNIOS}. Cygnus A is surrounded by a weak cocoon shock, driven
by the expanding lobes, extending $33$ kpc north and $74$ kpc west of
the cluster center \citep{1994CARILLI, 2018SNIOS}, with Mach numbers
ranging between $1.18- 1.66$ across the shock.

\begin{figure*}[!ht]
\centering
 \includegraphics[width=1.1\linewidth]{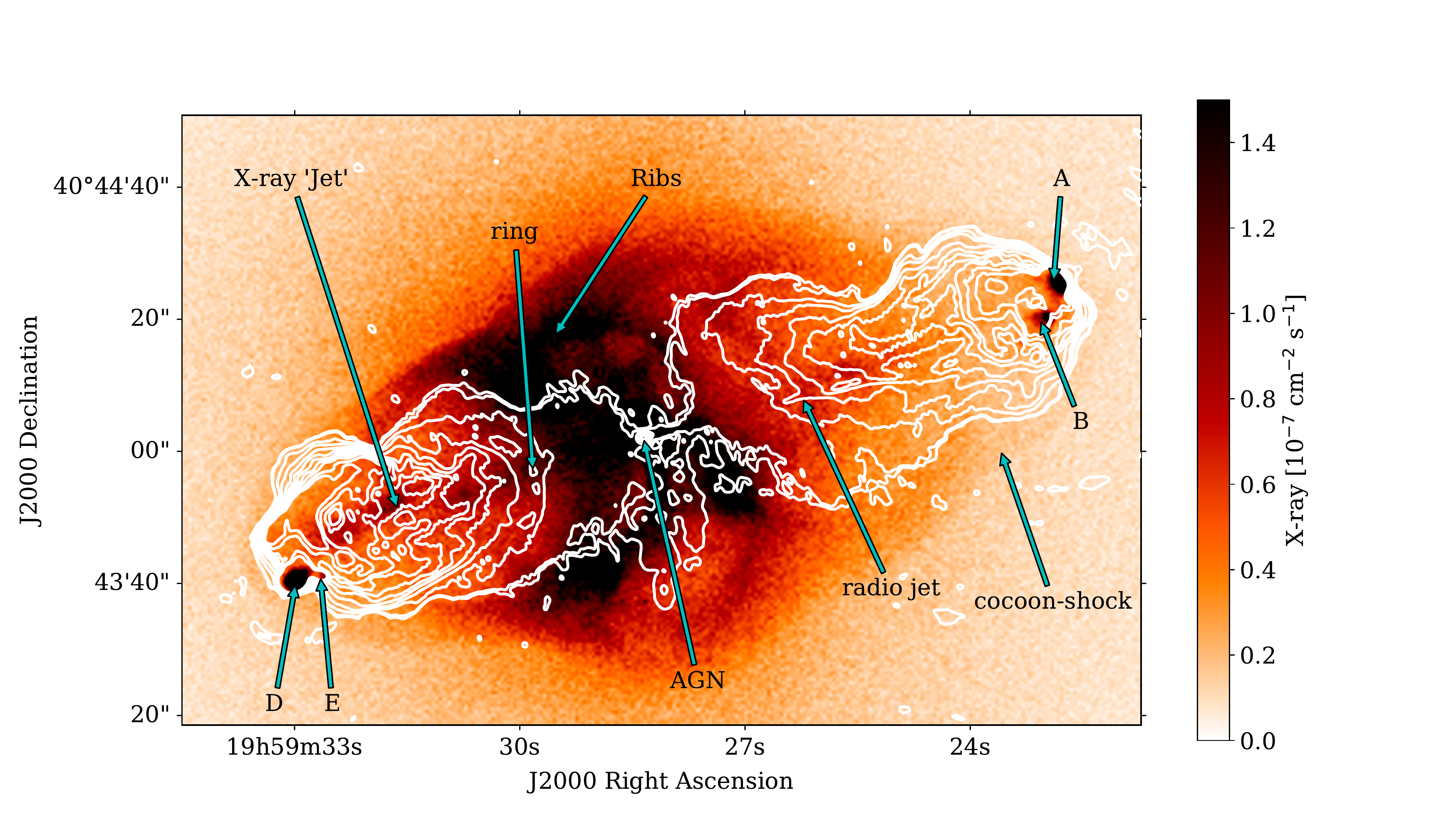}
 \caption{\textit{Chandra} X-ray data ($0.5-7$ keV) of Cygnus A
   superimposed by the total intensity contour image ($2.4$ GHz and
   $1\arcsec$) of our new wideband data. The contour levels are
   $0.005$, $0.01$, $0.05$, $0.1$, $0.2$, $0.3$, $0.4$, $0.5$, $0.6$,
   $0.8$, $1.0$ Jy beam$^{-1}$. Shown are the radio features: the hot spots
   A, B, D and E at the ends of the lobes, the ring-like feature at
   the tails of the eastern lobe, the radio jet in the western lobe,
   and the central AGN. Also shown are the X-ray features: the X-ray
   jet, cocoon shock (the sheath) and the ribs
   \citep{2018DEVRIES,2018SNIOS,2018DUFFY}. The off-source noise of
   the radio image is $1.7$ mJy beam$^{-1}$. \label{fig:Ilabels}}
\end{figure*}

Figure \ref{fig:Ilabels} shows the total intensity contours at $2$ GHz
with $1\arcsec$ resolution superimposed on the \textit{Chandra} X-ray
image in the energy interval $0.5-7$ keV.  The well-known structures
in the radio and X-ray are labeled: the central nucleus, the lobes,
the four hotspots and the radio jet, as well as filamentary structures
across the lobes, a ring-like structure in the tail of the eastern
lobe \citep{1984PERLEY}, and the X-ray `jet' \citep{2018DEVRIES},
cocoon-shock \citep{2018SNIOS} and ribs \citep{2018DUFFY}.

Cygnus A is significantly polarized with typical polarizations of
$\lesssim 40\%$, and as high as $70\%$ in the lobes and hot spots at
high resolution \citep{1989CARILLI}.  Polarimetric observations of
Cygnus A showed large Faraday rotations across its lobes, asymmetry in
the polarization properties of the lobes, as well as significant
decreases in fractional polarization with decreasing frequency at low
resolutions \citep{1966SLYSH, 1971MITTON, 1979DREHER, 1984ALEXANDER}.
The high resolution study by \citet{1987DREHER} revealed typical
rotation measures, $RM$, ranging between $-4000$ \radm and $ +3000$
\radm across the lobes, with gradients in the $RM$ distribution of
$300$ \radm arcsecond$^{-1}$ in most parts of the lobes, and with a
few places having gradients up to $1000$ \radm arcsecond$^{-1}$. These
large gradients are responsible for the low polarization seen in the
early low-resolution studies.  The primary focus since then has been
to understand the origin of these depolarizations and extraordinary
Faraday rotations. The locations considered were our own galaxy, the
X-ray cluster, the cocoon-shock, or a mixed thermal/synchrotron gas
region located at the boundary of the lobes, or within the lobes
themselves.

At the low galactic latitude of Cygnus A, $b = 5.8^{\circ}$, our own
galaxy contributes no more than $ RM \sim 300$ \radm
\citep{1981SIMARD}, and consists of gradients not exceeding $180$
\radm between component separated by no more than $1^{\circ}$
\citep{1992CLEGG}. Thus it was concluded that the origin of the large
$RM$ and gradients must be local to the source
\citep{1987DREHER}. \citeauthor{1987DREHER}, noting the high
polarization at $6$ cm, the large total rotation of the electric
vector (up to $600^{\circ}$ at that wavelength), and the apparently
perfect linearity of the observed electric field position angle vs
$\lambda^2$ between $6$ and $2$ cm wavelength, concluded that the
great majority of the observed $RM$ must be due to an external medium,
and cannot originate from a mixed thermal/synchrotron region inside
the lobes.  If the origin of the $RM$ is distributed throughout the
cluster, the resulting magnetic fields were inferred to be $\sim 5$
$\mu$G.  If the origin were confined to the cocoon-shock, the fields
would be $\sim 20$ times higher.

A cluster origin for the majority of the $RM$ does not exclude smaller
contributions from the cocoon-shock. \citet{1988CARILLI}, discovering
a region of enhanced $RM$ in front of hotspot `D' of the western lobe,
suggested that both the cluster gas and the cocoon-shock contribute,
with the cocoon-shock responsible for $|RM| \leq 1000 $ \radm ordered
on scales $<5$ kpc while the cluster is responsible for $|RM| > 1000$
\radm ordered on scales $>\,20$ kpc.

\citeauthor{1987DREHER}'s results were based on only four wavelengths
spanning $\lambda$ $6$ cm to $2$ cm.  Subtle depolarization effects,
such as those due to turbulence or to boundary layer effects will not
be visible with such sparse sampling. Since depolarization effects due
to Faraday rotation are manifested at lower frequencies, investigation
of such mechanisms require observations at frequencies below 5 GHz,
with continuous frequency coverage. With the completion of the
wideband Jansky Very Large Array \citep[JVLA, ][]{2001PERLEY}, we have
now the capability to observe Cygnus A at lower frequencies than those
available to \citeauthor{1987DREHER}, -- notably, use of the new $2$
-- $4$ GHz receiver -- and with complete frequency coverage.

In this paper we present the results of our new wideband ($2$ -- $18$
GHz), full polarization and high-spectral resolution observations of
Cygnus A taken with the JVLA. The primary science goals of this study
are to determine the spatial and frequency dependence of the
depolarization, identify the physical structures and conditions
responsible for this depolarization, and to determine the structures
in the magnetic fields of the source and the surrounding cluster.

The results are presented in two papers: this paper presents the
observations and primary data products, and results from analysis of
the high frequency data. The second paper will present results from
our modeling of the wideband data, and physical interpretations from
this modeling.

This paper is organized as follows: The observations and calibration of the
data are presented in section \ref{sec:observations}; and the imaging
of the data in section \ref{sec:imaging}. Section \ref{sec:newdata}
presents our polarization results, followed by the results of a
high-frequency high-resolution Faraday rotation study in section
\ref{sec:faradayrotation}. In section \ref{sec:beamtest} we show the
results of applying the high frequency, high resolution model to the
low frequency data.  A summary and discussion are in section
\ref{sec:discussion}.

\section{Observations and Calibration}\label{sec:observations}
The observations were taken under VLA observing project 14B-336.
The details of these observations are shown in Table \ref{tab1}. This
project utilized all array configurations for four observing bands; 2
-- 4 GHz (`S-band'), 4 -- 8 GHz ('C-band'), 8 -- 12 GHz (`X-band'),
and 12 -- 18 GHz (`Ku-band'), providing complete frequency coverage
from $2$ to $18$ GHz.

\begin{table*}
\begin{center}
\caption{Observing log \label{tab1}}
 \begin{tabular}{c c c c c c}
 \hline \hline 
 Configuration & Date & Band & IAT Range & TOS \tablenotemark{a} & LST Range \\
  & & & [H:M] &  [min] & [H:M]\\
 \hline \hline 
D   & 2015 Nov 15 &  S  & 20:09 - 01:96 & 20 & 16:40 - 22:20\\
    & 2015 Nov 15 &  C  & 20:09 - 01:96 & 40 & 16:40 - 22:20\\
    & 2015 Nov 15 &  X  & 20:09 - 01:96 & 60 & 16:40 - 22:20\\
    & 2015 Nov 15 &  Ku & 20:09 - 01:96 & 95 & 16:40 - 22:20\\
\hline 

C   & 2014 Nov 03 &  S  & 18:10 - 04:40 & 105 & 14:00 - 00:15 \\
    & 2014 Nov 03 &  C  & 18:10 - 04:40 & 100 & 18:10 - 04:40 \\
    & 2014 Nov 03 &  X  & 18:10 - 04:40 & 100 & 18:10 - 04:40 \\
    & 2014 Nov 03 &  Ku & 18:10 - 04:40 & 125 & 18:10 - 04:40 \\

\hline 

B   & 2015 Apr 05 &  S  & 09:15 - 16:50 & 130 & 15:10 - 22:30 \\
    & 2015 Apr 12 &  C  & 10:00 - 17:35 & 130 & 16:25 - 23:30  \\
    & 2015 Apr 12 &  X  & 10:00 - 17:35 & 180 & 16:25 - 23:30 \\
    & 2015 Apr 05 &  Ku & 09:15 - 16:50 & 185 & 15:10 - 22:30 \\

\hline  

A   & 2015 Aug 15 &  S   & 03:00 - 10:30 & 390 & 16:40 - 00:00\\
    & 2015 Jul 15 &  C   & 04:25 - 11:55 & 370 & 17:00 - 00:15 \\
    & 2015 Jul 14 &  X1\tablenotemark{b}  & 04:25 - 11:42 & 350 & 16:40 - 00:00 \\
    & 2015 Aug 11 &  X2  & 01:00 - 08:25 & 365 & 15:10 - 22:30 \\
    & 2015 Jun 29 &  Ku  & 03:40 - 11:15 & 370 & 15:10 - 22:35 \\
\hline  
\hline
\end{tabular}
\end{center}
\tablenotetext{a}{Time spent on source.}
\tablenotetext{b}{The X-band
  The A-configuration observation on July 14 was repeated on August 11
  because of a system failure in the control computers which caused
  the cross-hand (polarization) data to be uncalibratable.  All
  cross-hand data from this observation were flagged.  The
  parallel-hand data were not affected, and were retained.}
\end{table*}

The correlator was configured to span the entire frequency range for
each band.  To facilitate data editing (primarily for RFI), the time
averaging was set to $2$ seconds, and the channel width to $2$ MHz for
C, X, and Ku bands, and to $1$ MHz at S band.  Following editing and
calibration, the data were time- and frequency-averaged, as described
below.

A standard observing regimen was utilized -- a nearby calibrator,
J2007+4029 was observed periodically to monitor system health and to
provide both complex gain and antenna cross-polarization
calibration. At S-band in D-configuration, the visibility from Cygnus
significantly perturbs observations of J2007+4029, so the more distant,
but weaker, calibrator J2023+5427, was utilized for that band and
configuration.

All data reduction was done using the Astronomical Image Processing
System ({\tt AIPS}) software package. The procedures followed were
standard, and only a brief summary is given below:
\begin{enumerate}
 \item To reduce spectral ringing due to RFI signals, the visibilities
   were Hanning smoothed.
\item Antennas which are shadowed at low observing elevations were identified and flagged.
\item Data corrupted by system failures or RFI were identified and removed.
\item Delay and bandpass calibration for each antenna was done using 3C286.
\item The flux densities of J2007+4029 or J2023+5427 were established
  by bootstrapping from 3C286.
\item The complex gains were computed from the calibrators and applied to all sources.
\item The cross-hand delays were found using 3C286, and applied to all sources. 
\item Antenna cross-polarization (D-terms) were computed using the calibrator observations, and applied.
\item The cross-hand phase (to set the position angle of the linearly
  polarized flux) was determined from observations of 3C286, and
  applied to all sources.
\end{enumerate}
Following these procedures the data were then averaged in time and
frequency to generate the databases actually used in the imaging.  The
time and frequency averaging employed varied with the observing band
as described below.

Time averaging reduces the visibility amplitude due to phase rotation
of a source which is offset from the phase tracking center. We adopted
a criterion of a maximum loss of $10\%$ on the longest baseline for a
point source offset by $70\arcsec$ -- the location of the western
hotspot with respect to the nucleus.  The resulting condition is
\begin{equation}
  \delta t =  \frac{\lambda}{4 \, B_{\text{max}} \,\omega \,\chi}
\end{equation}
where $B_{max}$ is the maximum baseline in meters, $\omega$ is the
earth's rotation rate in radians s$^{-1}$, $\chi$ is the offset in
rad, and $\lambda$ is the observing wavelength in meters.  For Cygnus
A, and the $35$-km maximum baseline, the integration time is $22$,
$11$, $7.1$ and $5$ seconds, for the S, C, X, and Ku-bands,
respectively. We utilized $10$ seconds for S and C, $8$ seconds for X,
and $6$ seconds for Ku bands.  Note that since the hotspots are well
resolved, the actual loss of brightness due to this effect will
be considerably less than the 10\% condition utilized.

There are two conditions which set the maximum tolerable channelwidth
-- chromatic aberration (bandwidth smearing), and polarization
intensity reduction due to the rotation measure.  We discuss these in
turn:

\begin{enumerate}
 \item Frequency averaging radially stretches the emission from a
   source, with increasing offset from the phase tracking center.  We
   applied the same condition as for time averaging -- that the
   longest baselines suffer a maximum of $10\%$ loss in amplitude for
   a point-source $70\arcsec$ from the nucleus. This results in the
   following:
\begin{equation}
\delta\nu =  \frac{c}{4 \, B_{\text{max}} \, \chi} 
\end{equation}
For the $70\arcsec$ offset of the western hotspot and the $35$-km
maximum baselines, the maximum channel width is $6.3$ MHz. We adopted
$8$ MHz.  We note again that as the hotspots are well resolved on the
longest baselines, the actual intensity loss is much less than the
10\% criterion.
\item As discussed in Section 2, a linearly polarized EM wave,
  propagating through a magnetized ionized medium has its plane of
  polarization rotated by
\begin{equation}
 \Delta\chi = RM \lambda^2
\end{equation}
radians.  Over a bandwidth $\delta\nu$, centered at frequency $\nu$,
the rotation of the plane of polarization is given by
\begin{equation}\label{eqn:rmavg}
\Delta \chi =  2\lambda^2RM\frac{\delta\nu}{\nu} 
\end{equation}
where it has been assumed that $\Delta\nu << \nu$.  The absolute value
of the maximum $RM$ in Cygnus A is known to be $\sim5000$ rad m$^{-2}$ \citep{1987DREHER}.
With this value, and utilizing the stringent condition of a maximum of
$10$ degrees rotation across a channel, the maximum channelwidth is
$1.5$ MHz at $2$ GHz, $5.2$ MHz at $3$ GHz, and $12.4$ MHz at $4$ GHz.
Hence, only at S-band is the polarization condition more stringent
than the chromatic aberration condition.  Thus, we adopted $2$ MHz for
the lower half of S-band ($2$ -- $3$ GHz), $4$ MHz for the upper half
of S-band ($3$ -- $4$ GHz), and $8$ MHz for all other bands.
\end{enumerate}

This data compression process resulted in a database for each spectral
window (spanning $64$ MHz at S-band, and $128$ MHz in all other bands)
-- hence $32$ channels for the lower half of S-band, and $16$ channels
for all other bands in each spectral window and each configuration --
a total of $498$ databases.  These were then combined over
configuration, resulting in 144 databases, each containing the data
from all configurations for a single spectral window, from which the
images were made.

Although Cygnus A has a very nearby calibrator, standard phase
calibration alone results in image dynamic range (ratio of the
brightest component to the rms noise) of less than $1000$ , far less
than that needed for detailed imaging analysis.  Fortunately, the high
flux density of Cygnus A, combined with the `sharp' features of the
nucleus and hotspots, permits application of the technique of
self-calibration \citep{1981CORNWELL}.  To accomplish this, total
intensity images of Cygnus A were generated from each of these
decimated databases for each spectral window, utilizing a few central
channels.  These images were then used to self-calibrate the data to
remove residual phase and gain fluctuations inherent in the process of
calibration with an external calibrator.  During this process, care
was taken to ensure the unresolved nucleus was registered in the
central (phase-tracking) cell, and that the total flux density of
Cygnus A (following correction for the primary beam) equaled the
values published by \citet{2017PERLEY}.  We believe the phase
registration is accurate to $5$\% of the full resolution at all bands,
and that the flux density matches the \citet{2017PERLEY} scale to
better than $2$\%.  The resulting images showed imaging dynamic ranges
up to $30,000$:$1$ -- more than an order of magnitude better than the
initial images.

\section{Imaging}\label{sec:imaging}
Imaging was done using the {\tt AIPS} software package, using the task
{\tt IMAGR}. We made Stokes $I$, $Q$ and $U$ image cubes using the
multiscale cleaning deconvolution algorithm \citep{
  2003GREISEN, 2008CORNWELL, 2009GREISEN}. The number of frequency planes utilized to
prevent depolarization from the high $RM$ or from chromatic
aberration, for each frequency band is shown in Table \ref{tab2}. In
total, $1184$ frequency planes were used.

\begin{table}
 \centering
 \caption{The number of frequency planes in each band utilized to
   avoid Faraday depolarization. \label{tab2}}
 \begin{tabular}{ c c c c c }
 \hline 
 Band & $\nu$-interval & $\Delta \nu$  &  $N_{avg}$ & $N_{planes}$\\
 &  [GHz] & [MHz]  &  & \\
 \hline  \hline 
 $\mathrm{S_{lo}}$ & 2-3 & 2 & 1 & 512\\
 $\mathrm{S_{hi}}$ & 3-4 & 4 & 1 & 256\\
 $\mathrm{C_{lo}}$ & 4-6 & 8 & 1 & 256\\
 $\mathrm{C_{hi}}$ & 6-8 & 32 & 4 & 64\\
 $\mathrm{X_{lo}}$ & 8-10 & 64 & 8 & 32\\
 $\mathrm{X_{hi}}$--Ku & 10-18 & 128 & 16 & 64 \\ 
 \hline \hline 
 \end{tabular}
\end{table}

The image cubes were made at two standard resolutions: i)
$0.75\arcsec$ resolution -- corresponding to the highest resolution
available at $2$ GHz (which thus includes all frequencies) and ii)
$0.30\arcsec$ which corresponds to the highest resolution at $6$ GHz. All the images were primary beam corrected using the AIPS task 'PBCOR'.

The polarized intensity $P$ is a complex number, with amplitude given
by $|P|=\sqrt{Q^2+U^2}$, and phase by
\begin{equation}\label{eqn:polA}
 \chi = \frac{1}{2} \, \text{tan}^{-1} \bigg(\frac{U}{Q}\bigg).
\end{equation}
The amplitude was corrected for Ricean bias using the Maximum
Likelihood approximation:
\begin{equation}
 P_{\text{corr}} \approx P - \frac{0.5\, \sigma^2_{P}}{P} 
\end{equation}
where $\sigma_P$ is the error in $P$ \citep{1986KILLEN}. We derived
fractional polarizations by taking the following ratio
\begin{equation}\label{eqn:p}
p=\frac{P_{\text{corr}}}{I}. 
\end{equation}
Errors associated with these quantities were derived assuming standard
propagation of error formulae with the noise in $Q$, $U$ and $I$
estimated in an off-source region of each map. At 0.30$\arcsec$ the off-source noise ranges between 0.18 mJy beam$^{-1}$ and 0.5 mJy beam$^{-1}$ in Stokes $Q$, and $U$, and 0.3 mJy beam$^{-1}$ and 1 mJy beam$^{-1}$ in Stokes $I$. At 0.75$\arcsec$ 
the off-source noise ranges between 0.6 mJy beam$^{-1}$ and 1.6 mJy beam$^{-1}$ for Stokes $Q$ and $U$ images, and 0.8 mJy beam$^{-1}$ and 6 mJy beam$^{-1}$ for Stokes $I$ images.

We computed Faraday spectra, $\tilde{F}$, for every line of sight using
the RM-synthesis technique as employed by \citet{2005BRENTJENS}. We
compute $\tilde{F}$ by summing the fractional polarized
emission, $p$, over all $N$ channels as follows:
\begin{equation}
 \tilde{F} (\phi) = \Bigg (  \sum_{n=1}^{N} W_n \Bigg)^{-1}
 \sum_{n=1}^{N} \tilde{p}_n e^{-2i \phi (\lambda^2_n -\lambda^2_0)},
\end{equation}
 where $\lambda^2_n$ is the wavelength squared of channel $n$, $W_n =
 W(\lambda^2_n)$ is the weight of each channel, $\tilde{p}_n = W_n
 p_n$, $\lambda^2_0$ is a weighted offset in wavelength-squared space,
 and $\phi$ is Faraday depth, given by
 
 \begin{equation}\label{eqn:rm}
  \phi = 810\int\limits_{z}^{0}n_e\mathbf{B}\cdot\mathrm{dz}
  \quad \quad [\mathrm{rad\, m}^{-2}].
\end{equation}
Here, $\mathbf{B}$ is the magnetic field in $\mu$Gauss, $n_e$ is the
electron density\footnote{Strictly speaking this is the difference
  between the electron and positron densities: $n_--n_+$.  Here we
  assume the cluster gas is comprised of electrons and protons.} in
$\mathrm{cm}^{-3}$, and $z$ is the pathlength in kpc. The derived
$\tilde{F}$ is a `dirty' spectrum; the true $F(\phi)$ convolved by the
rotation measure transfer function, RMTF or $R(\phi)$:
\begin{equation}
 R(\phi) = \Bigg (  \sum_{n=1}^{N} W_n \Bigg)^{-1}
 \sum_{n=1}^{N} W_n e^{-2i \phi (\lambda^2_n -\lambda^2_0)}.
\end{equation}

\begin{figure}
 \centering
 \begin{minipage}[b]{1\linewidth}
    \includegraphics[width=1.0\linewidth]{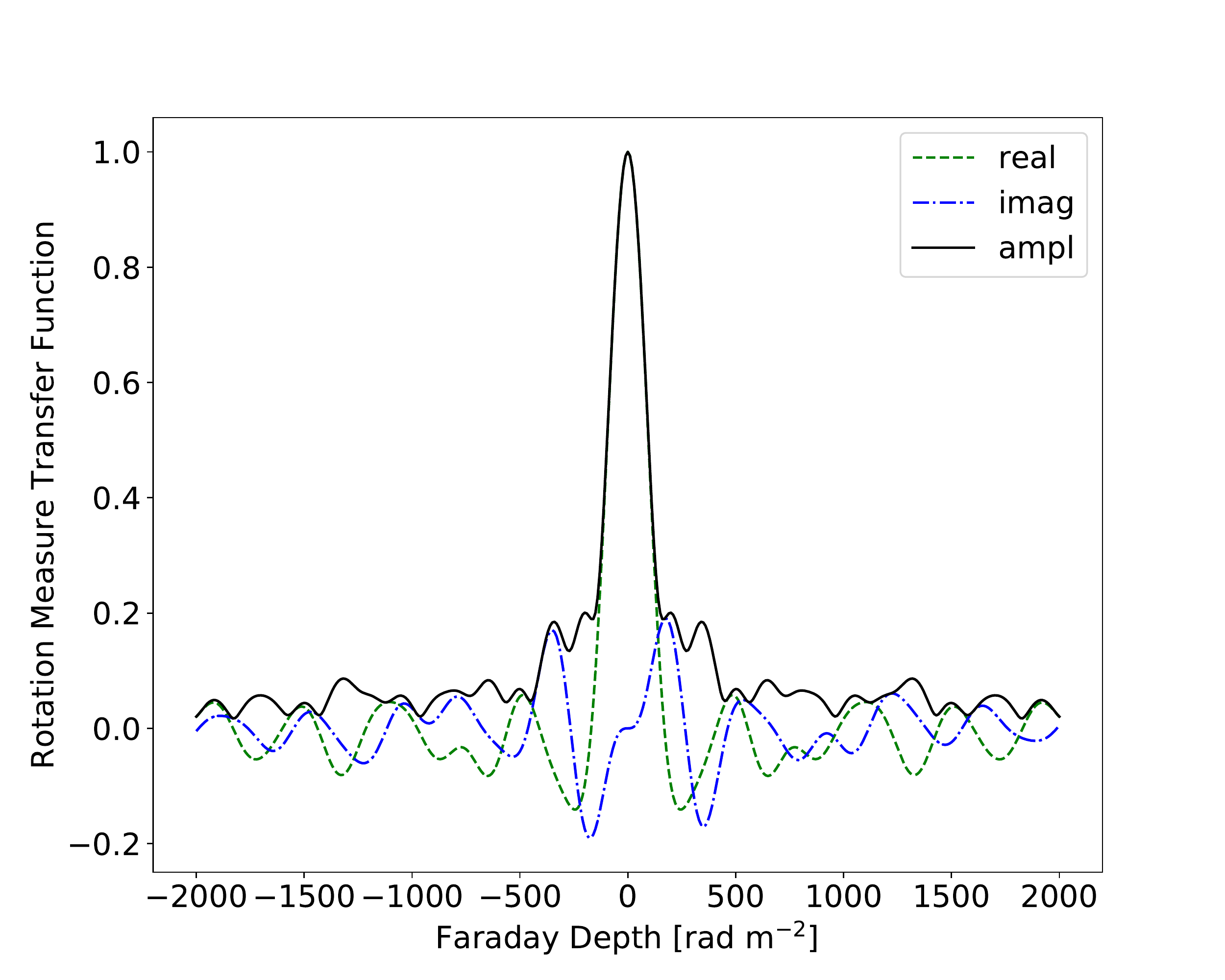}
  \end{minipage}
  \begin{minipage}[b]{1\linewidth}
    \includegraphics[width=1.0\linewidth]{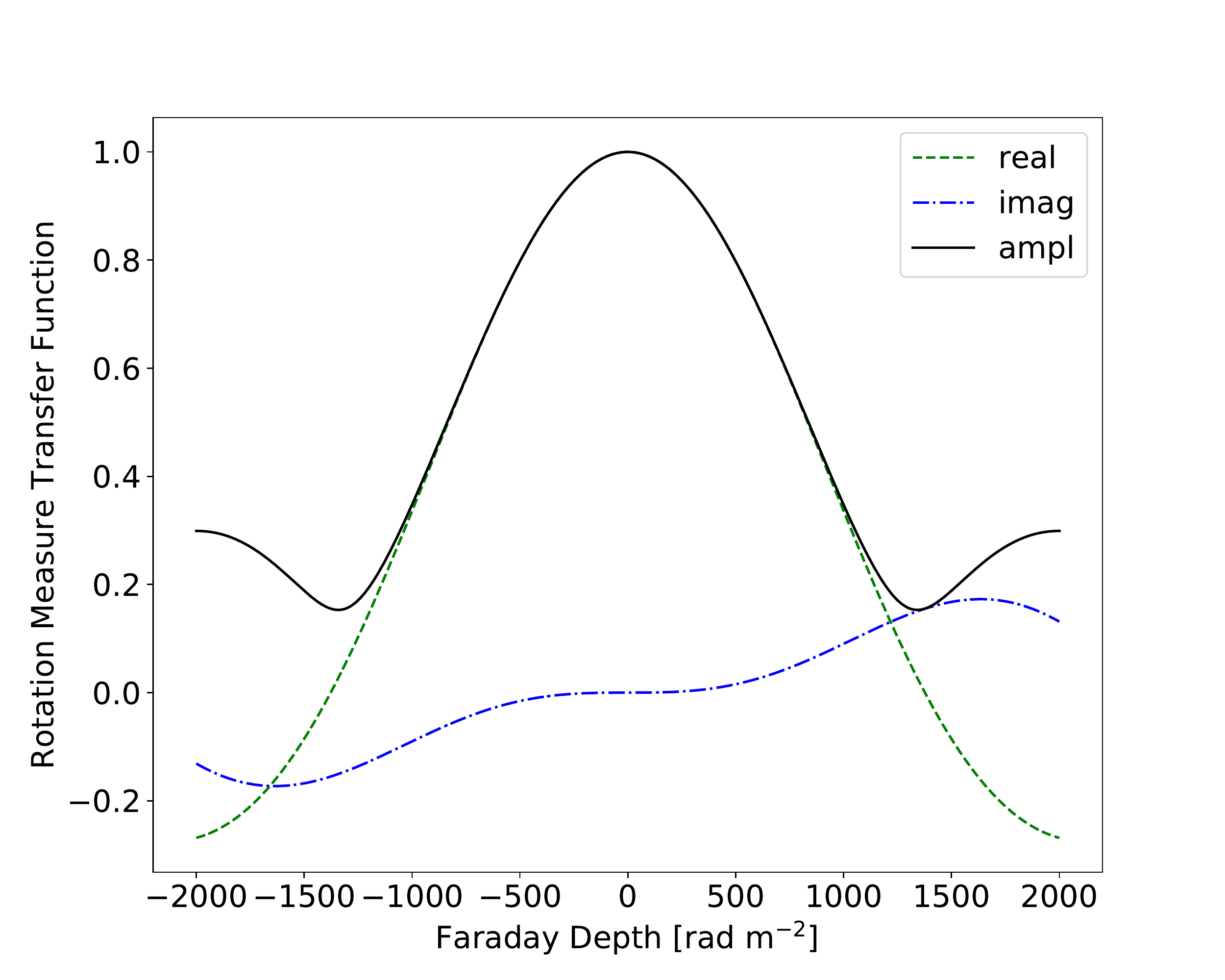}
  \end{minipage}
 \caption{Rotation measure transfer functions of our data. Top is
   2 - 18 GHz data, and bottom is 6 - 18 GHz. \label{fig:rmtf}}
\end{figure}

The resolution of the RMTF is 
\begin{equation}
 \delta \phi \approx \frac{3.8}{\Delta \lambda^2} \quad \text{rad m$^{-2}$}, 
\end{equation}
where $\Delta \lambda^2 = \lambda^2_{\text{max}} -
\lambda^2_{\text{min}}$ is the width in the observed
wavelength-squared space \citep{2009SCHNITZELER}. For our data,
$\delta\phi$ $\sim$175 rad m$^{-2}$ for 2 - 18 GHz, and $\delta \phi$
$\sim$1700 rad m$^{-2}$ for 6 - 18 GHz. Figure \ref{fig:rmtf} shows
the rotation measure transfer function for our data.

The Faraday spectra shown in this paper were deconvolved using the
cleaning algorithm presented by \citet{2009HEALD}. We utilized both
RM-synthesis and RM-clean provided in the AIPS task `TARS'. Our $\lambda^2$
sampling is non-uniform -- resampling the data on a uniform grid did
not alter the results in any significant way. We used uniform
weighting, $W=1$, for all channels, and defined $\lambda^2_0$ as a
weighted mean of the observed $\lambda^2$.

The maximum Faraday depth we can observe without significant
attenuation is given by the phase rotation across a channel at the
lowest frequency (see Eq. \ref{eqn:rmavg}).  An RM of 25000 rad m$^{-2}$ across
a 2 MHz channel width at a frequency of 2 GHz will cause a rotation of
1 radian.  Thus, the maximum RM in Cygnus A of $\sim$ 5000 rad m$^{-2}$ will
suffer no appreciable attenuation.  As noted by \citet{2005BRENTJENS},
the short wavelength limit of the observations will limit the width of
Faraday depths that can be recovered by Faraday synthesis for any
given line of sight.  Our minimum $\lambda^2$ of 3$\times$10$^{-4}$
m$^2$, would attenuate by 50\% Faraday structures wider than 10000 rad m$^{-2}$
-- far wider than those expected in Cygnus A.

\section{Polarization Imaging of Cygnus A}\label{sec:newdata}
Polarized emission of a source undergoing Faraday rotation can be used
to study the properties of the medium causing the rotation.
Distinguishing between various depolarization mechanisms can be
difficult, and requires high resolution, low frequency observations.
In this section, we show the polarization characteristics of Cygnus A
as a function of resolution and frequency.

\subsection{Polarization as a Function of Frequency}\label{polfreq}
\begin{figure*}
\centering
   \begin{minipage}[b]{0.43\linewidth}
    \includegraphics[width=1.10\linewidth]{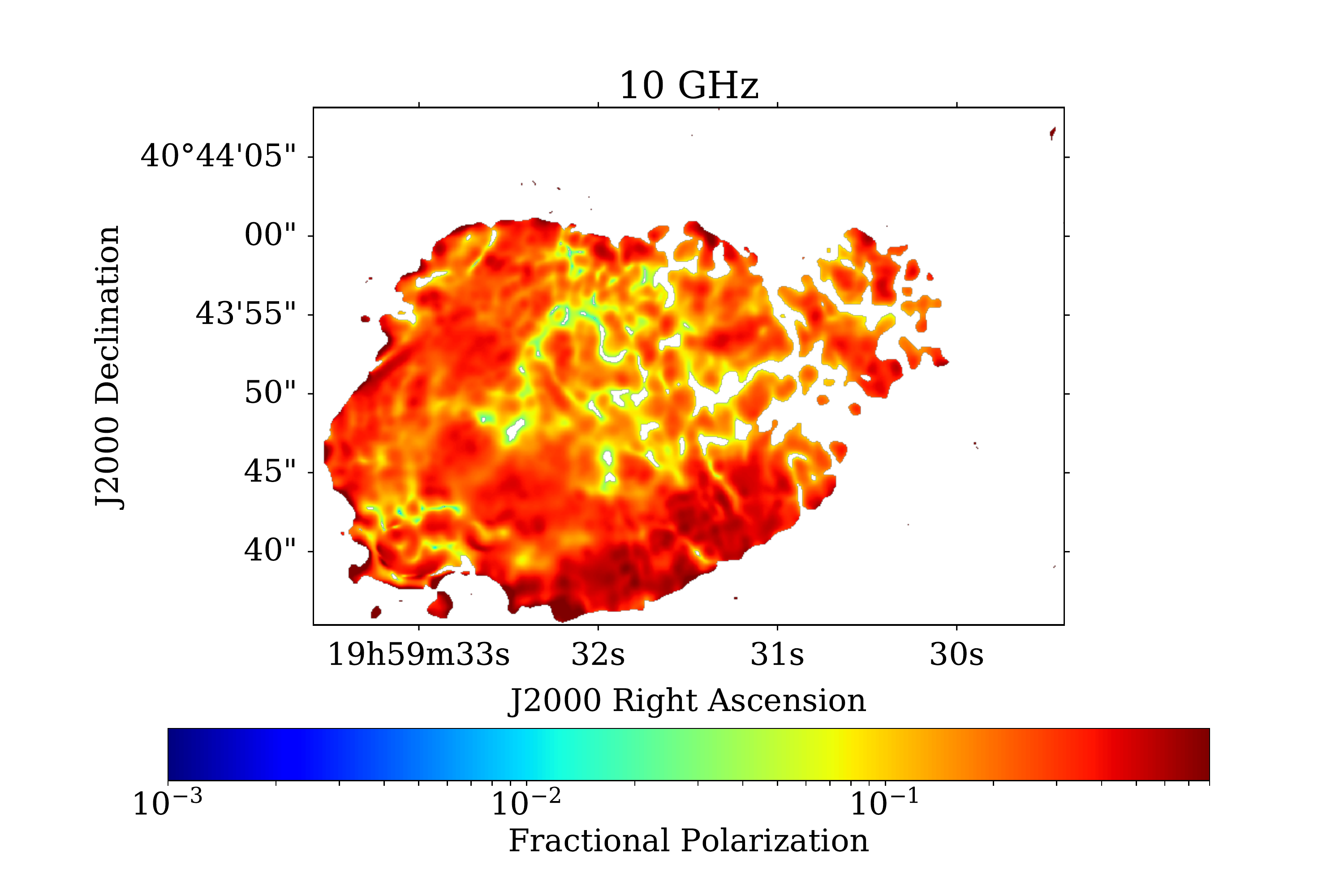}
  \end{minipage}
  \begin{minipage}[b]{0.43\linewidth}
    \includegraphics[width=1.10\linewidth]{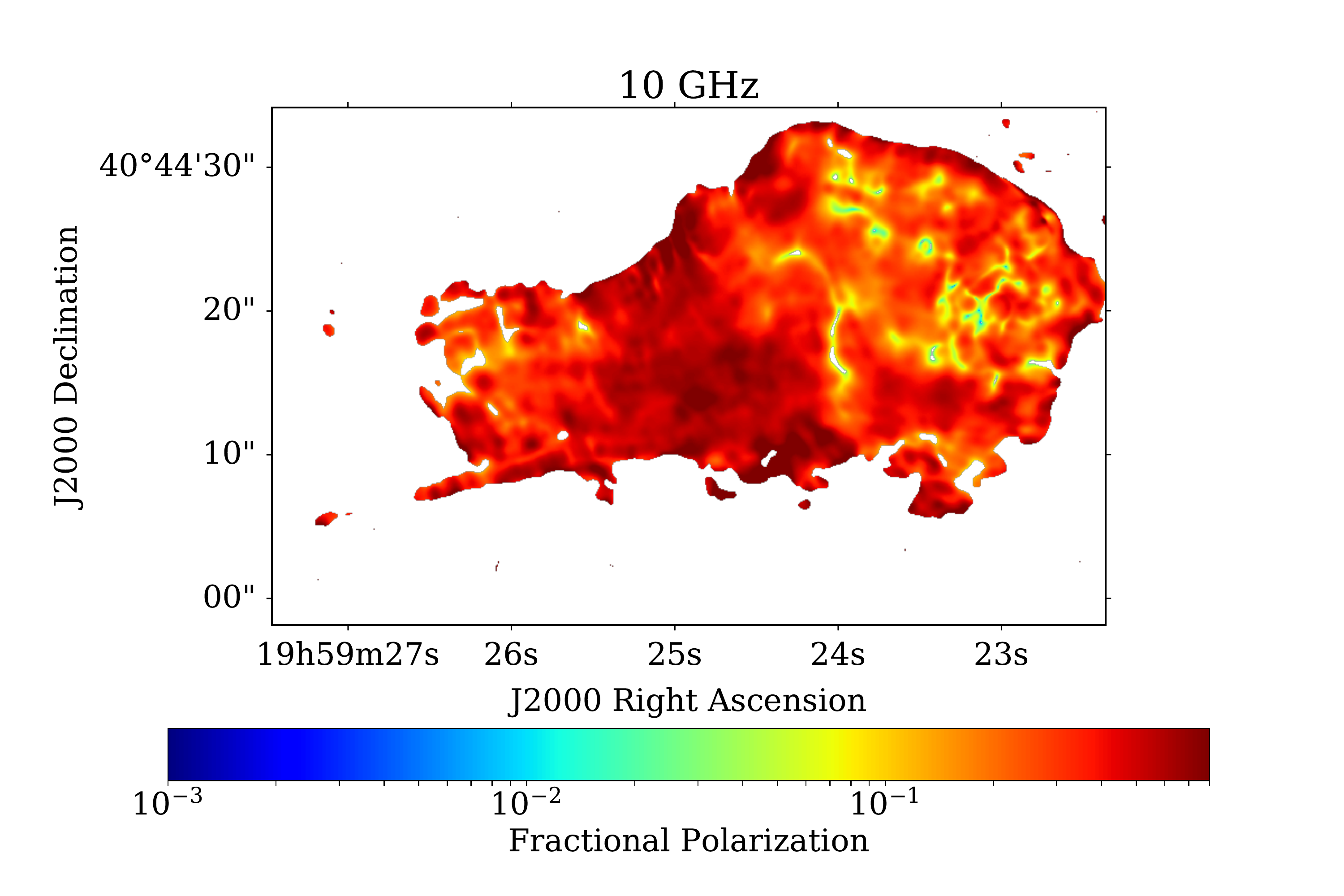}
  \end{minipage}
   \begin{minipage}[b]{0.43\linewidth}
    \includegraphics[width=1.10\linewidth]{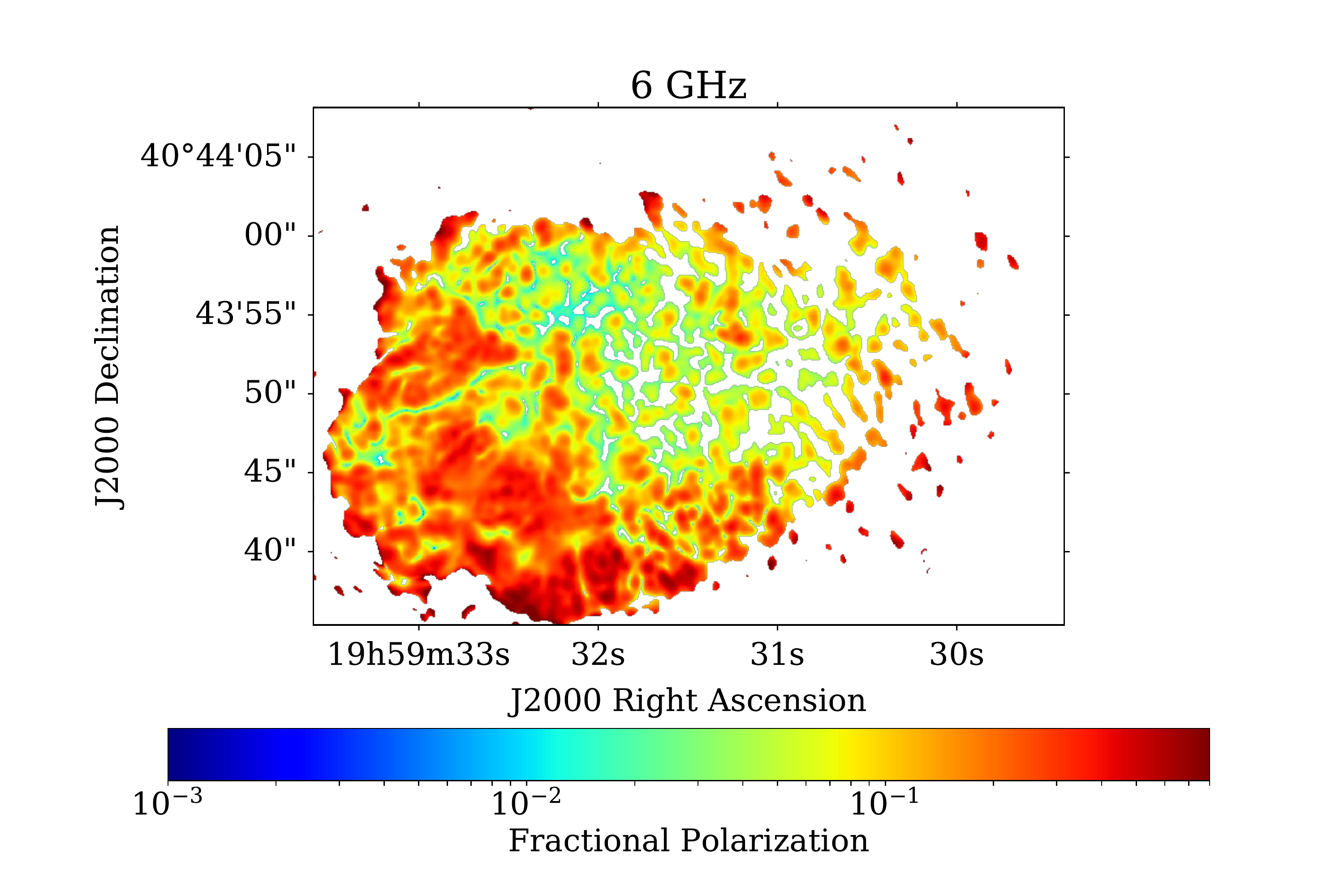}
  \end{minipage}
  \begin{minipage}[b]{0.43\linewidth}
    \includegraphics[width=1.10\linewidth]{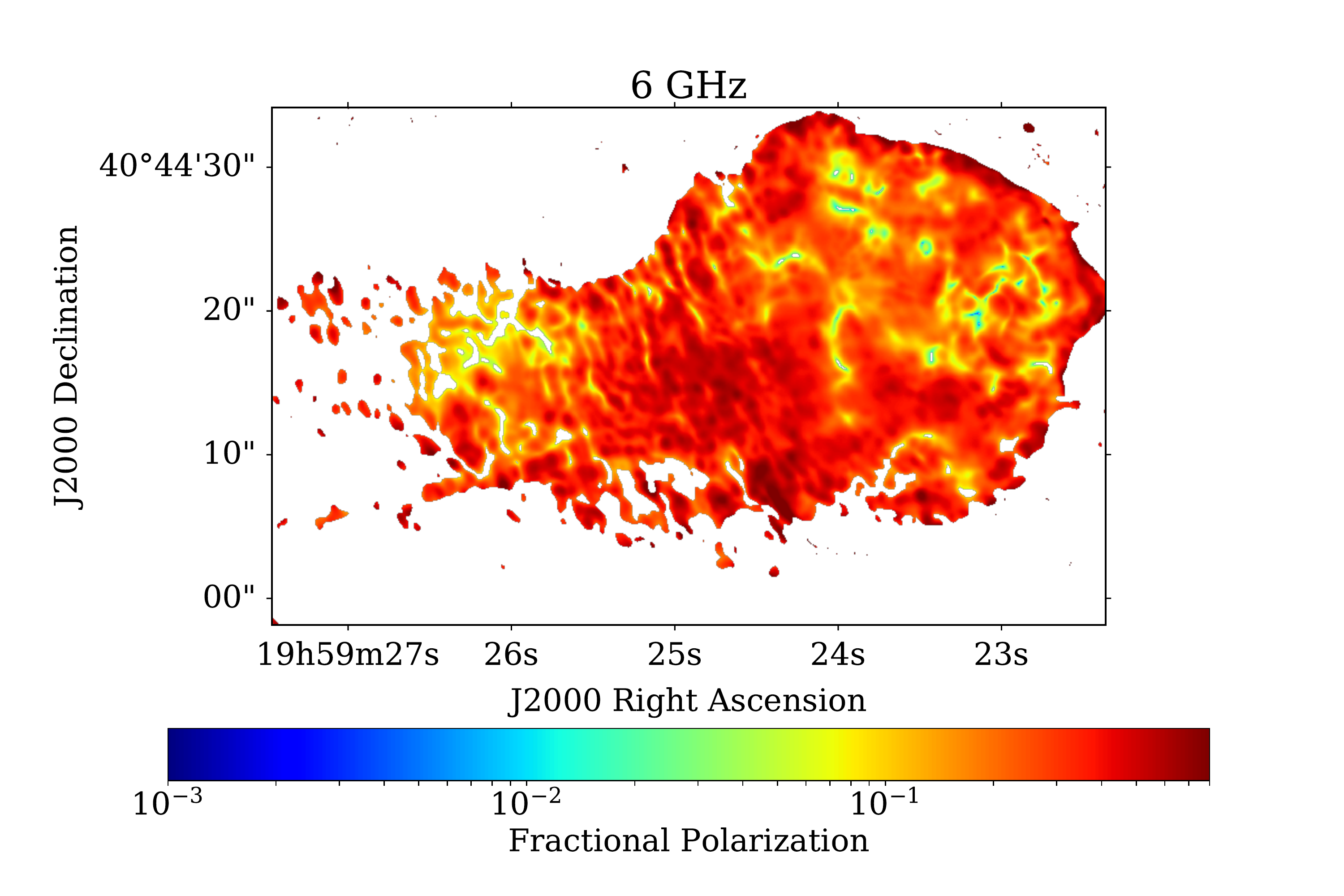}
  \end{minipage}
     \begin{minipage}[b]{0.43\linewidth}
    \includegraphics[width=1.10\linewidth]{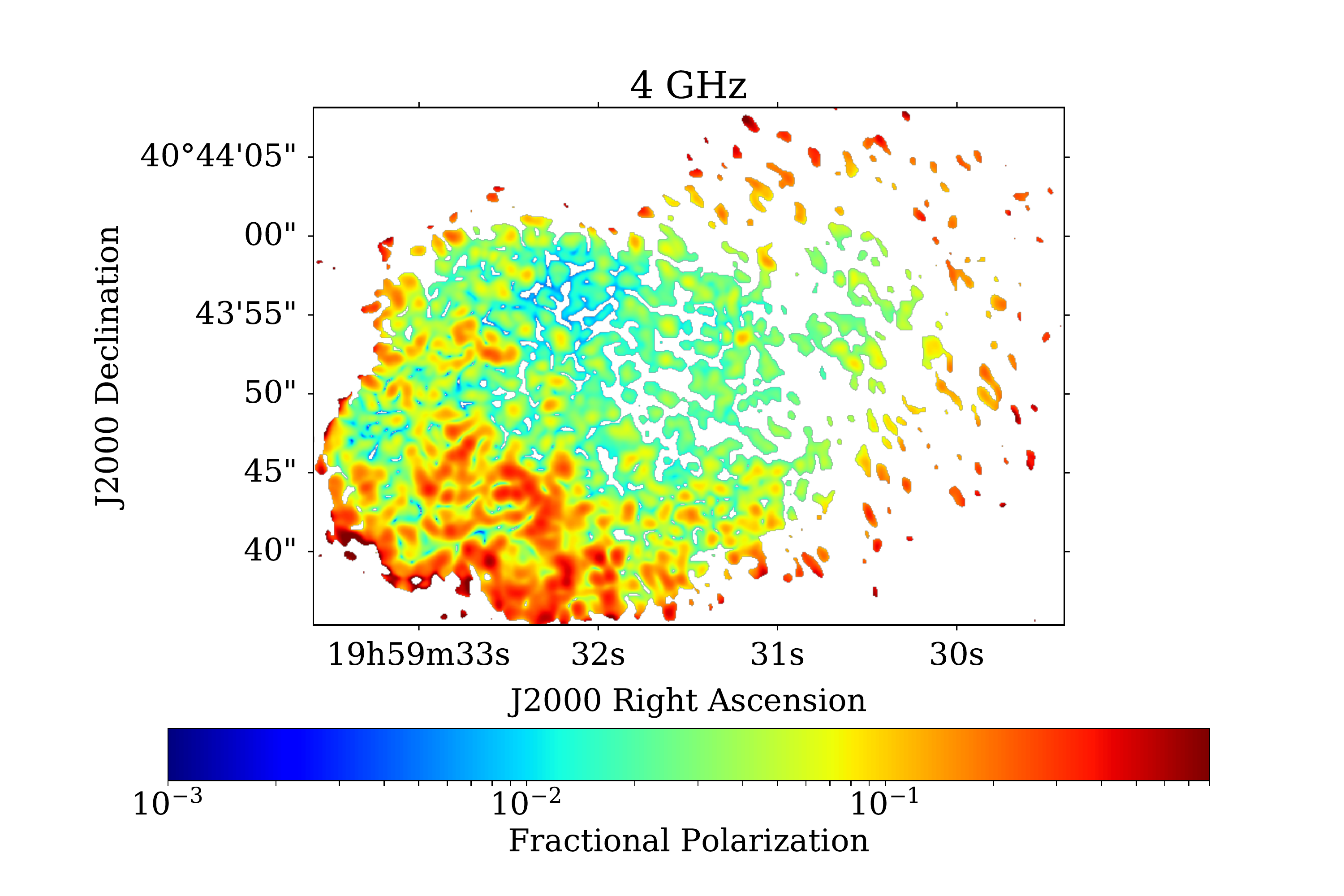}
  \end{minipage}
  \begin{minipage}[b]{0.43\linewidth}
    \includegraphics[width=1.10\linewidth]{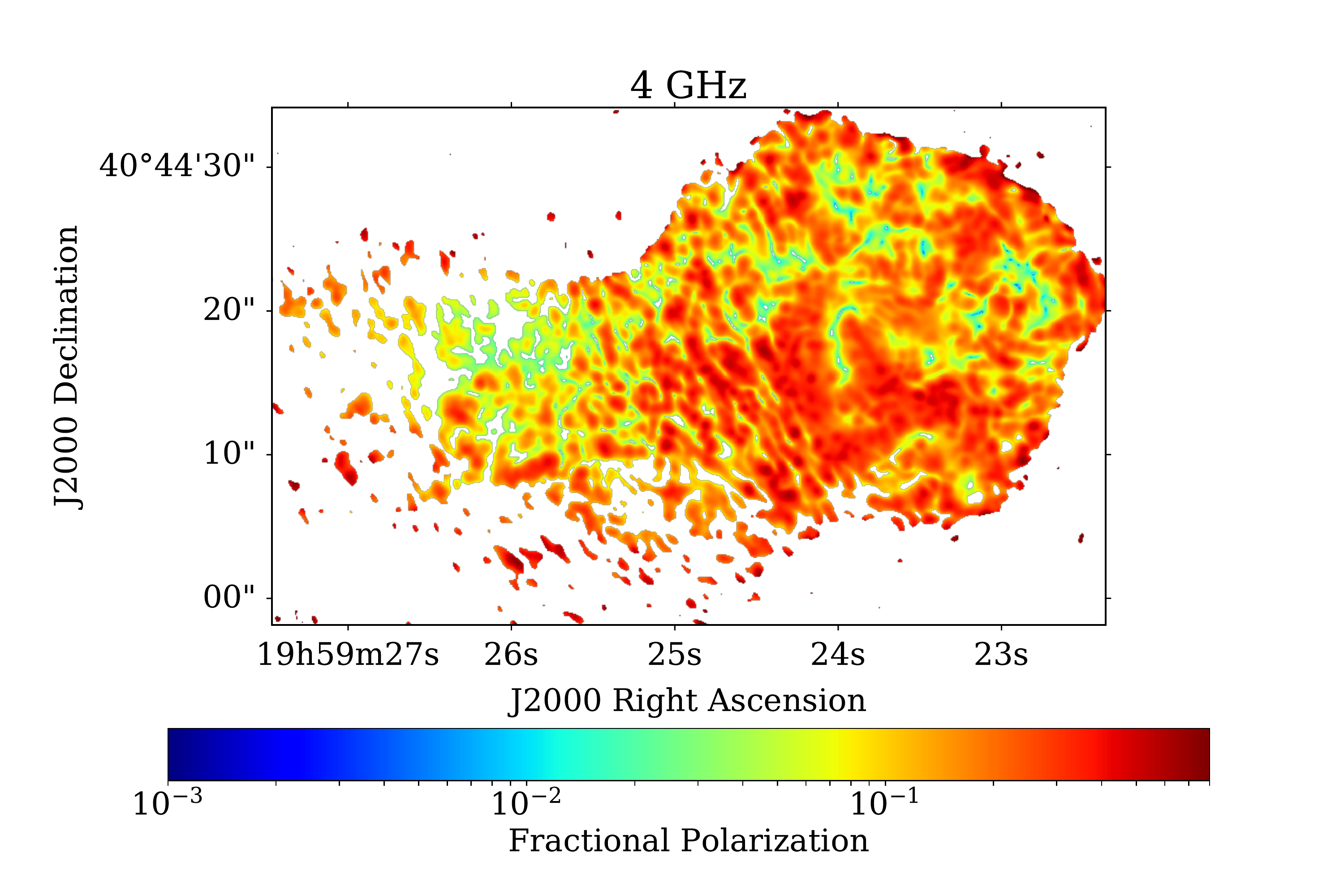}
  \end{minipage}
     \begin{minipage}[b]{0.43\linewidth}
    \includegraphics[width=1.10\linewidth]{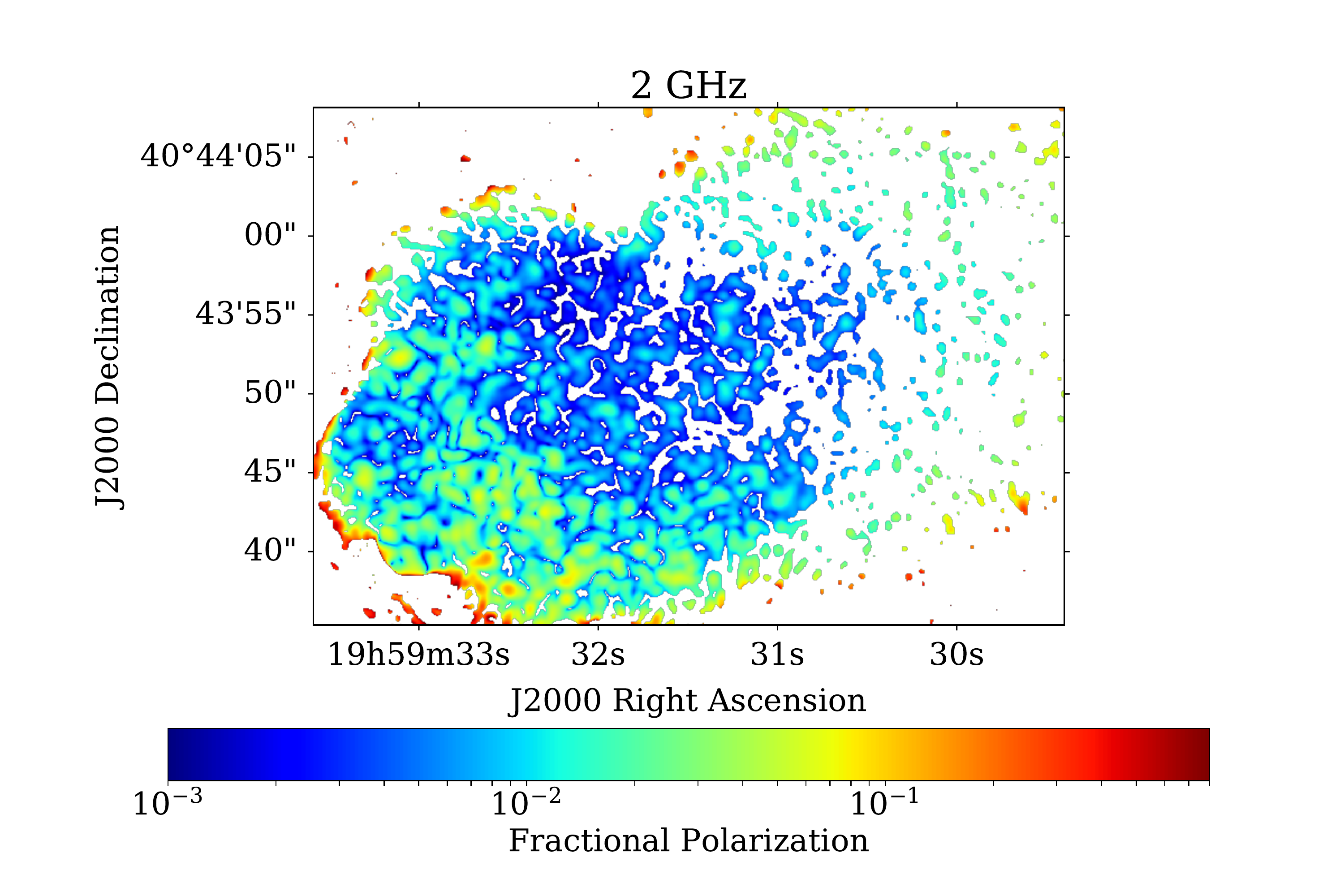}
  \end{minipage}
  \begin{minipage}[b]{0.43\linewidth}
    \includegraphics[width=1.10\linewidth]{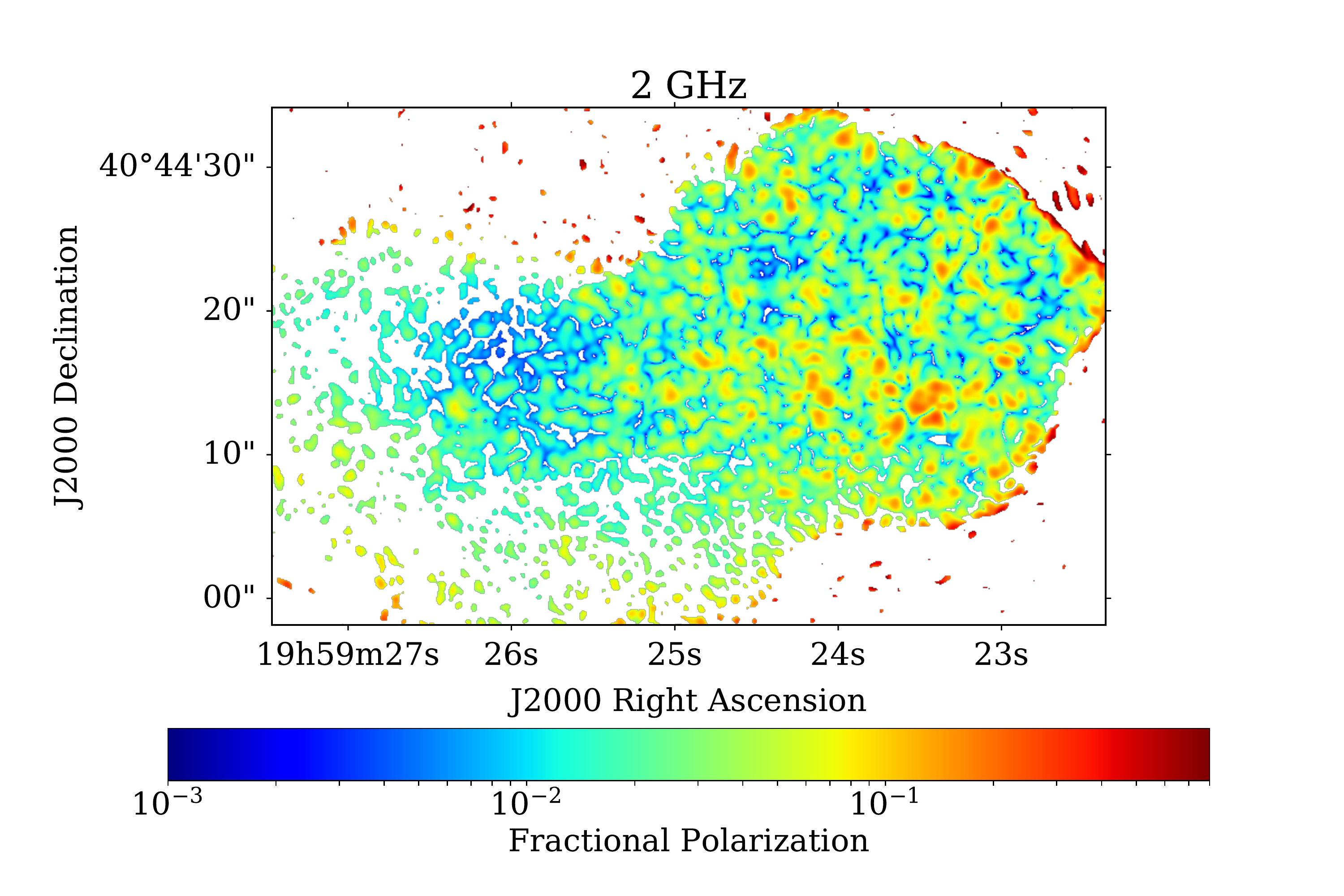}
  \end{minipage}
 \caption{Fractional polarization maps of Cygnus A at $0.75\arcsec$
   resolution at four different frequencies: $10$ GHz, $6$ GHz, $4$
   GHz and $2$ GHz. The color-bar ranges between 0.1\% and 80\%. Only
   pixels with $(\sigma_p/p) < 0.5$ are shown. There is a global
   decrease in fractional polarization with decreasing frequency. The
   eastern lobe depolarizes at higher frequency than the western.  The
   inner regions (near the nucleus) of the lobes depolarize more
   rapidly than the outer regions. \label{fig:polarizationmaps}}
  \end{figure*}

Figure \ref{fig:polarizationmaps} shows the $0.75\arcsec$ resolution
maps of the fractional polarization across Cygnus A radio lobes at 10
GHz, 6 GHz, 4 GHz and 2 GHz.  We show only those pixels with relative
error in the fractional polarization, $(\sigma_p/p) \le 0.5$. As seen
in the figure, the fractional polarization of the lobes decreases
significantly with decreasing frequency.  Note that the eastern lobe
depolarizes at higher frequencies than the western, and for both
lobes, the depolarization is more rapid in the inner regions -- those
closer to the nucleus.  The spatial distribution of the fractional
polarization of the lobes is very uniform at high frequencies, with
typical values of 20\%, with some regions as high as 70\%, and becomes
clumpy on scales of $\sim 1$ kpc at low frequencies -- most notably at
2 GHz.  The fractional polarization at 2 GHz is less than 10\% for
virtually all lines of sight at this resolution.  The emission from
the bright hotspots, and from the jet, depolarizes in the same manner
as that from the adjacent lobes.  There is no correlation between the
underlying source brightness or structure and the nature of the
depolarization.

The strong depolarization, and the very turbulent appearance of the
low frequency fractional polarization images suggests that the
depolarization may be related to partially-resolved fluctuations in
the Faraday depth of the surrounding medium.  To investigate this, we
determined the dependence of the polarization as a function of
frequency over our full bandwidth ($2-18$ GHz) at $0.75\arcsec$
resolution.  This task is made challenging by the sheer quantity of
information -- over the solid angle subtended by the lobes, there are
over $\sim 3000$ independent lines of sight. Thus, we can only show a
few representative examples.

In order to efficiently identify specific lines of sight, we have
defined a relative coordinate system, centered on the galaxy nucleus,
with units of tens of milliarcseconds. Positive is to the west and
north, and negative to the east and south. For example, a line of sight
with coordinates (2444, -1024) is $22.44\arcsec$ west and
$10.24\arcsec$ south of the nucleus. We analyze only those locations
with error in the fractional polarization $ < 10\%$ at $8$ GHz.  This
resulted in a total of $2096$ independent lines of sight.
 
\begin{figure*}
 \centering
    \begin{minipage}[b]{1\linewidth}
  \centering
    \includegraphics[width=0.6\linewidth]{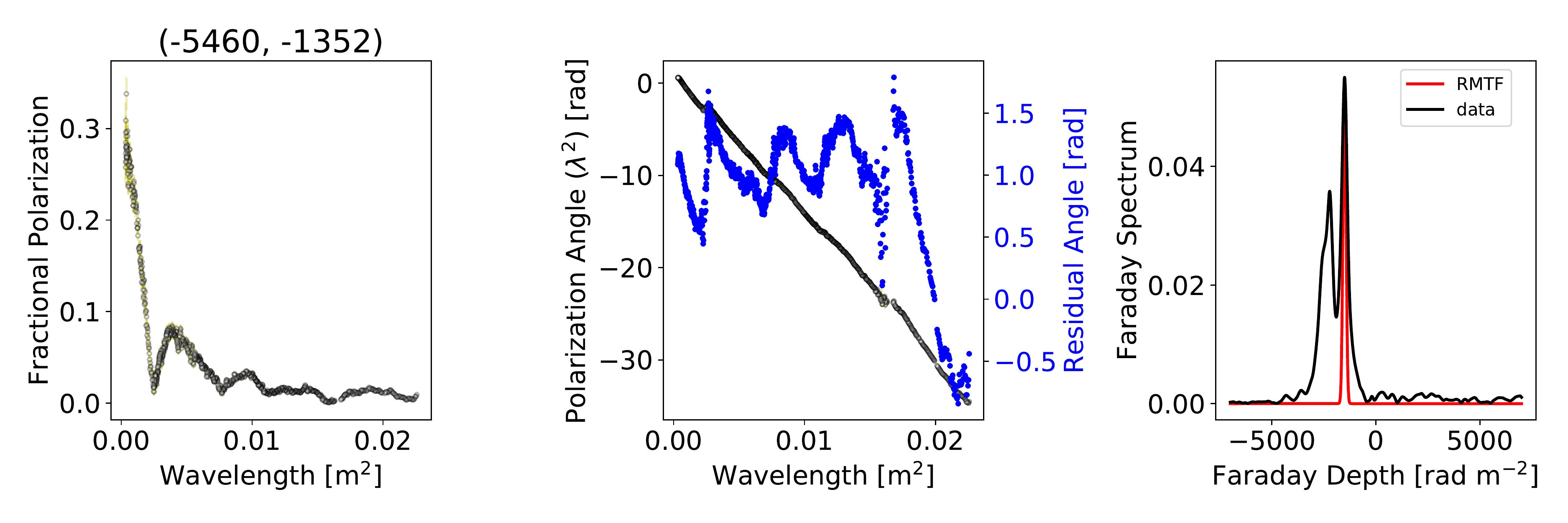} 
  \end{minipage}
  \begin{minipage}[b]{1\linewidth}
   \centering
     \includegraphics[width=0.6\linewidth]{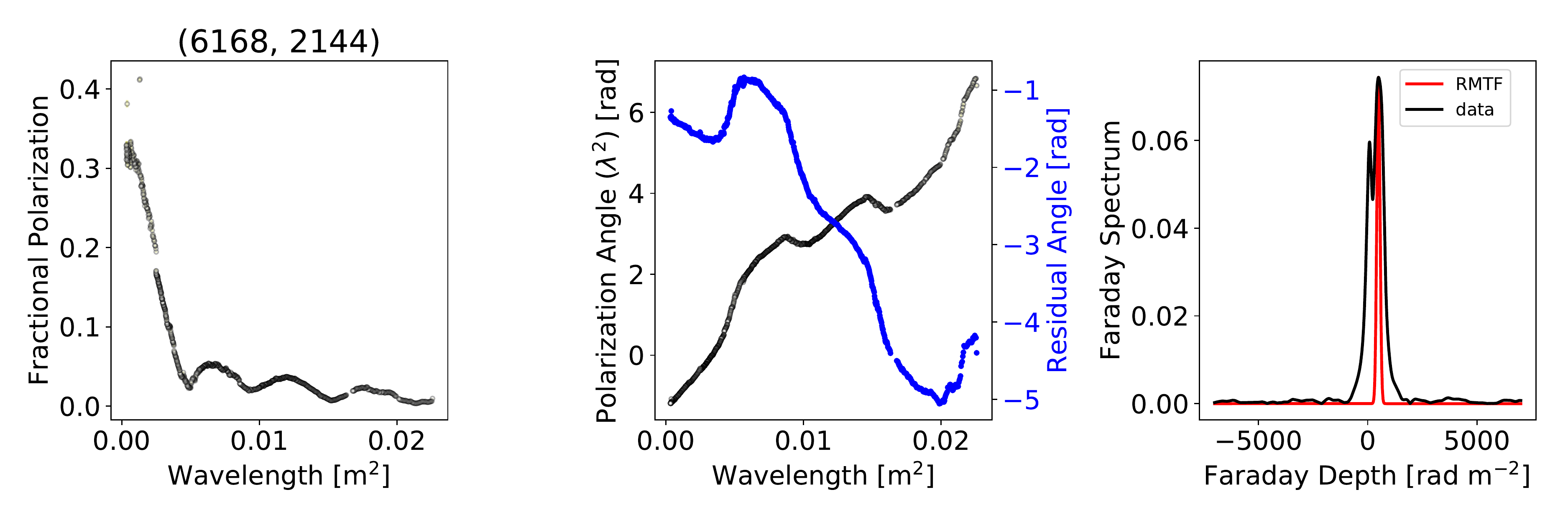}  
   \end{minipage} 
       \begin{minipage}[b]{1\linewidth}
    \centering
    \includegraphics[width=0.6\linewidth]{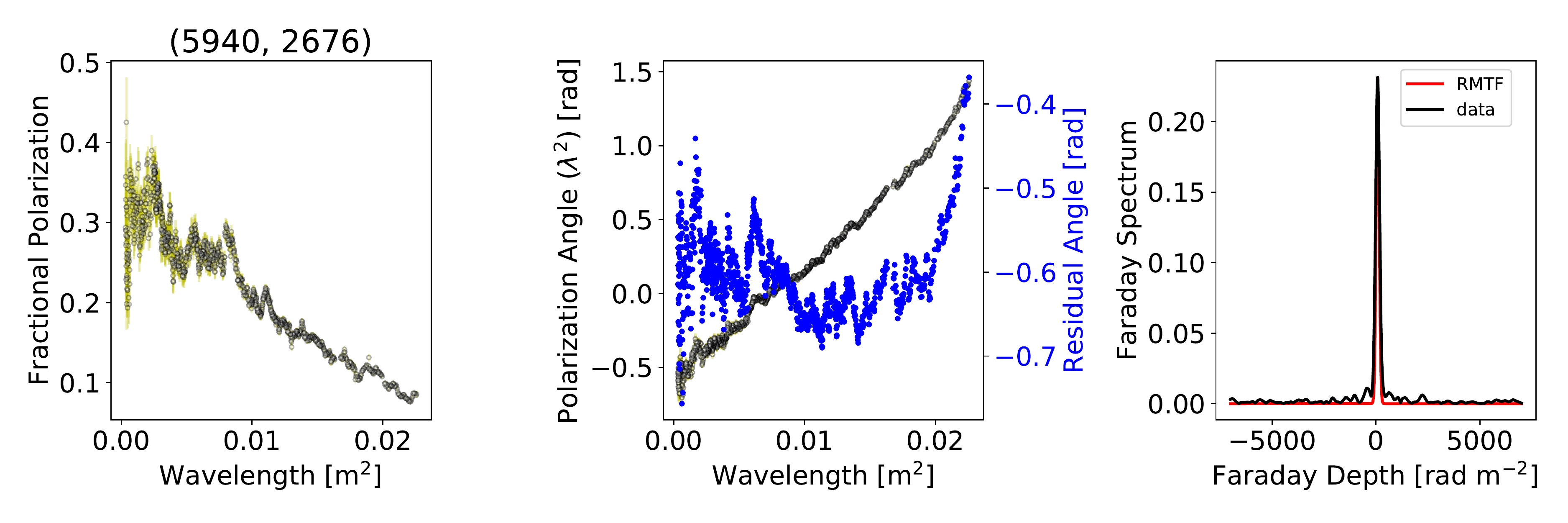}  
  \end{minipage}
  \begin{minipage}[b]{1\linewidth}
    \centering
     \includegraphics[width=0.6\linewidth]{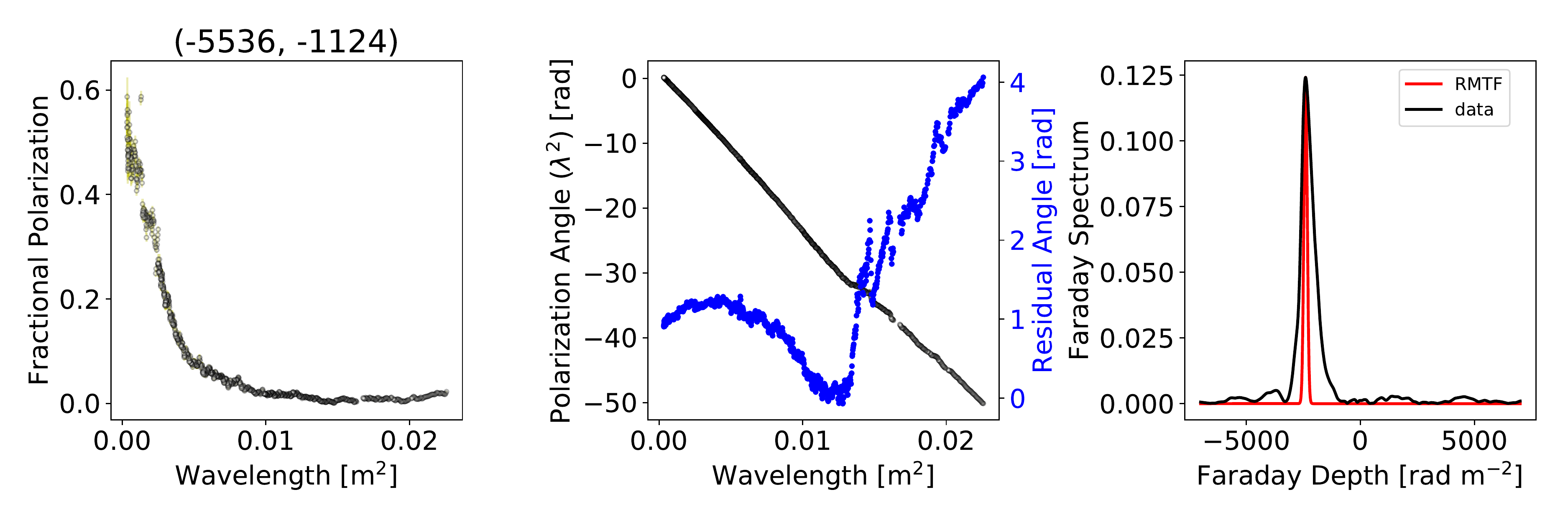} 
   \end{minipage} 
       \begin{minipage}[b]{1\linewidth}
    \centering
    \includegraphics[width=0.6\linewidth]{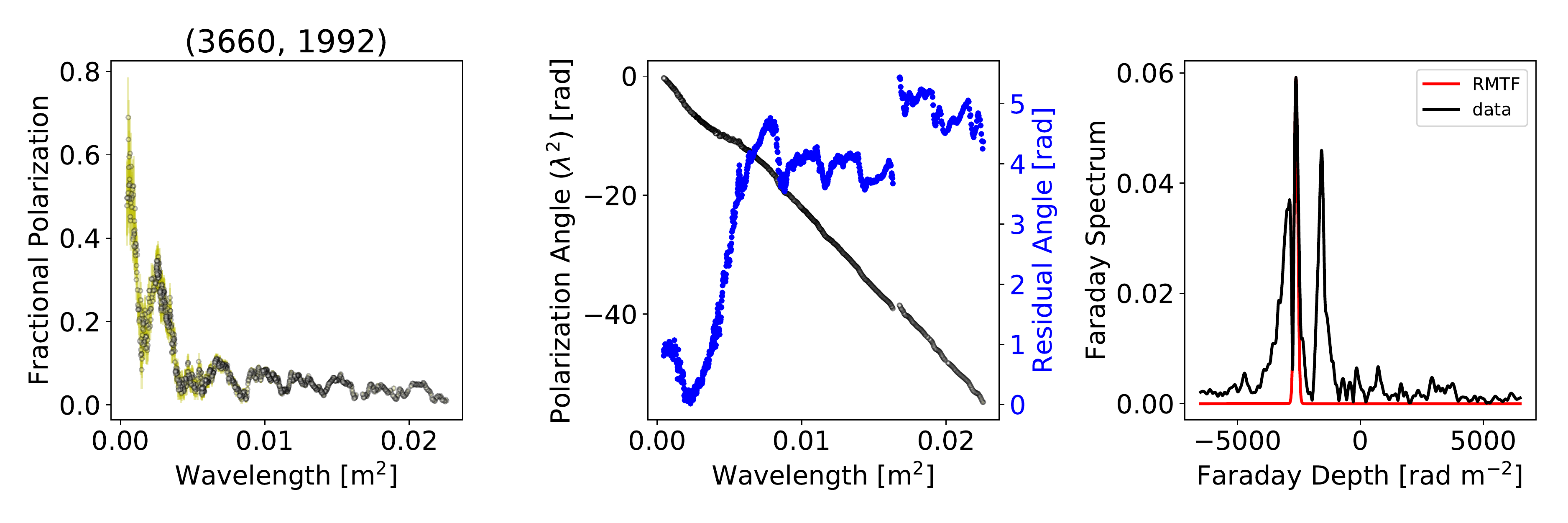} 
  \end{minipage}
  \begin{minipage}[b]{1\linewidth}
     \centering
     \includegraphics[width=0.6\linewidth]{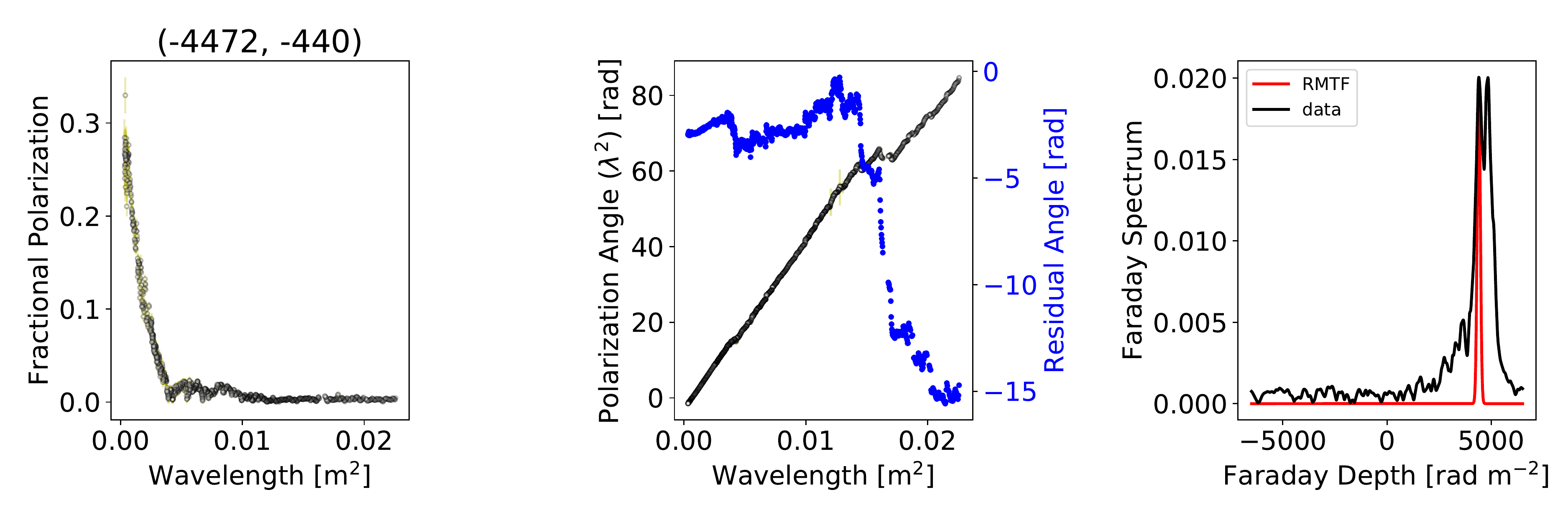}
   \end{minipage} 
  \caption{Fractional polarization from 2 to 18 GHz at
    $0.75\arcsec$ resolution for six representative locations.  The
    left column shows the fractional polarization amplitude, the middle column
    the observed position angle (black) and the residual position
    angle (blue) after the dominant peak in the Faraday spectrum was removed. The right panel shows the amplitude of the Faraday
    spectrum (in black) with a Gaussian of width equal to that of the
    rotation measure transfer function (RMTF) superposed in red.
    \label{fig:los}}
\end{figure*}

Figure \ref{fig:los} shows these depolarization functions for six
representative lines of sight. Each row of three panels displays fractional polarization (Eq. \ref{eqn:p}) in the left panel, and the polarization position
angle in the middle panel, all as functions of $\lambda^2$.  The
right panel shows the amplitude of the Faraday spectrum, superimposed in red is the Gaussian with width equal to that of the 
rotation measure transfer function (RMTF). The RMTF is shifted and scaled to match the
location and amplitude of the maximum amplitude of the Faraday
spectrum. Each line of sight is labeled according to the coordinate
offsets defined above.

The fractional polarization of all lines of sight decreases
significantly with increasing $\lambda^2$. For some lines of sight,
the decrease is smooth, while for most, there is considerable
structure in the depolarization, ranging from sharp nulls and peaks as
shown in the top two rows of the figure, to more complex
behaviors, as shown in the bottom two rows.

The polarization angle as a function of $\lambda^2$ of the data (shown
in black) is plotted together with the residual angle (shown in blue)
obtained by removing $RM_{\text{peak}}\lambda^2$ of the peak of the
Faraday spectrum. The residual polarization angle, $\chi_{\text{res}}$, shows significant
deviations; we find that 38\% of the lines of sight have $0 < \chi_{\text{res}} < \pi$, 22\% have $\pi < \chi_{\text{res}} < 3\pi$, 13\% have  $3\pi < \chi_{\text{res}} < 6\pi$, 12\% have $6\pi < \chi_{\text{res}}$, and the remaining 15\% were noisy or the n$\pi$-ambiguity couldn't be corrected. Lines of sight showing small deviations are usually associated with simple
smooth decaying fractional polarizations with $\lambda^2$ or
equivalently, with a single, nearly unresolved peak, similar to that
shown on the third row of Fig. \ref{fig:los}. The nonlinearities are
more prominent when the fractional polarization vs. $\lambda^2$
behavior has structure such as in the oscillatory and complex cases.

The Faraday spectra show interesting structures -- some lines of sight
show a simple single peak, some have multiple peaks, and many lines of
sight show a rather wide `base', indicating a wide range of Faraday
depths within the resolving beam. As expected, the depolarization
curves with pronounced oscillations show multiple-peaked spectra, with
the separation of the peaks typically $\lesssim 500$ \radm, and a
maximum separation of $\sim 1500$ \radm (approximated by eye). The detailed characterization of Faraday spectra will be presented in paper 2.
 
To investigate whether these different depolarization functions have
any spatial relationship, we generated a simple classification scheme
based on the observed depolarization behavior. By viewing each
fractional polarization as a function of $\lambda^2$ (left panel of Fig. \ref{fig:los}), we classified by eye the lines of sight into three categories as follows:

\begin{enumerate}
 \item `Sinc-like' decay: the behavior in the decreasing fractional
   polarization as a function of increasing $\lambda^2$ is
   approximately that of a sinc function: $p \propto
   \sin(K\lambda^2)/K\lambda^2$. The minima are sharp, and occur at
   nearly constant intervals, while the peaks decrease in amplitude
   towards large $\lambda^2$. Examples are shown in the top two rows
   in Fig. \ref{fig:los}.
 \item Smooth decay: the fractional polarization decreases smoothly
   with increasing $\lambda^2$ within measurement errors. Examples are
   shown in the third and fourth rows in Fig. \ref{fig:los}.
 \item Complex decay: the depolarization displays a combination of the
   above, or more complex behavior in the fractional polarization
   vs. $\lambda^2$. We further classified complex lines of sight into two sub-classes: 'complex-oscillatory' and 'complex non-oscillatory'. The former resemble sinc-like pattern than smooth decay but the fractional polarization doesn't approach 0 across $\lambda^2$ or the intervals are not periodic, or both, and the latter resemble smooth decay more than sinc-like but the decay is not completely smooth. See the last two rows in Fig. \ref{fig:los} for complex-oscillatory and non-oscillatory, respectively.
\end{enumerate}

The fraction of lines of sight in each of these three classes are given the second column of Table \ref{tab2}. Moreover, column 3 to 7 gives a fraction of lines of sight with deviations in polarization angle (the residual angle, $\chi_{\text{res}}$) in each of the intervals: low ($0 < \chi_{\text{res}} < \pi$), med ( $\pi < \chi_{\text{res}} < 3\pi$), high ( $3\pi < \chi_{\text{res}} < 6\pi$), and limit( $6\pi < \chi_{\text{res}}$), as well as a fraction of those that couldn't be classified (denoted as 'N/A'). Most lines of sight are complex/intermediate -- with the majority of the complex resembling smooth decays. In terms of the residual $\chi$, we find that decaying behavior without oscillations tend to have small deviations, while oscillatory structure results in slightly higher deviations.

\begin{table*}
 \caption{A fraction in \% of lines of sight in different classes.\label{tab2}}
 \begin{tabular}{lc  cccccccccc cccc}
 \hline

Class Name & Total & \multicolumn{5}{c}{$\chi$- deviations} & & \multicolumn{4}{c}{p($\lambda^2$) vs. $\theta$} & &\multicolumn{3}{c}{Predictions}\\
\cline{3-7} \cline{9-12} \cline{14-16}
& & Low & Med & High & Limit & N/A & & class A & class B & class C & N/A & & Good & Approx & N/A\\

  \hline 
sinc-like               & 11 &30&42&17&7 & 4 & & 36 & 16 & 46 &2& & 8 & 73 & 19\\
smooth                  & 25 &70&14&8 &4 & 4 & &35 &22 &40 &3& &13 & 72 & 15\\
complex osc             & 20 &25&32&22&15& 6 & &43&18 &36 & 3& & 19 & 72& 9\\
complex non-osc         & 30 &40&24&14&13& 9 & &38&20 &40 &2&& 17& 70& 13\\
complex (both)          & 50 &33&28&18&14& 7 & & 40 &19 &38 &3& &18 & 71&11\\
&&&&& &&&&& && & &\\
All classifiable        & 86 &41 &28 &15 & 10& 6 & &38 &19 & 40.5 &2.5 & & 14& 72& 14\\
Not class               & 14 & - &- &- &- &- & & -& -&- &- & & -&-&-\\

\hline  
 \end{tabular}
\end{table*}

\begin{figure*}
\centering
  \begin{minipage}[b]{1\linewidth}
       \includegraphics[width=1.12\linewidth]{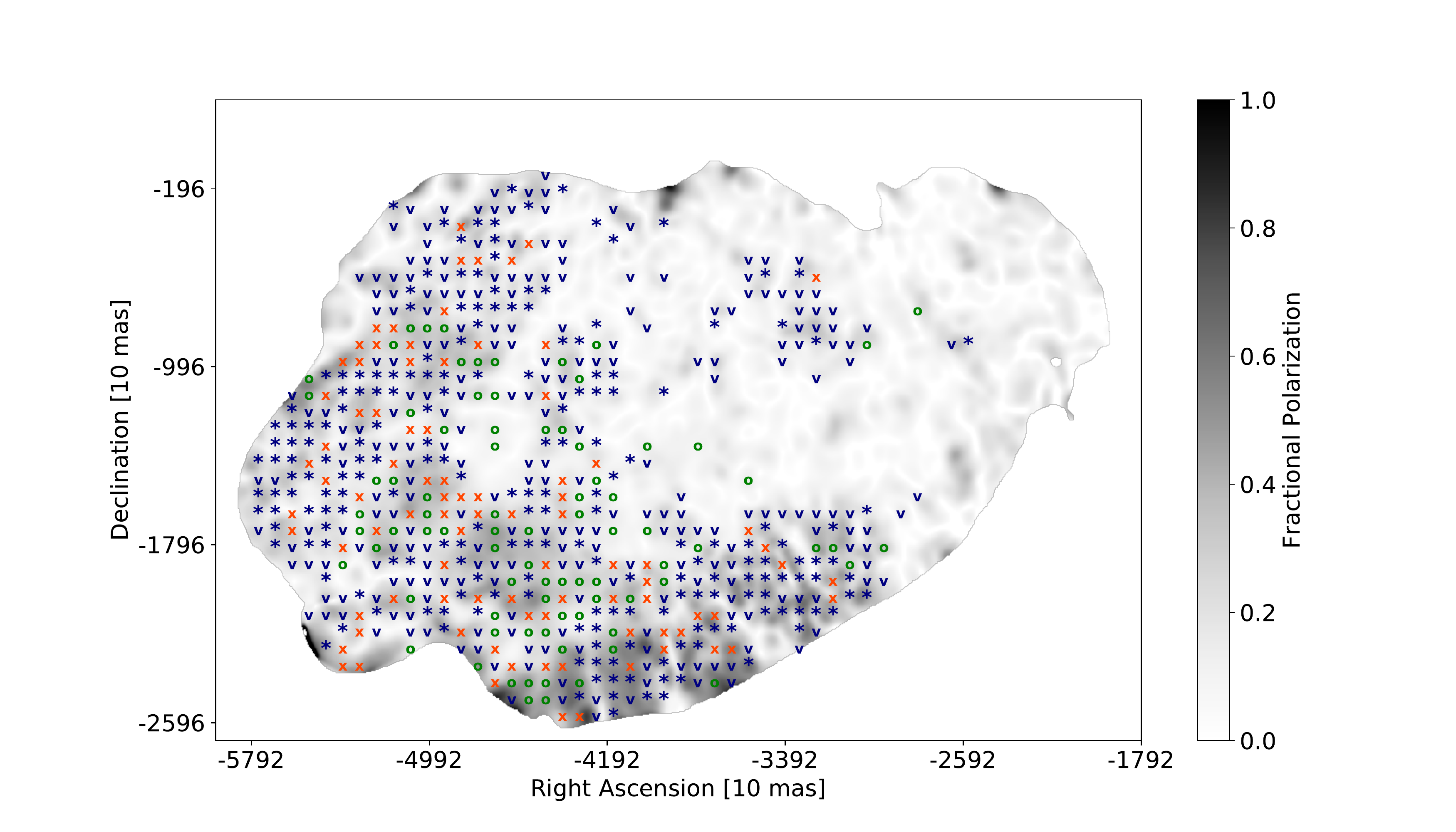}
     \end{minipage}
     
     \begin{minipage}[b]{1\linewidth}
       \includegraphics[width=1.1\linewidth]{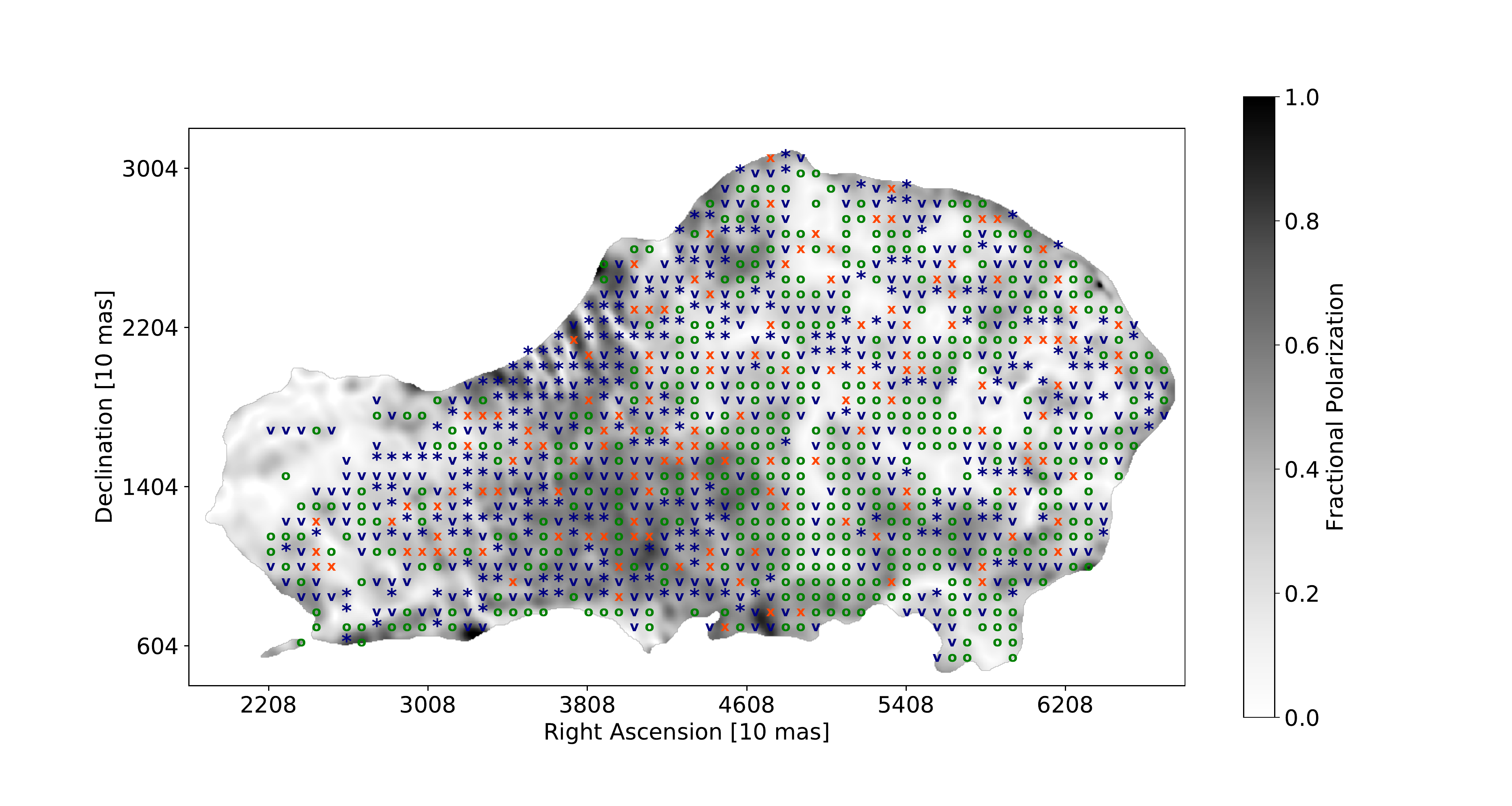}
     \end{minipage}
     \caption{Lines of sight across each lobe classified based on the
       behavior of their fractional polarization vs. $\lambda^2$. The  sinc-like behaviors are shown in red-orange with an `x' symbol),
       smooth decay in green ('o' symbol) and complex decay in navy
       ($\ast$ symbol for complex-oscillatory and 'v' for complex non-oscillatory). The lines of sight are separated by the        resolution beam of $0.75\arcsec$. These classes show no obvious
       large-scale spatial correlations. The majority ($50\%$) of the
       lines of sight show complex behavior, $25\%$ show smooth decay,
       and $11\%$ sinc-like.
     \label{fig:los-imshow}}
   \end{figure*}

Figure \ref{fig:los-imshow} shows a color-coded display of the
distribution of these classifications across the lobes: lines of sight
with sinc-like behavior are shown in red-orange (symbol x), smooth
decay in green (o symbol), and complex decay in navy ($\ast$ symbol for oscillatory-like complex decay, and 'v' symbol for non-oscillatory). The figure shows there are
no clear spatial relationships for these behaviors, yet they are not
completely random as adjacent cells are more likely to show the same
behavior.  These correlated `clumps' are typically of a few kpc scale.
There is a tendency for more complex behavior in the eastern lobes,
and relatively fewer smooth decay and sinc-like behaviors. The
distribution across the the western lobe, on the other hand, consists
mostly of smooth decay particularly at the extremes of the lobes, and
complex decay.

It should be noted that these classifications are only
valid for this frequency span ($2-18$ GHz) and $0.75\arcsec$
resolution, and that the classes are likely to change with a change in
spectral coverage or resolution.

\subsection{Polarization as a Function of Spatial Resolution}\label{polresol}

The depolarization behavior shown in Section \ref{polfreq} can be
explained by randomized fluctuations in the foreground Faraday screen
on scales less than the $0.75\arcsec$ resolution.  If this is the
case, then the fractional polarization will in general decrease as the
resolution decreases at any given frequency.

\begin{figure*}[!ht]
\centering
  \begin{minipage}[b]{0.43\linewidth}
    \includegraphics[width=1.10\linewidth]{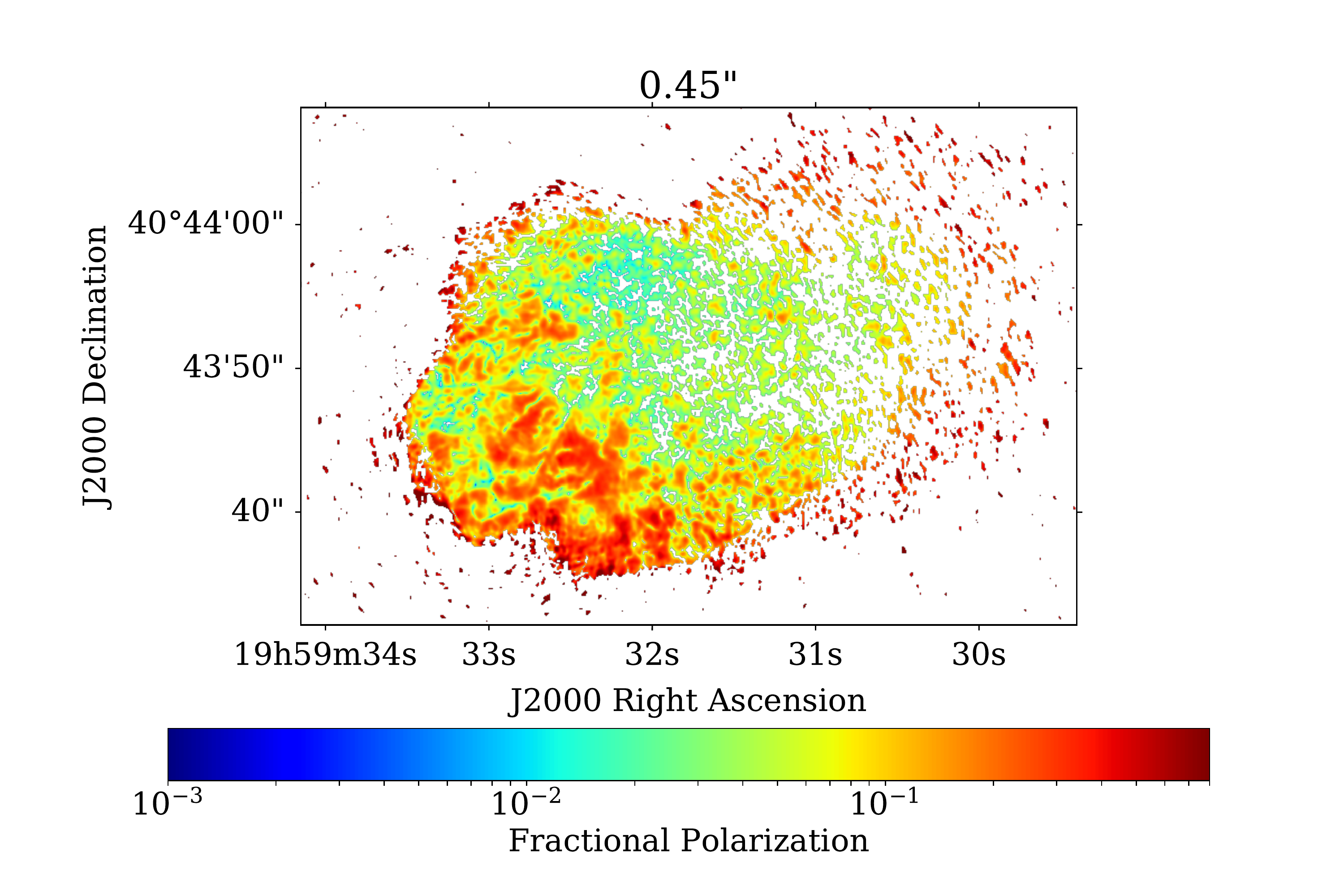}
  \end{minipage}
    \begin{minipage}[b]{0.43\linewidth}
    \includegraphics[width=1.10\linewidth]{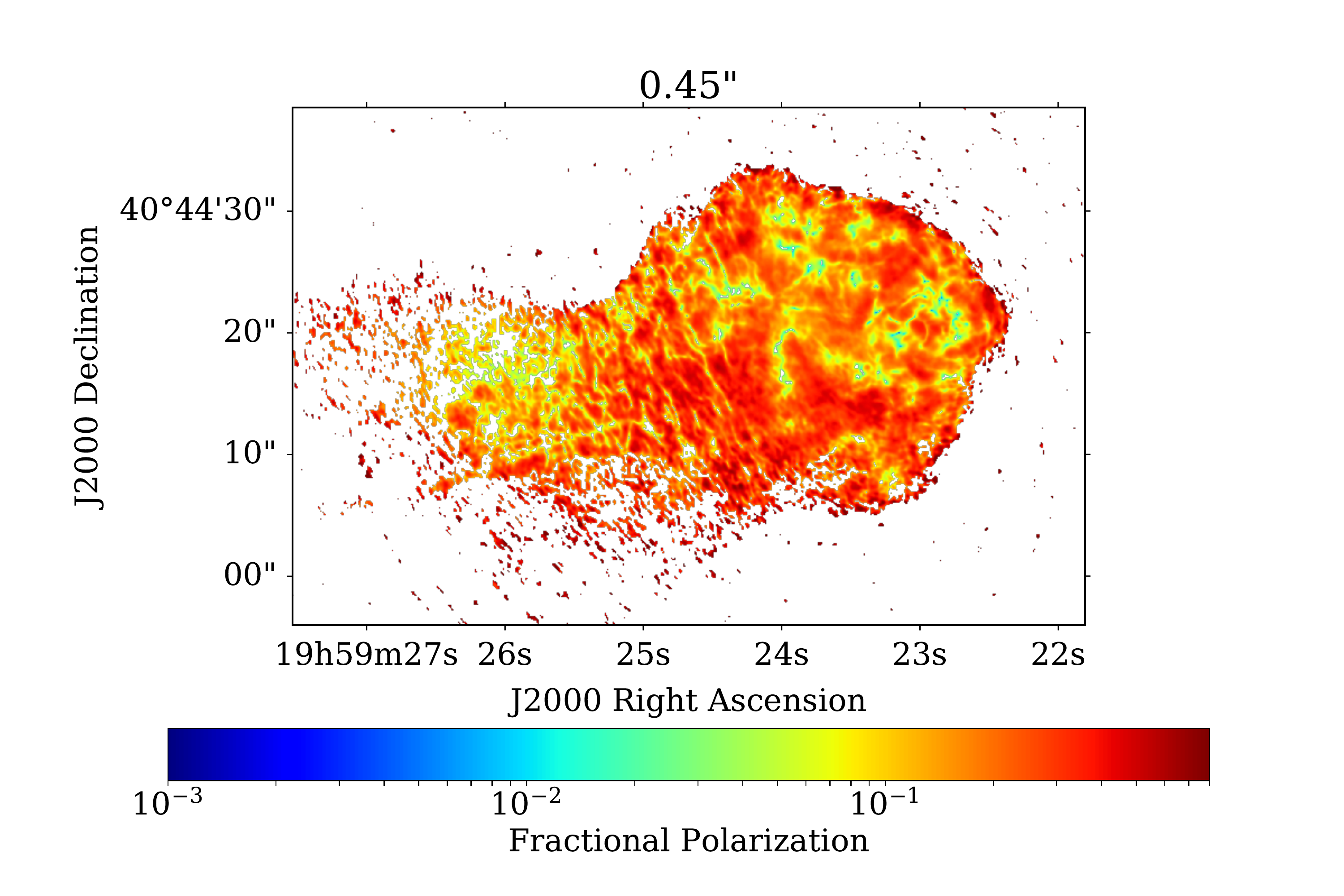}
  \end{minipage}
  \begin{minipage}[b]{0.43\linewidth}
    \includegraphics[width=1.10\linewidth]{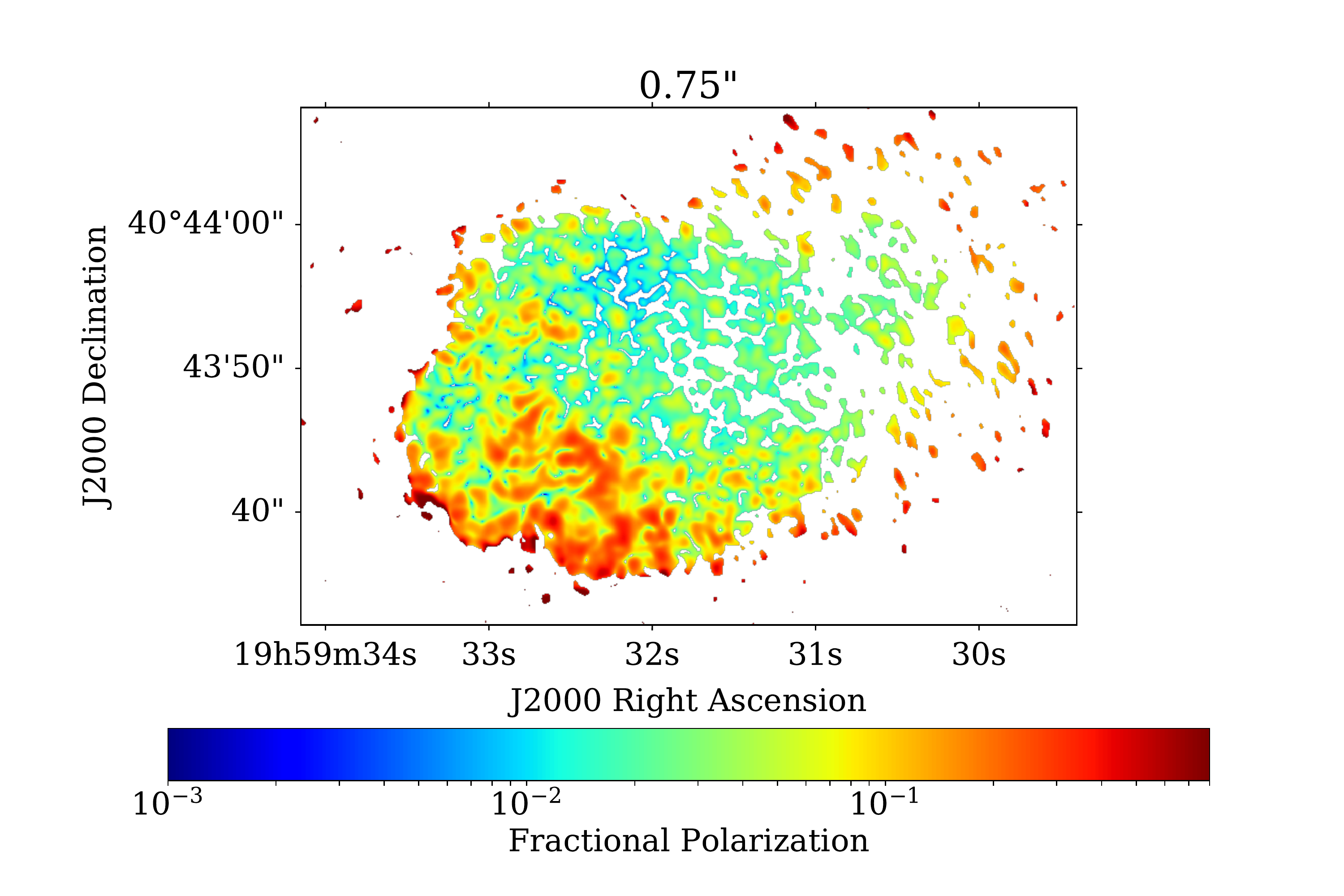}
  \end{minipage}
    \begin{minipage}[b]{0.43\linewidth}
    \includegraphics[width=1.10\linewidth]{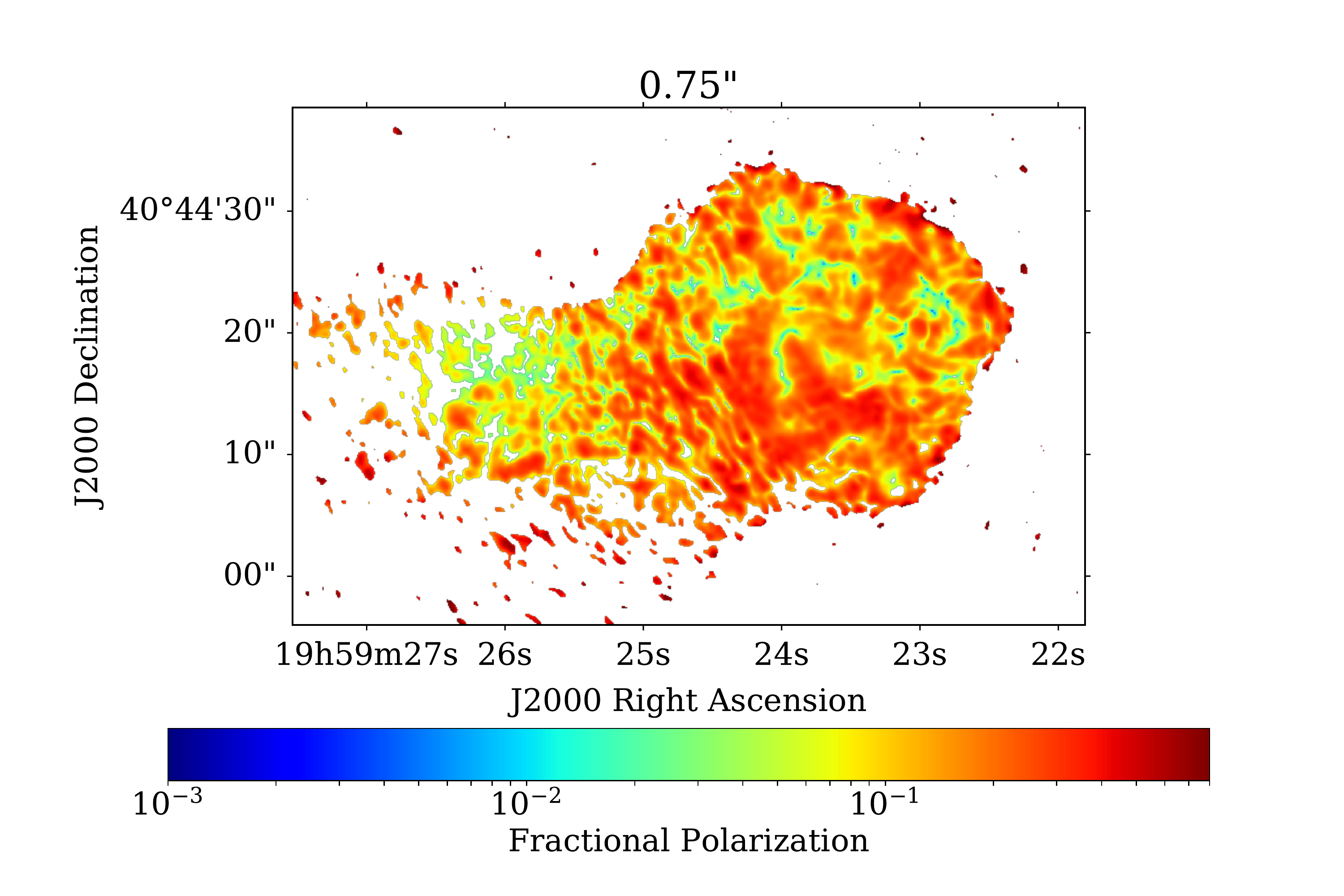}
  \end{minipage}
   \begin{minipage}[b]{0.43\linewidth}
    \includegraphics[width=1.10\linewidth]{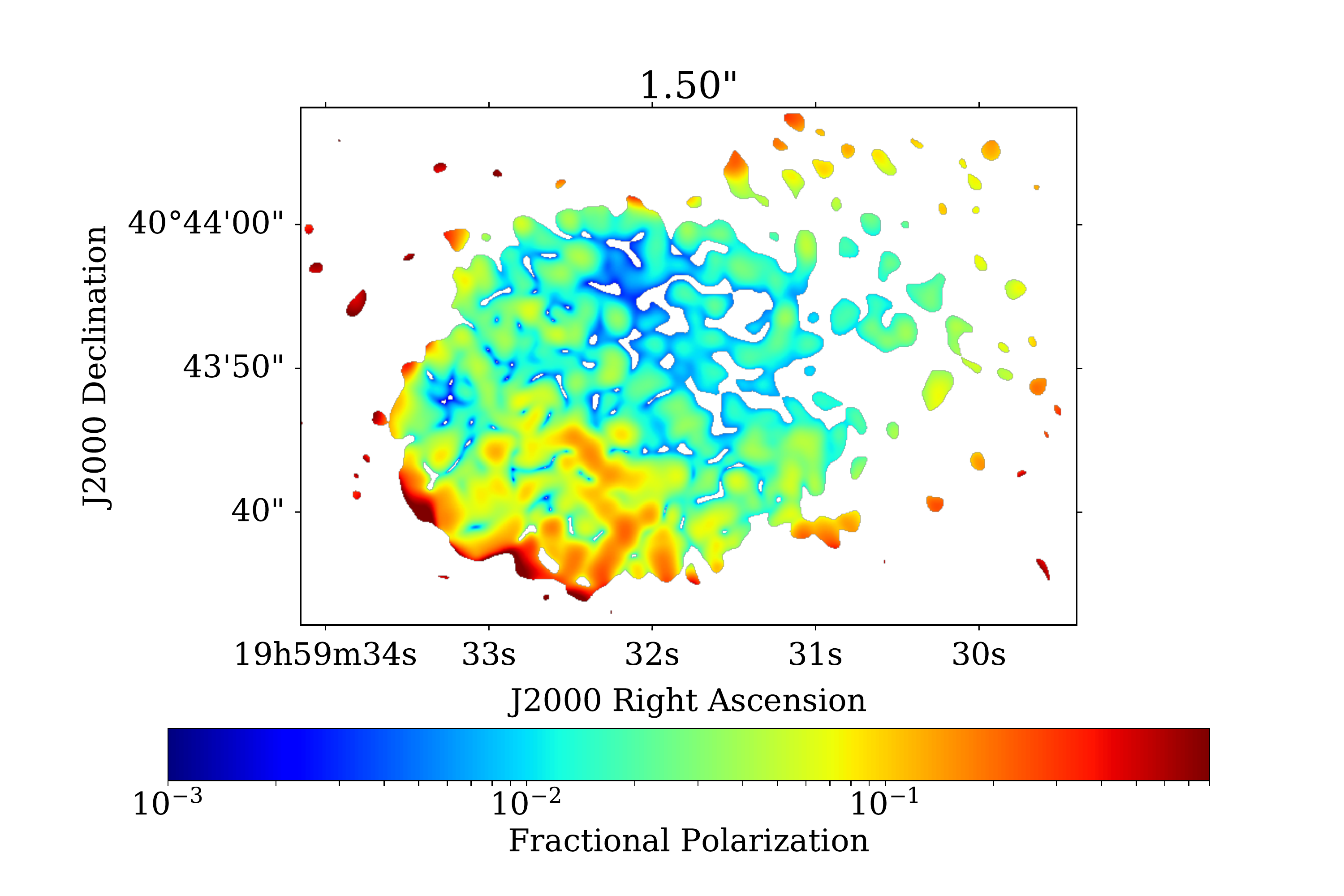}
  \end{minipage}
    \begin{minipage}[b]{0.43\linewidth}
    \includegraphics[width=1.10\linewidth]{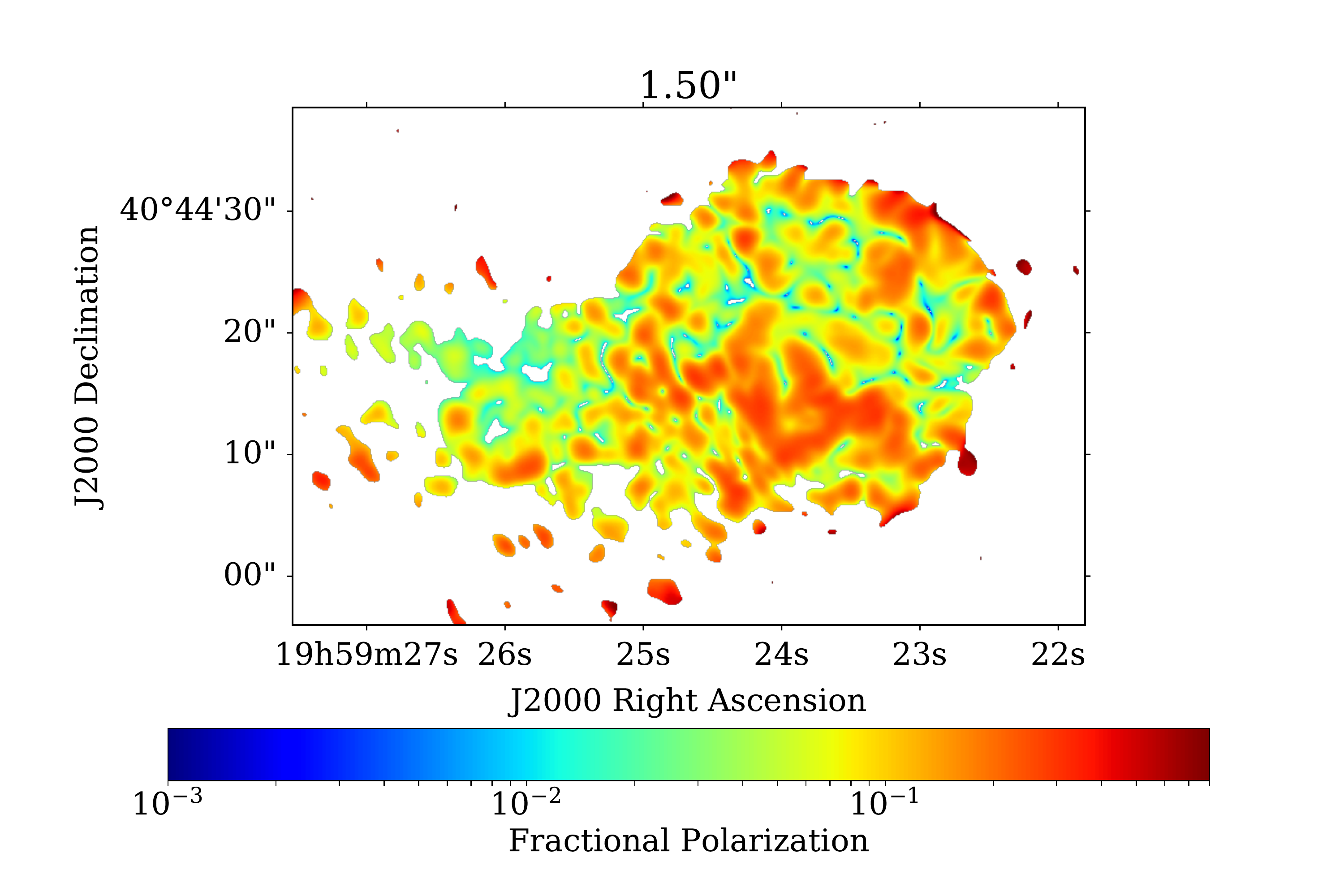}
  \end{minipage}
  \begin{minipage}[b]{0.43\linewidth}
    \includegraphics[width=1.10\linewidth]{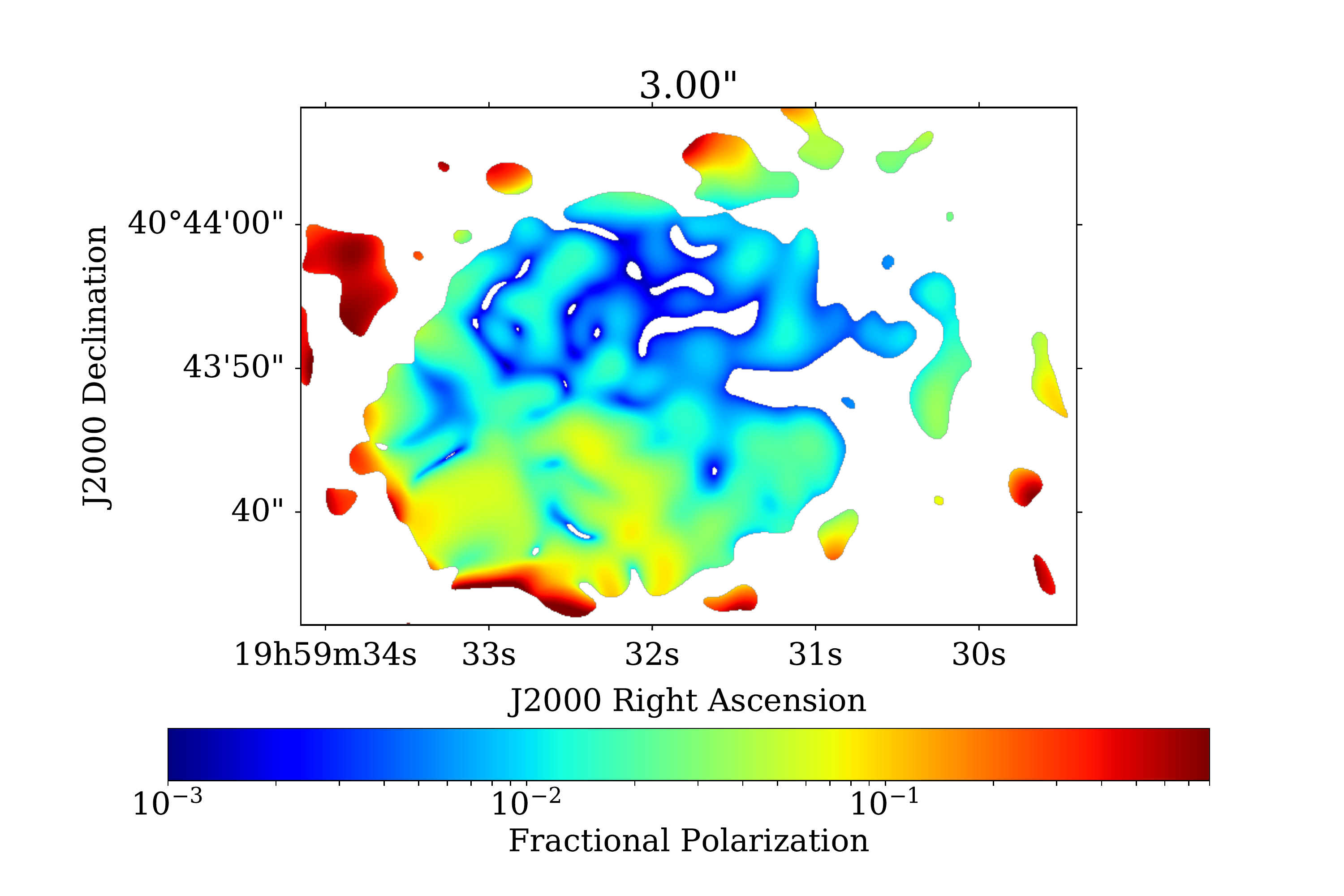}
  \end{minipage}
    \begin{minipage}[b]{0.43\linewidth}
    \includegraphics[width=1.10\linewidth]{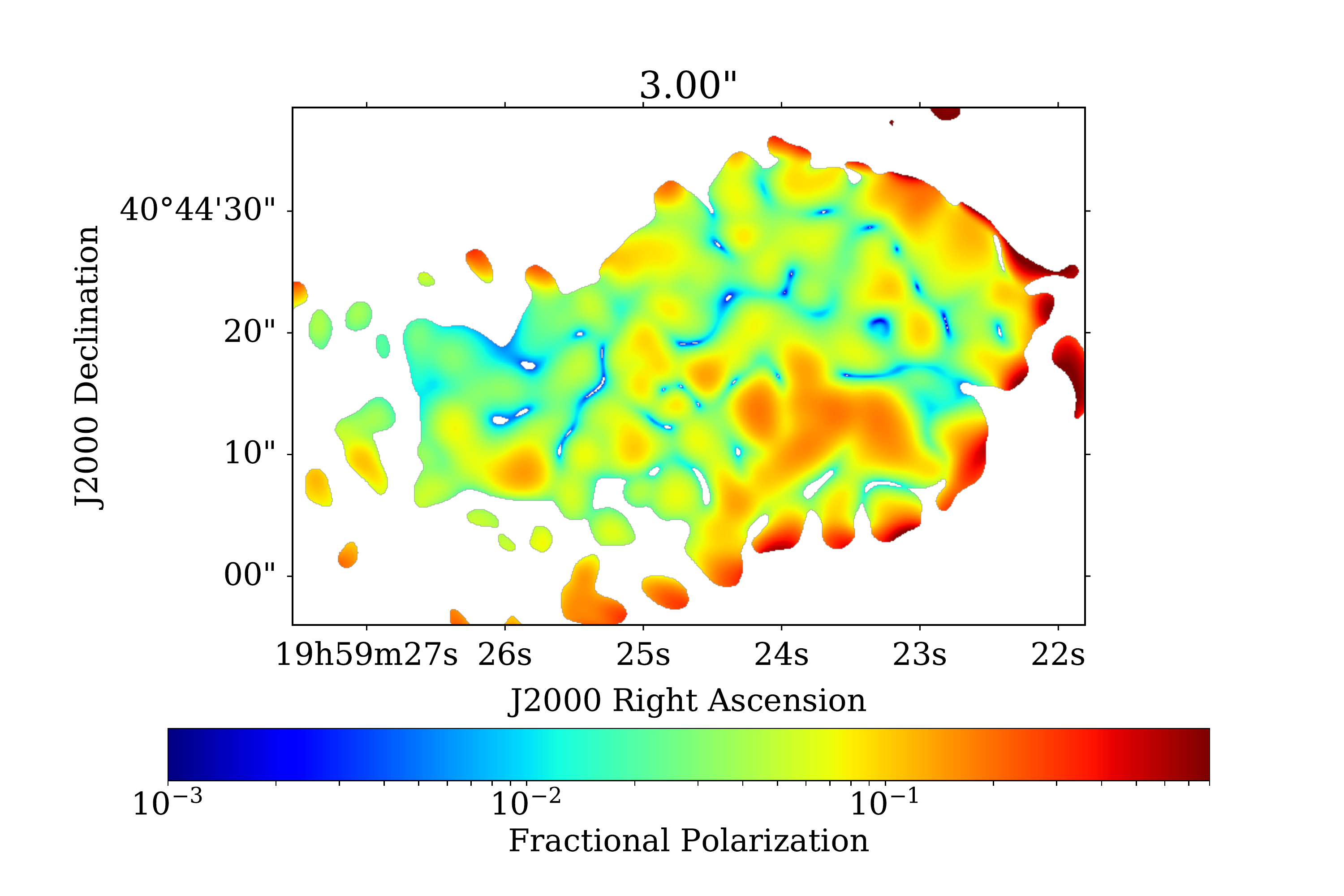}
  \end{minipage}
 \caption{Fractional polarization maps at $4.0$ GHz at four different
   resolutions: $0.45\arcsec$ (top row), $0.75\arcsec$ (second row),
   $1.50\arcsec$ (third row), and $3\arcsec$ (bottom row).  The
   color-bar ranges between 0.1\% and 80\%. Only pixels with
   $(\sigma_p/p) < 50\%$ are shown. The fractional polarization
   decreases as the resolution degrades. This depolarization is higher
   in the eastern lobe especially in the region close to the
   galaxy. \label{fig:resolutionmaps1}}
  \end{figure*}

Figure \ref{fig:resolutionmaps1} shows the fractional polarization at
$4.0$ GHz of Cygnus A at four resolutions -- $0.45\arcsec$,
$0.75\arcsec$, $1.50\arcsec$ and $3.00\arcsec$. The resolution
$0.45\arcsec$ is the highest we can obtain at this frequency. We show
only pixels with $(\sigma_p/p)<50\%$. As expected, the fractional
polarization decreases as the resolution degrades, most notably in the
inner regions of the eastern lobe.  The figure shows the effects of wavelength-dependent
beam depolarization, as expected if the external medium is
differentially rotating the polarized emission within the resolution
element, as discussed below.

If the depolarization is due to transverse fluctuations in the
foreground screen, then at a resolution corresponding to smallest
significant angular scale of the fluctuations, these structures will
be resolved out, at which point the observed fractional polarization
will be that of the source itself.  If this intrinsic value can be
established, any fractional polarization changes with frequency must
then be due to intermixed thermal and synchrotron gas of the source
itself, allowing an estimate of the thermal gas content.  This is the
only method known to us which can clearly discriminate between
internal depolarization (i.e., within the lobes or a mixed boundary
layer), and external depolarization (due to unresolved structures in
the propagating medium).

\begin{figure}
 \center
   \begin{minipage}[b]{0.45\linewidth}
    \centering
    \includegraphics[width=1.1\linewidth]{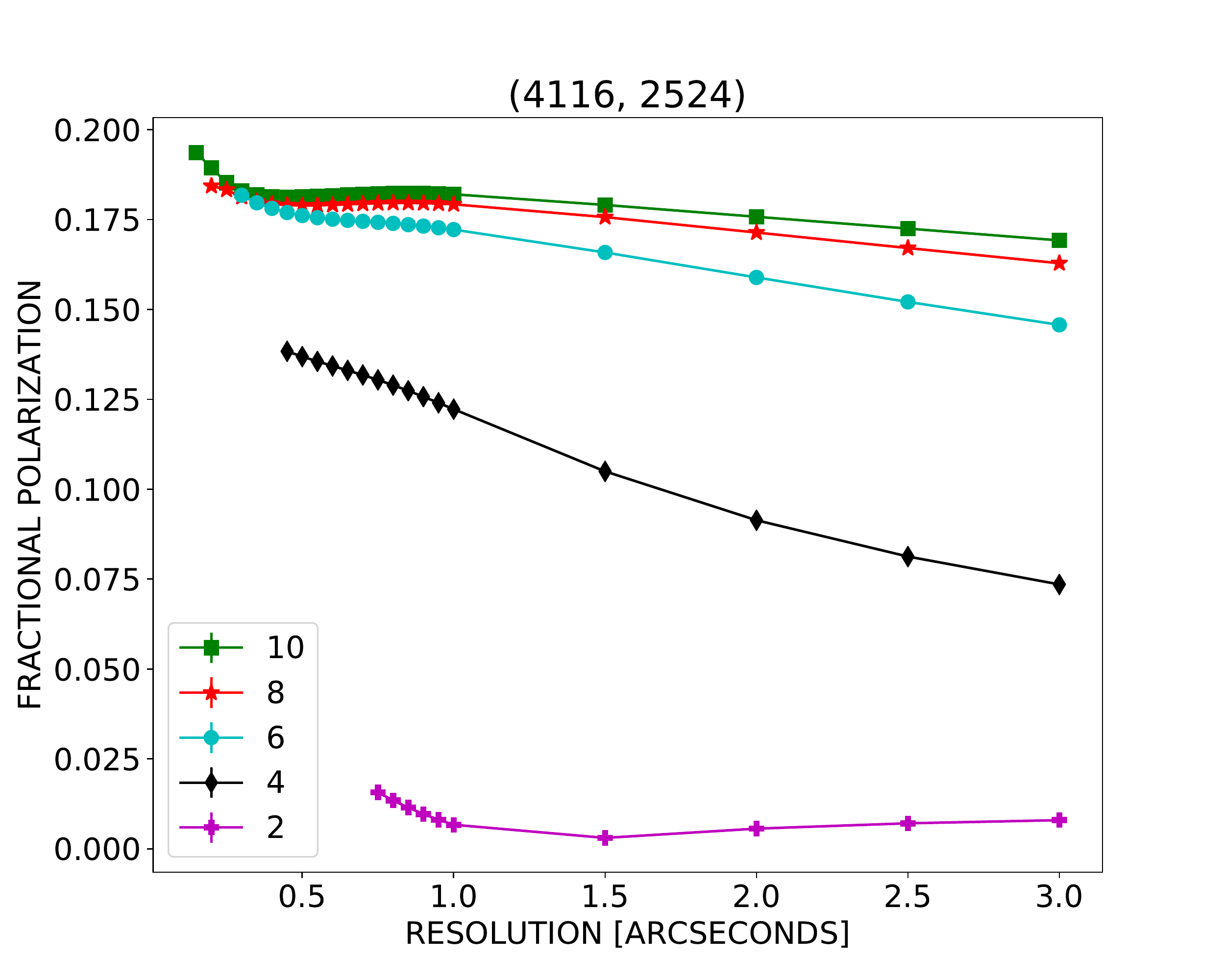} 
  \end{minipage}
     \begin{minipage}[b]{0.45\linewidth}
    \centering
    \includegraphics[width=1.1\linewidth]{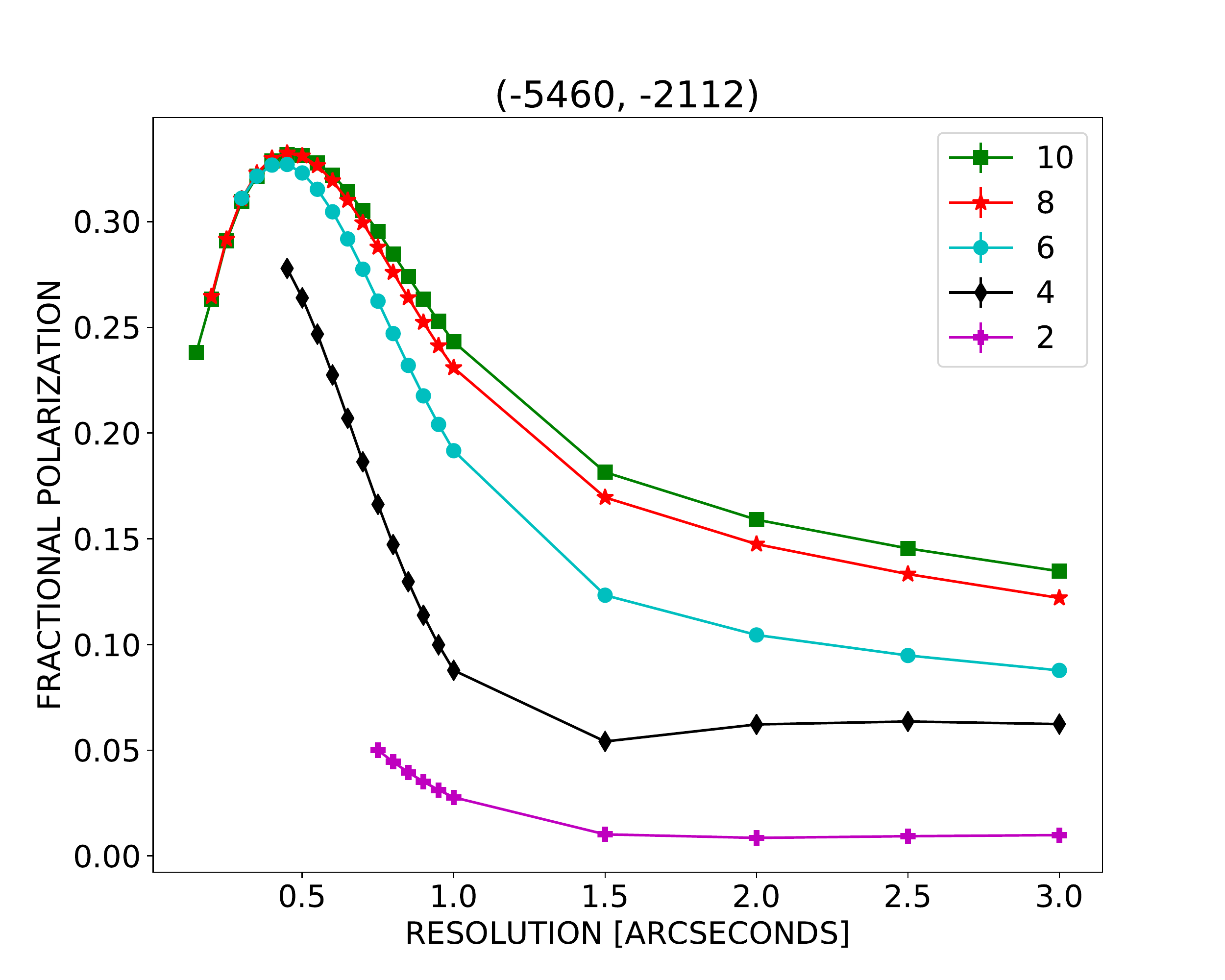}
  \end{minipage}
  
     \begin{minipage}[b]{0.45\linewidth}
    \centering
    \includegraphics[width=1.1\linewidth]{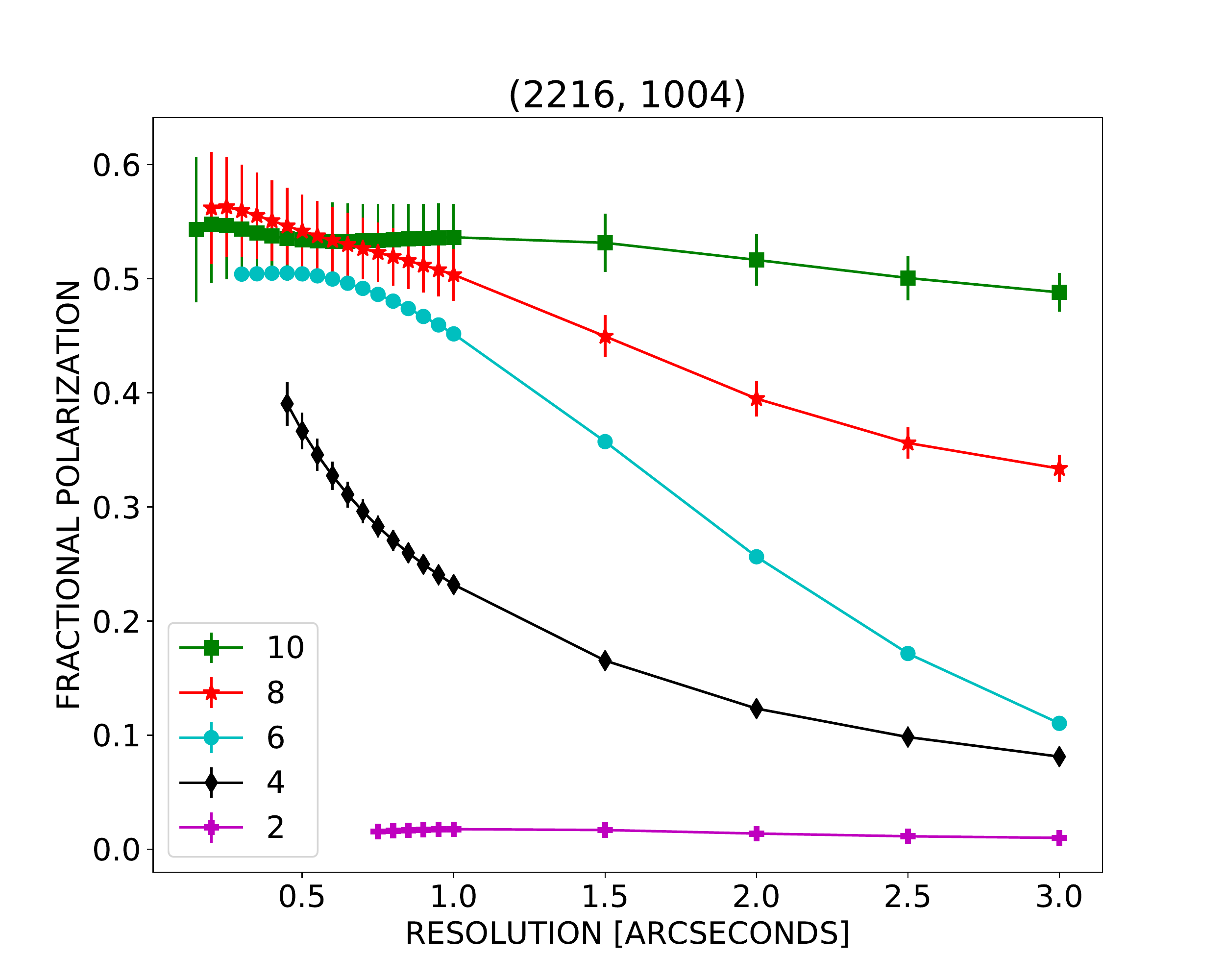}
  \end{minipage}
       \begin{minipage}[b]{0.45\linewidth}
    \centering
    \includegraphics[width=1.1\linewidth]{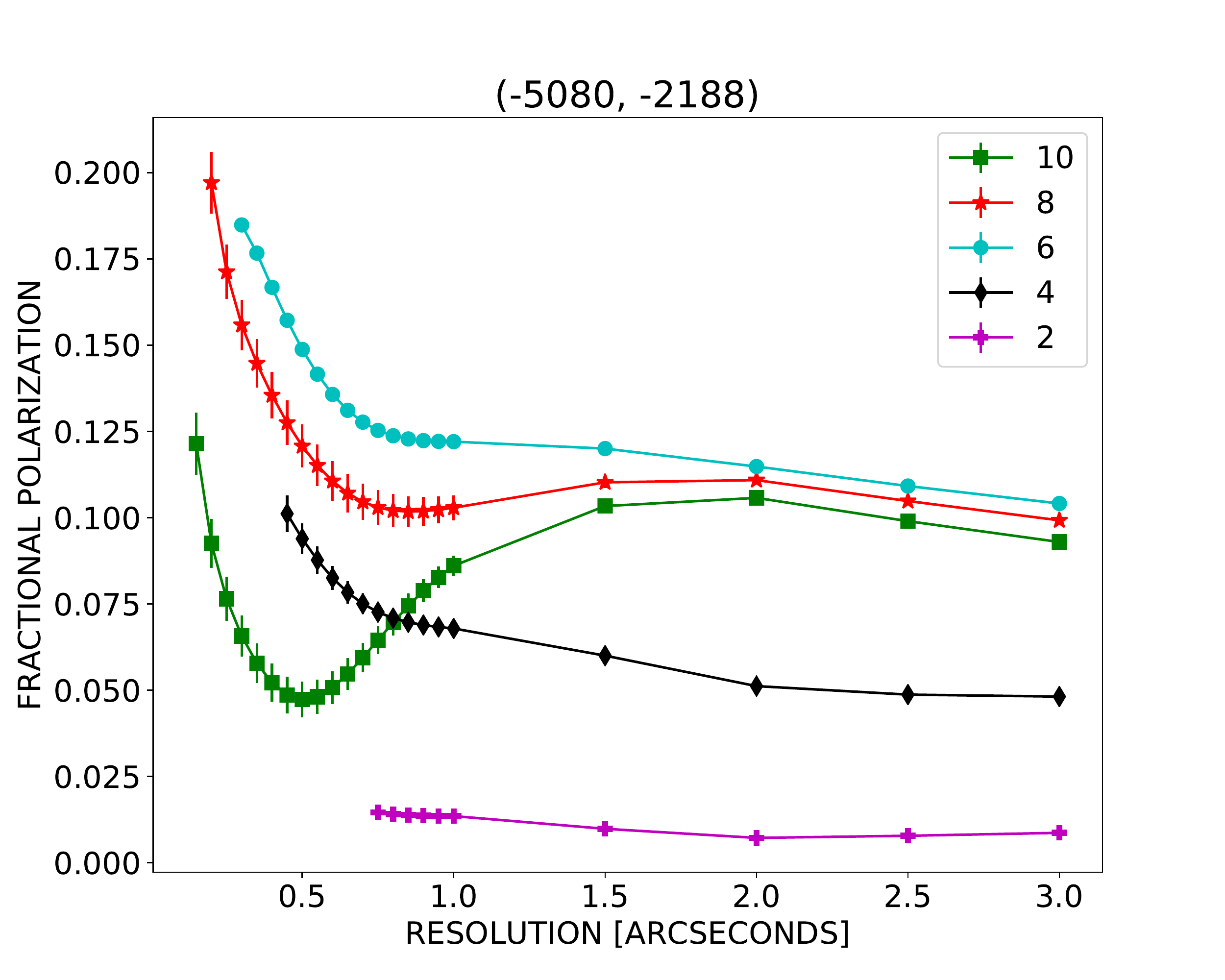}
  \end{minipage}
  
     \begin{minipage}[b]{0.45\linewidth}
    \centering
    \includegraphics[width=1.1\linewidth]{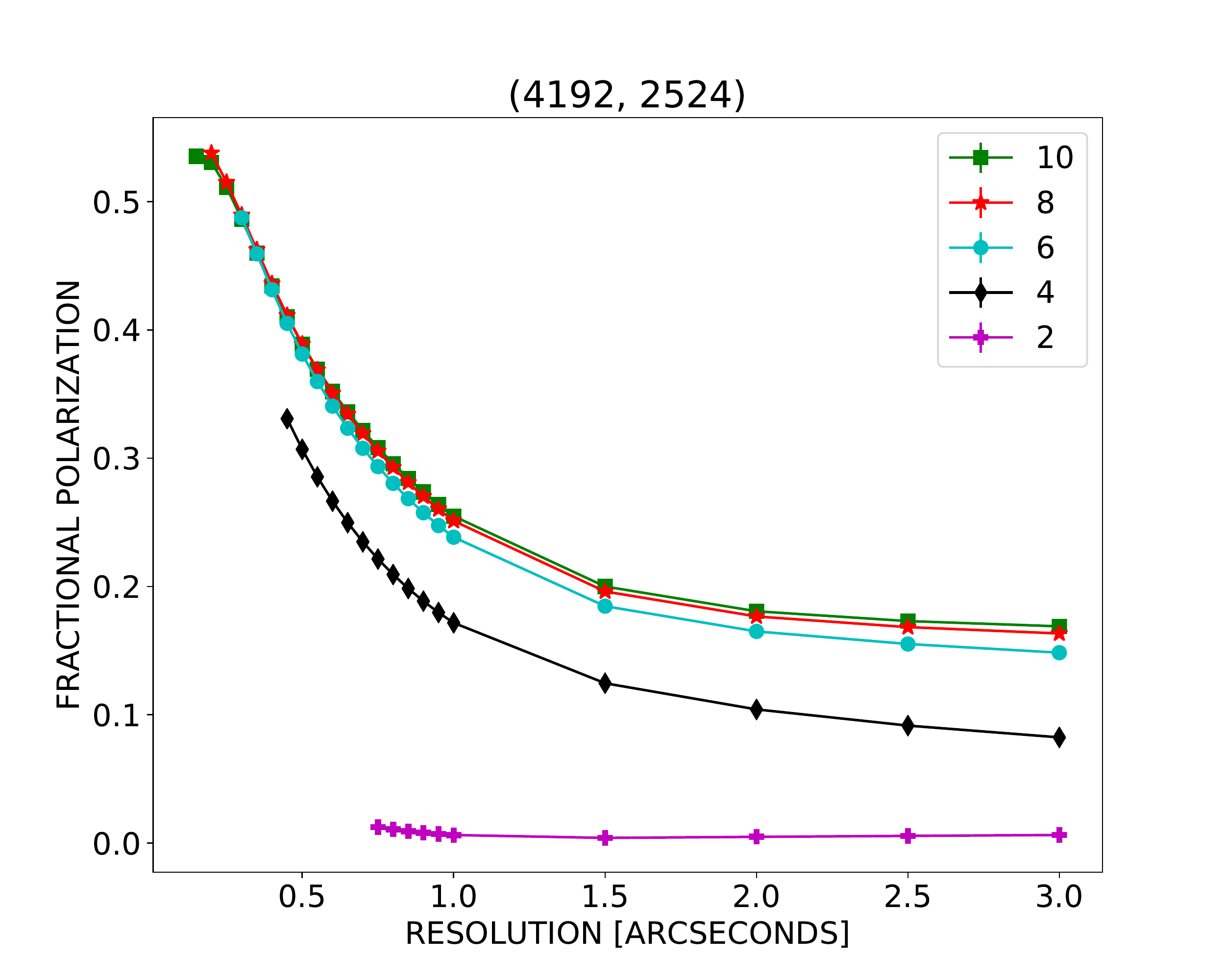}
  \end{minipage}
   \begin{minipage}[b]{0.45\linewidth}
    \centering
    \includegraphics[width=1.1\linewidth]{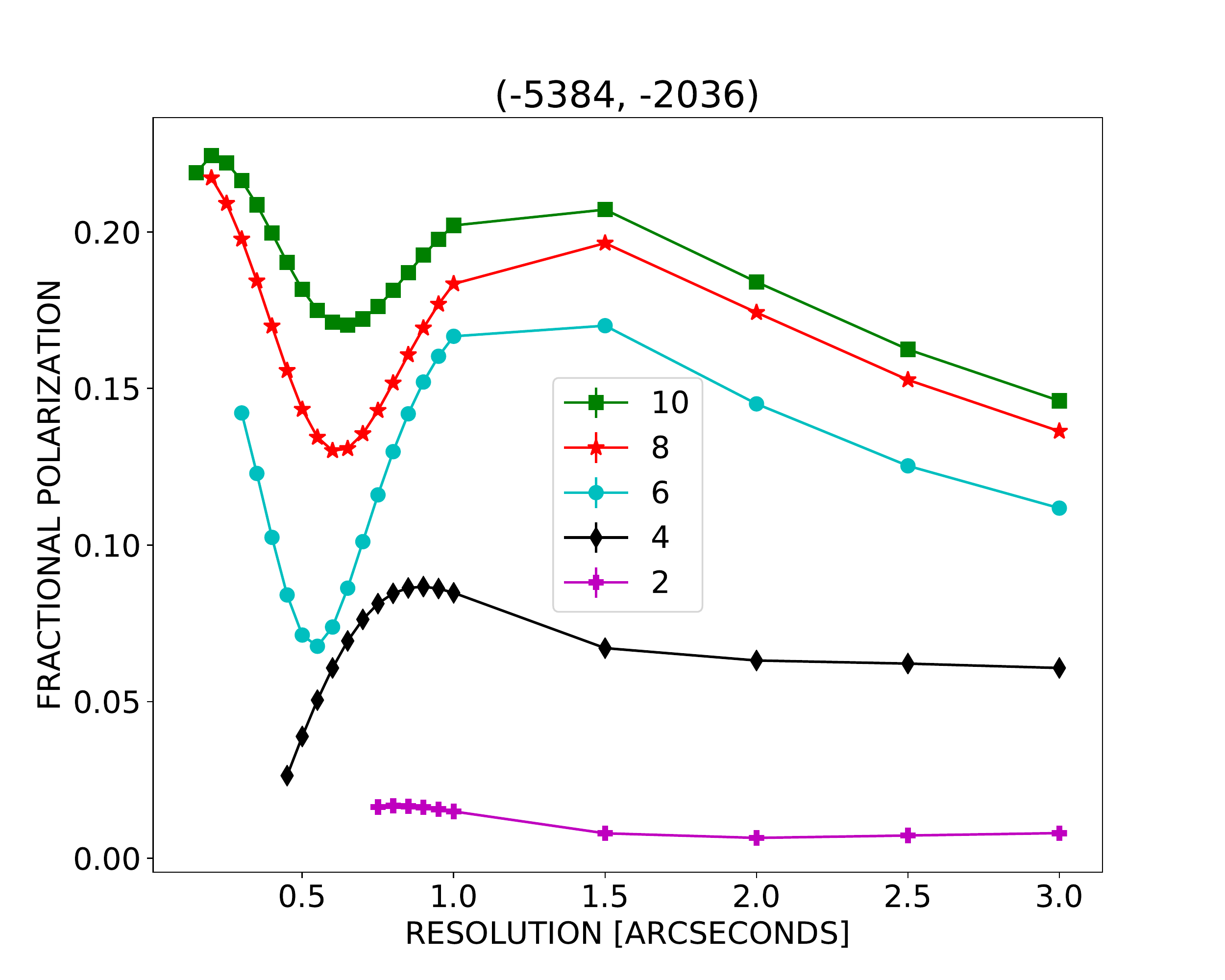}
  \end{minipage}
  \caption{Fractional polarization as a function of resolution. The
    frequencies shown are: $10$ GHz (green), $8$ GHz (red), $6$ GHz
    (blue), $4$ GHz (black) and $2$ GHz (purple). The complex
    behavior, and the lack of a rise in the low-frequency plots,
    demonstrates considerable beam-related depolarization remains at
    the highest resolutions available. \label{fig:LoS_Resolution}}
\end{figure}

Here we describe our efforts to determine the intrinsic value of the
polarization by plotting the fractional polarization as a function of
resolution ($\theta$) at five different frequencies, for $2096$ lines of sight.
Figure \ref{fig:LoS_Resolution} shows a few examples of the typical
behavior.  Each of the six panels shows the fractional polarization at
$2$, $4$, $6$, $8$, and $10$ GHz as a function of resolution, from $3
\arcsec$ to the highest permitted by the diffraction limit for given
frequency, for different lines of sight through the lobes chosen to
display the range of behavior.  The left side shows three lines of
sight where the behavior is close to the expected -- a smooth rise in
polarization with a flattening at sub-arcsecond resolutions.  Note,
however, that only in the top two is there any evidence for asymptotic
behavior at the highest resolution.  And even for these, the critical
low frequency polarizations needed to establish whether there is
internal depolarization do not have sufficient resolution to determine
what the asymptotic values are.  

The right hand panels show much more complex behavior, with structures
in the depolarizing behavior which are likely due to structure of the
fields in the emitting medium (i.e., the lobes), or the external
depolarizing medium. We classified the lines of sight into three groups: i) those with fractional polarization which increases monotonically at least across four frequencies (class A), ii) those with non-monotonic behavior (class B) at four frequencies -- this class also includes lines of sight with fractional polarization which decreases at higher resolutions, iii) and those whose fractional polarization vs. resolution behavior that differs for across the frequencies (class C). For example, left panel of Fig. \ref{fig:LoS_Resolution} all fall into class A, the top two plots on the right panel fall into class B, and bottom-right plot fall into class C. This classification was done by eye. Column 7 to 11 of Table \ref{tab2} gives the fraction of lines of sight in each of the above classes for different decaying behavior. We find that the same fraction (40\%) of lines of sight are found in class A and C. Moreover, the classes do not seem to depend on the whether the line of sight is sinc-like, smooth decay, or complex.

It is clear that much higher resolutions -- by at
least an order of magnitude -- are needed to be able to resolve out
the beam depolarization effects, and permit an estimate of the true
source polarization at the critical lower frequencies where
depolarization phenomena are manifest. These plots emphasize that even
with the VLA and its sub-kpc resolution, the responsible depolarizing
mechanisms and underlying structures cannot be uniquely identified. A
much larger instrument will be needed to finally clarify the processes
responsible.

\section{Modeling the $RM$ Screen at $0.3\arcsec$ Resolution}\label{sec:faradayrotation}

In the previous section, we showed the result that the
depolarization of the majority of the lines of sight are non
monotonic, showing fluctuating, and in some cases oscillating,
fractional depolarization with increasing wavelength.  For all lines
of sight, there is strong overall depolarization, resulting in almost
complete depolarization by a frequency of $2$ GHz. This strongly
suggests `beam depolarization' due to the turbulent medium on
scales less than $0.7\arcsec$ is an important contributor to the
depolarization.

The higher frequency data provide much higher resolution -- at $6$
GHz, the limiting resolution is $0.3\arcsec$. Noting that the
oscillatory behavior in the depolarization is restricted to the lower
frequencies -- it is not seen above $6$ GHz in all the $2000$ 
independent lines of sight -- we
can determine the properties of the Faraday screen with the high
resolution, high frequency data alone, and use this to predict the
lower frequency, lower resolution data which displays the complex
depolarization behavior. A good match between such a prediction and
the observations would be strong evidence that the foreground screen
alone is primarily responsible for the majority of the depolarization. 
That is, the depolarization is dominantly beam-related.

To test this idea, we fitted simple models incorporating random
unresolved fluctuations in a depolarizing screen to our high frequency
high resolution data.  We used the {\tt
  LMFIT} \footnote{https://lmfit.github.io/lmfit-py/} software package
to perform a nonlinear least-squares fitting of our data to the
following model:
\begin{equation}\label{eqn:tofit}
 p = p_0 e^{2i\chi_0} e^{2i RM \lambda^2 - 2\sigma^2 \lambda^4},
\end{equation}
where $RM$ is the Faraday depth due to any large-scale magnetic field,
$p_0$ and $\chi_0$ are the intrinsic 'zero-wavelength' fractional
polarization and polarization angle respectively, and $\sigma$ is the
Faraday dispersion due  small scale, random, Gaussian fluctuations in
Faraday depths within clouds of fixed size \citep{1966BURN}:
\begin{equation}\label{eqn:sigma}
  \sigma = 812n_tB_td\sqrt{N} \quad \quad [\mathrm{rad\, m}^{-2}]
\end{equation}
where $N=L/d$ is the number of turbulent cells of size $d$ lying along
the path length $L$, $n_t$ is the electron density in the cell, and
$B_t$ is the magnetic field of the cell.

Note however that the dispersion parameter can also describe the spread
due a linear gradient across a Gaussian beam - the first order term in
a Taylor series expansion defined in Eq. 2 of \citet{2008LAING}.

\begin{figure}
 \centering
   \begin{minipage}[b]{1\linewidth}
     \includegraphics[width=1.0\linewidth]{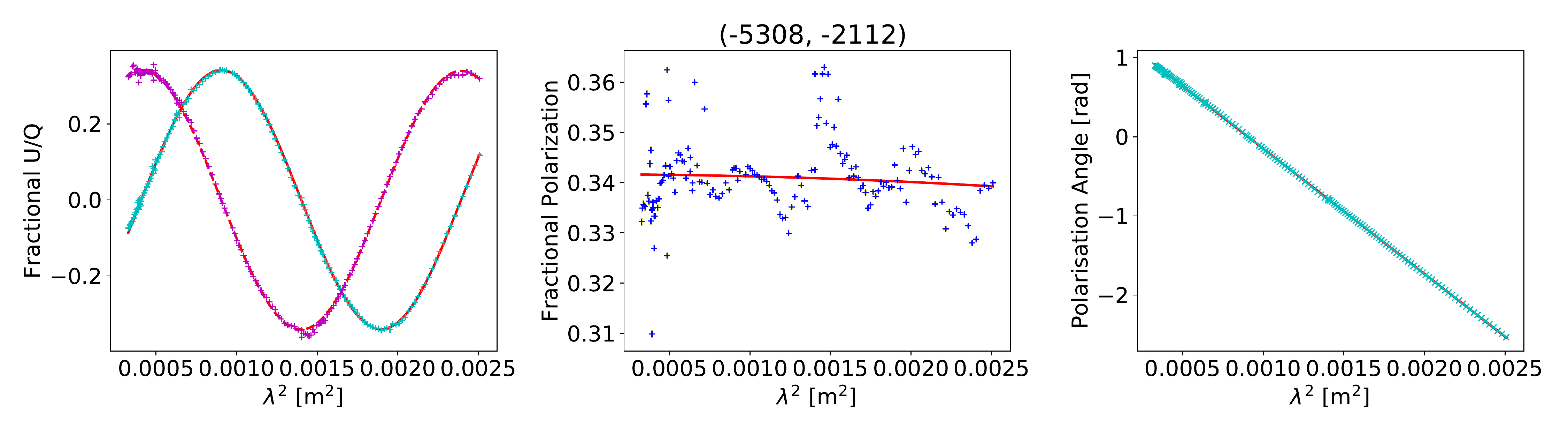}
   \end{minipage}
   \begin{minipage}[b]{1\linewidth}
     \includegraphics[width=1.0\linewidth]{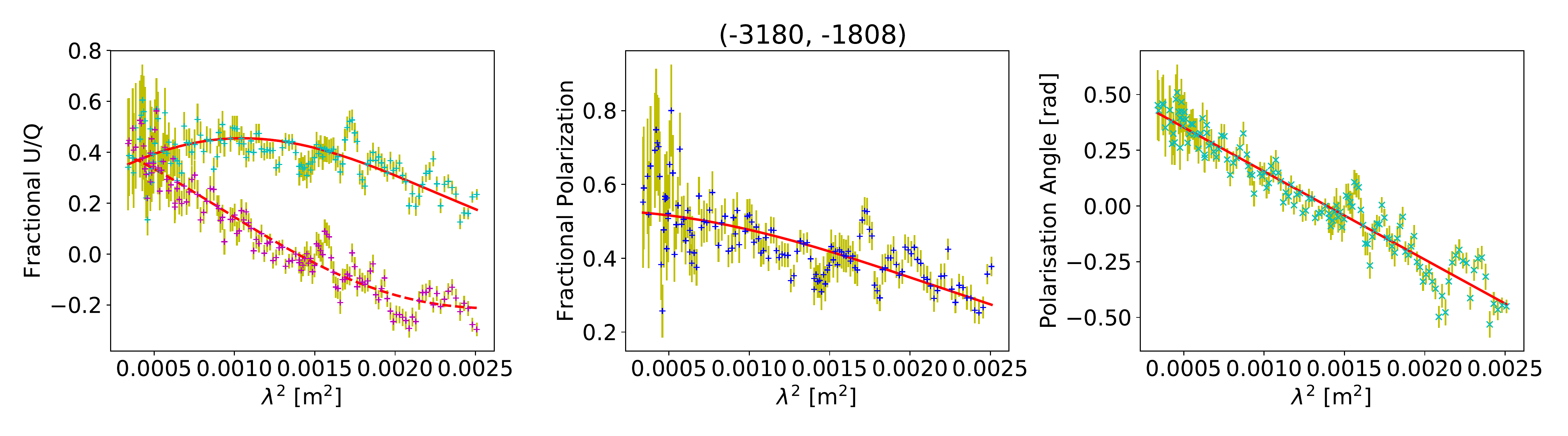}
   \end{minipage}
   \begin{minipage}[b]{1\linewidth}
     \includegraphics[width=1.0\linewidth]{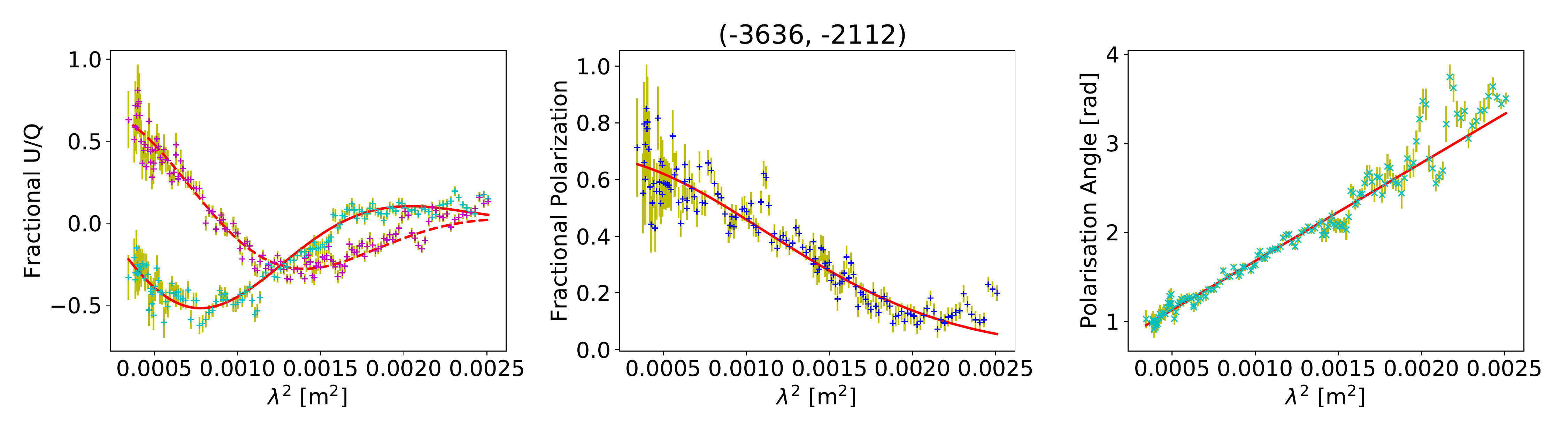}
   \end{minipage}
   \begin{minipage}[b]{1\linewidth}
     \includegraphics[width=1.0\linewidth]{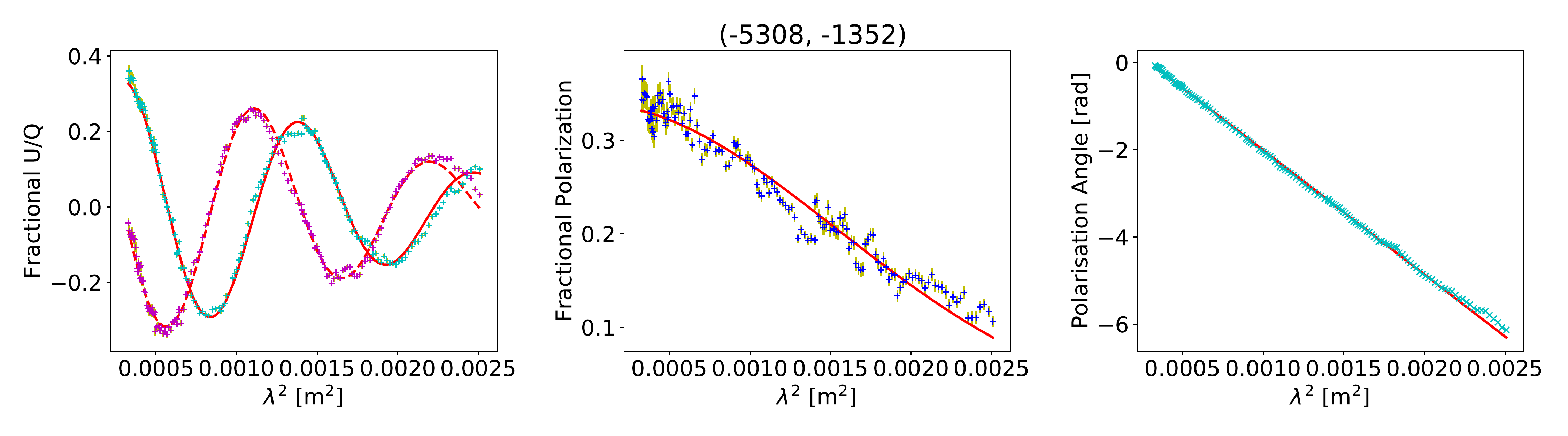}
   \end{minipage}
      \begin{minipage}[b]{1\linewidth}
     \includegraphics[width=1.0\linewidth]{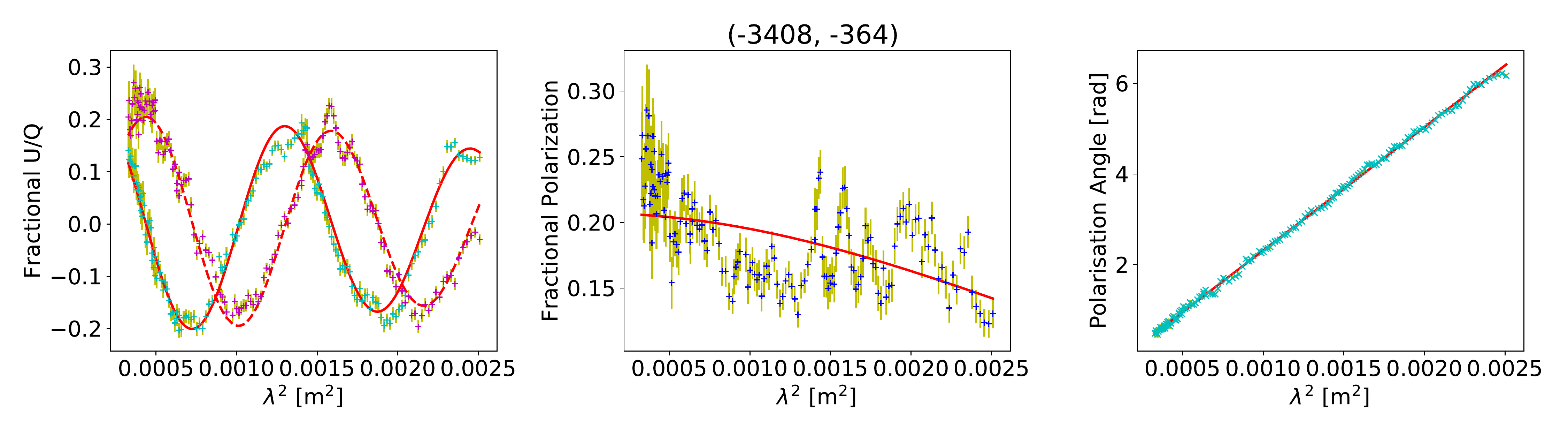}
   \end{minipage}
   \begin{minipage}[b]{1\linewidth}
     \includegraphics[width=1.0\linewidth]{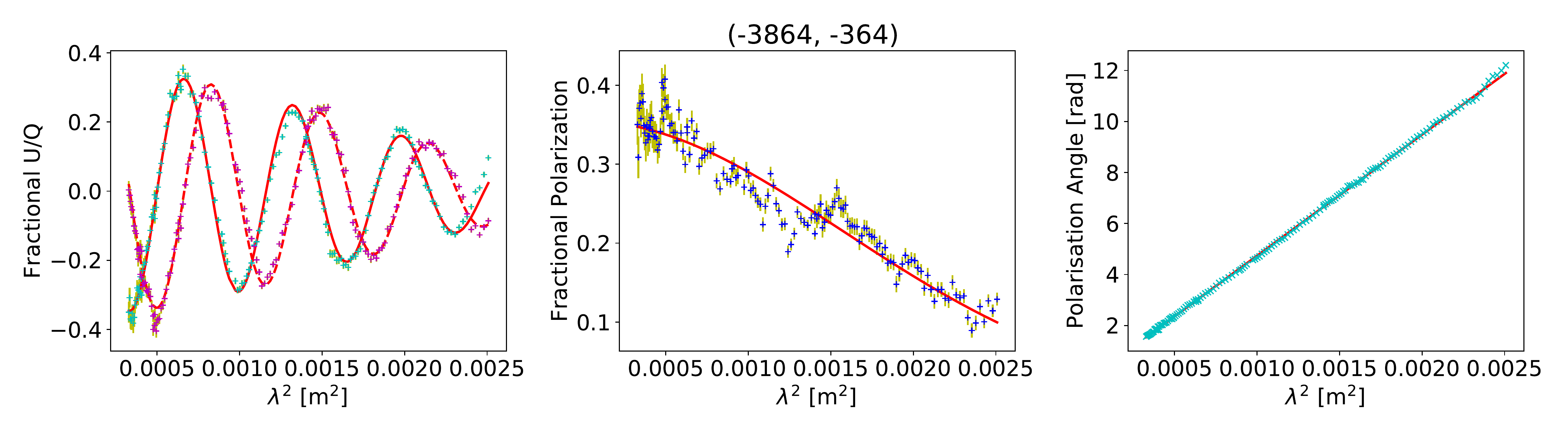}
   \end{minipage}
   \caption{Examples of a simple depolarization screen fitted
     to the $6-18$ GHz, $0.3\arcsec$ data of Cygnus A. The left column
     plots show the fractional Stokes $Q/I$ and $U/I$, the middle
     column shows the fractional polarization and right column shows
     the polarization angle, all as a function of $\lambda^2$. The
     data are predicted very well by this simple depolarization
     model. \label{fig:fitexample}}
 \end{figure}

We performed fitting without weighting so that we do not disadvantage
the high frequency data since they are of low signal-to-noise ratios
and are relatively less represented due to frequency averaging.  We
considered only pixels with an intensity in the $2$ GHz image
$5\times$ the off-source noise. Figure \ref{fig:fitexample} shows a
few examples of the fitting of this model to the data, indicating a
reasonable fit to the data, as judged by eye.

The modeling returns values of the intrinsic (zero-wavelength)
fractional polarization and polarization position angle, the Faraday
depth ($\phi$), (here equal to the RM, as there are no significant
deviations from $\lambda^2$), and the Faraday dispersion, $\sigma$,
which describes the overall decline in fractional polarization between
18 and 6 GHz, at $0.3\arcsec$ resolution.  Figure
\ref{fig:fittedparams} shows images of the intrinsic fractional
polarization across the two lobes (top row), the rotation measure
(middle row), and the Faraday dispersions (bottom row). These are
discussed in the next section. The left panels show the eastern lobe,
the right hand panels the western lobe. We consider only pixels with
$\sigma_{RM}/RM < 0.2$ and $\sigma_{p_0}/p_0 < 0.2$, where
$\sigma_{RM}$ and $\sigma_{p_0}$ are errors derived from the
fits. This step ensures that we remove pixels which are less reliable.

 \subsection{Intrinsic Fractional Polarization} 
The derived intrinsic fractional polarization maps indicate that
Cygnus A is highly polarized in nearly all areas, with typical values
between 15\% and 45\%, and some as high as 70\%. Even the inner
portions of the eastern lobe -- which depolarizes the most rapidly --
show 'zero-wavelength' polarizations similar to all other parts of the
source. These are consistent with the 10 GHz map shown in
Fig. \ref{fig:polarizationmaps}. These results indicate that the low
polarization seen at intermediate wavelengths are not intrinsic to the
source. The low fractional polarization regions across the lobes seen
at lower frequencies and resolutions are clearly largely a result of
unresolved fluctuations in a magnetized foreground screen.

\subsection{Rotation Measure}
The rotation measure maps presented here are much more detailed than
those derived by \citet{1987DREHER}. These new results extend the $RM$
mapping into the inner region of the eastern lobe, and the tail of the
western lobe. Both areas reveal extremely high $RM$s, resulting in an
increased maximum $RM$ range: $-4500$ and $+6400$ rad m$^{-2}$ in the
eastern lobe, and $-5000$ and $+3000$ in the western lobe. More striking is that the inner large rotation measures have opposite signs suggesting field reversal in the surrounding material.
In most
regions the $RM$ gradients are typically a few $100$ rad m$^{-2}$
arcsec$^{-1}$, with a few regions having gradients as high as $\sim
1000$ \radm arcsec$^{-1}$.  The $RM$ changes sign on 3--20 kpc scales,
and this change in sign seems to occur along the source axis,
particularly in the eastern lobe.  The $RM$s are relatively smooth
across the western lobe, and in general increase with decreasing
radial distance from the AGN.  These confirm the results, made with
more limited data, given by \citet{1996PERLEY}. The $RM$s in the
eastern lobe are relatively chaotic at this resolution, with the
$|RM|$ showing no clear radial dependence but instead occurring in
bands of high and low $RM$, similar to those seen in M84, 3C 353,
0206+35, and 3C 270 \citep{2011GUIDETTI}.  Note that, with the
exception of the enhanced $RM$ arc surrounding hotspot B found by
\citet{1988CARILLI}, there is no evident correlation between the $RM$
values and the source brightness or with any of the evident structures
in total intensity (i.e., the hotspot, jet, or filaments).  This
supports the interpretation that the observed $RM$ structures
originate outside the source of the emission. 

 \subsection{Random $RM$ Fluctuations}
\begin{figure*}
 \centering
    \begin{minipage}[b]{0.45\linewidth}
   \includegraphics[width=1.15\linewidth]{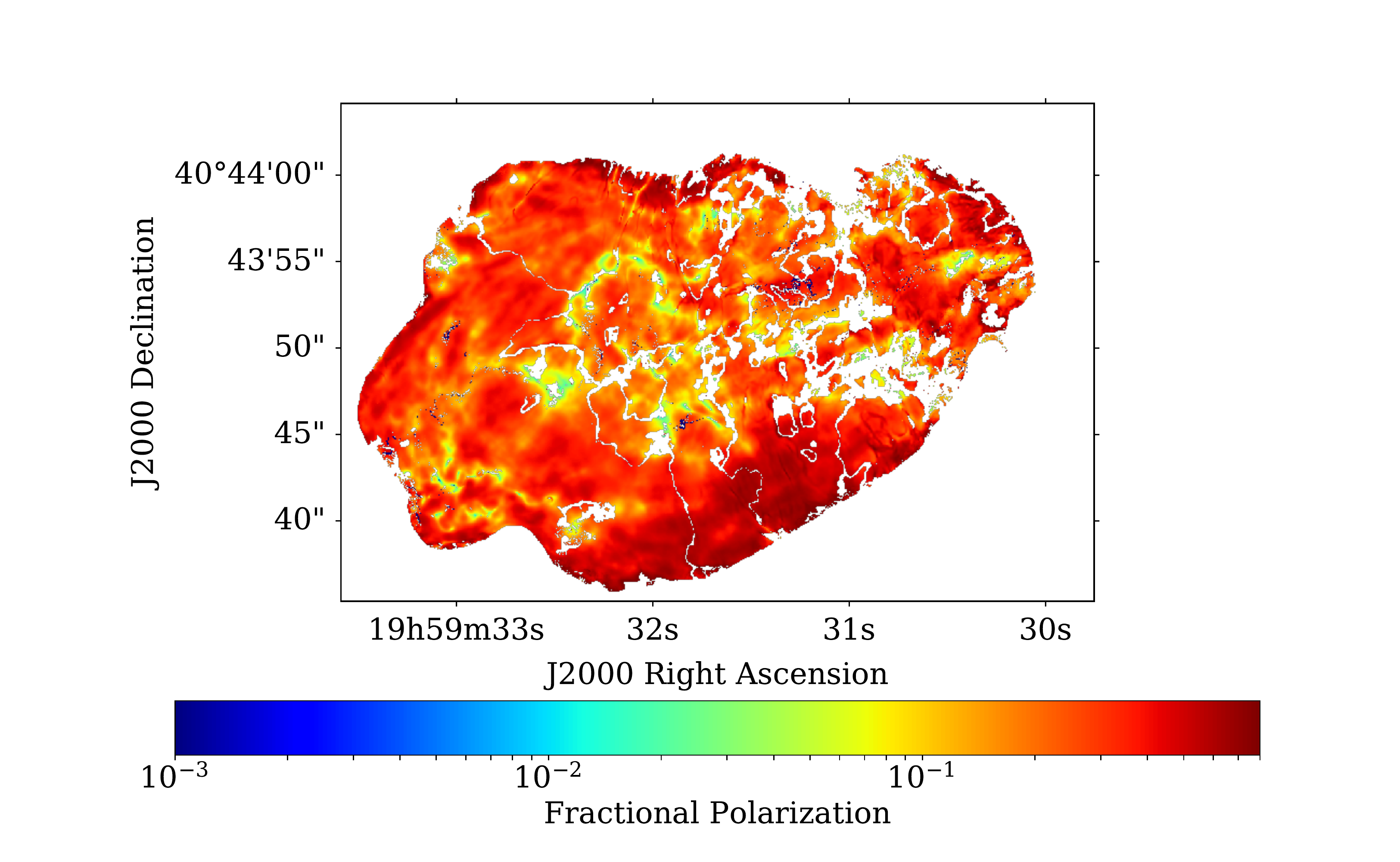}
    \end{minipage}
    \begin{minipage}[b]{0.45\linewidth}
   \includegraphics[width=1.15\linewidth]{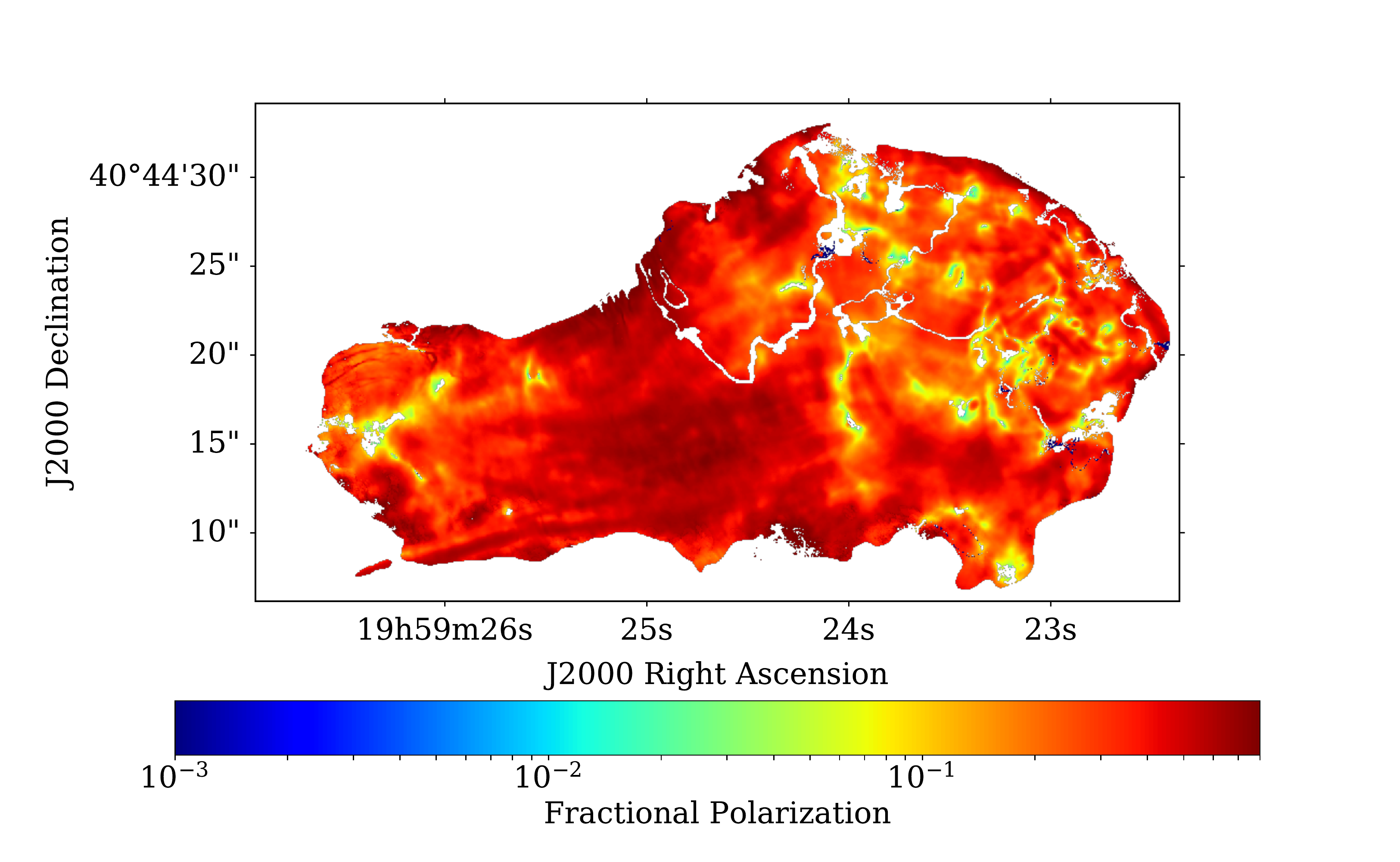}    
    \end{minipage}
    \begin{minipage}[b]{0.45\linewidth}
   \includegraphics[width=1.1\linewidth]{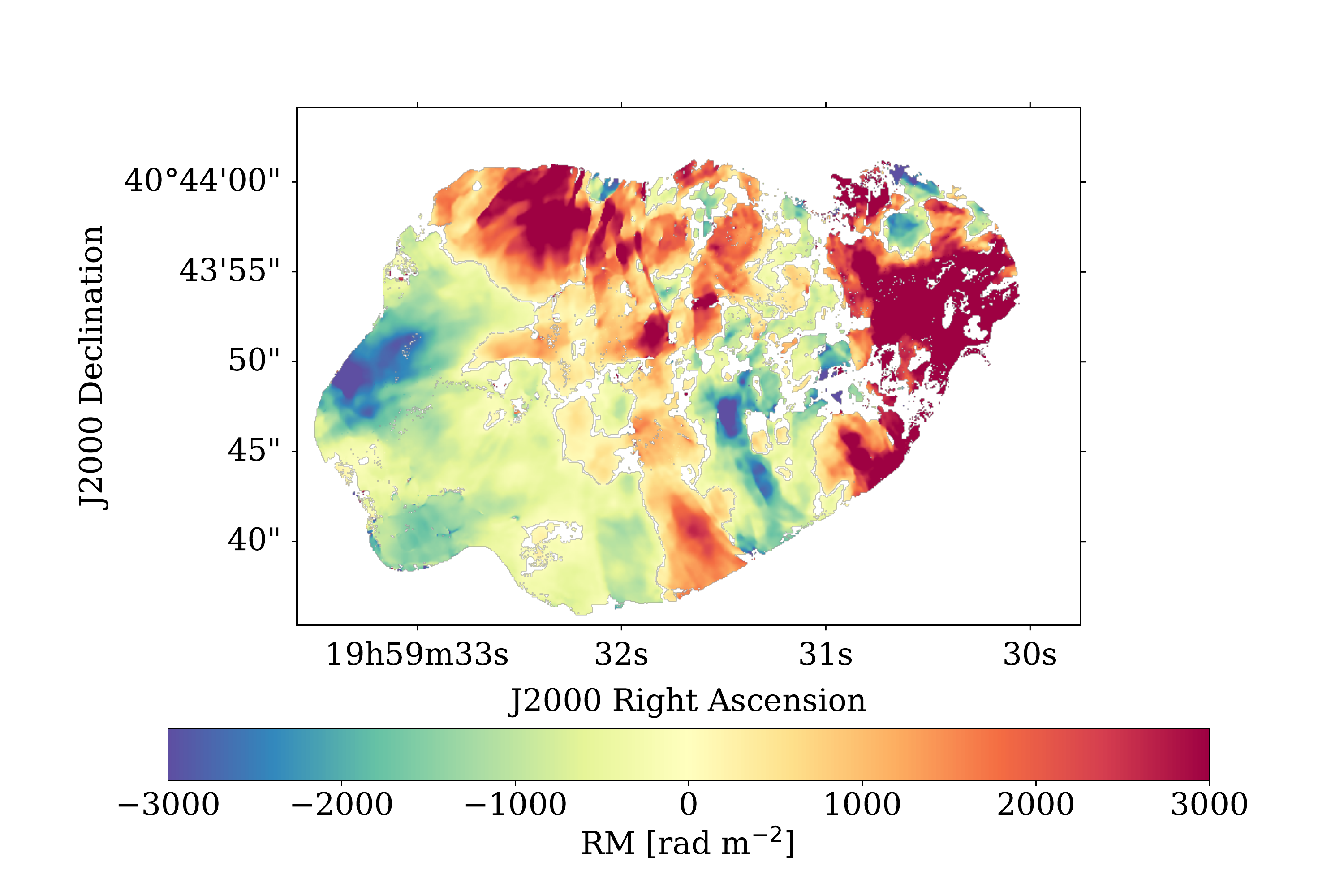}
    \end{minipage}
    \begin{minipage}[b]{0.45\linewidth}
   \includegraphics[width=1.1\linewidth]{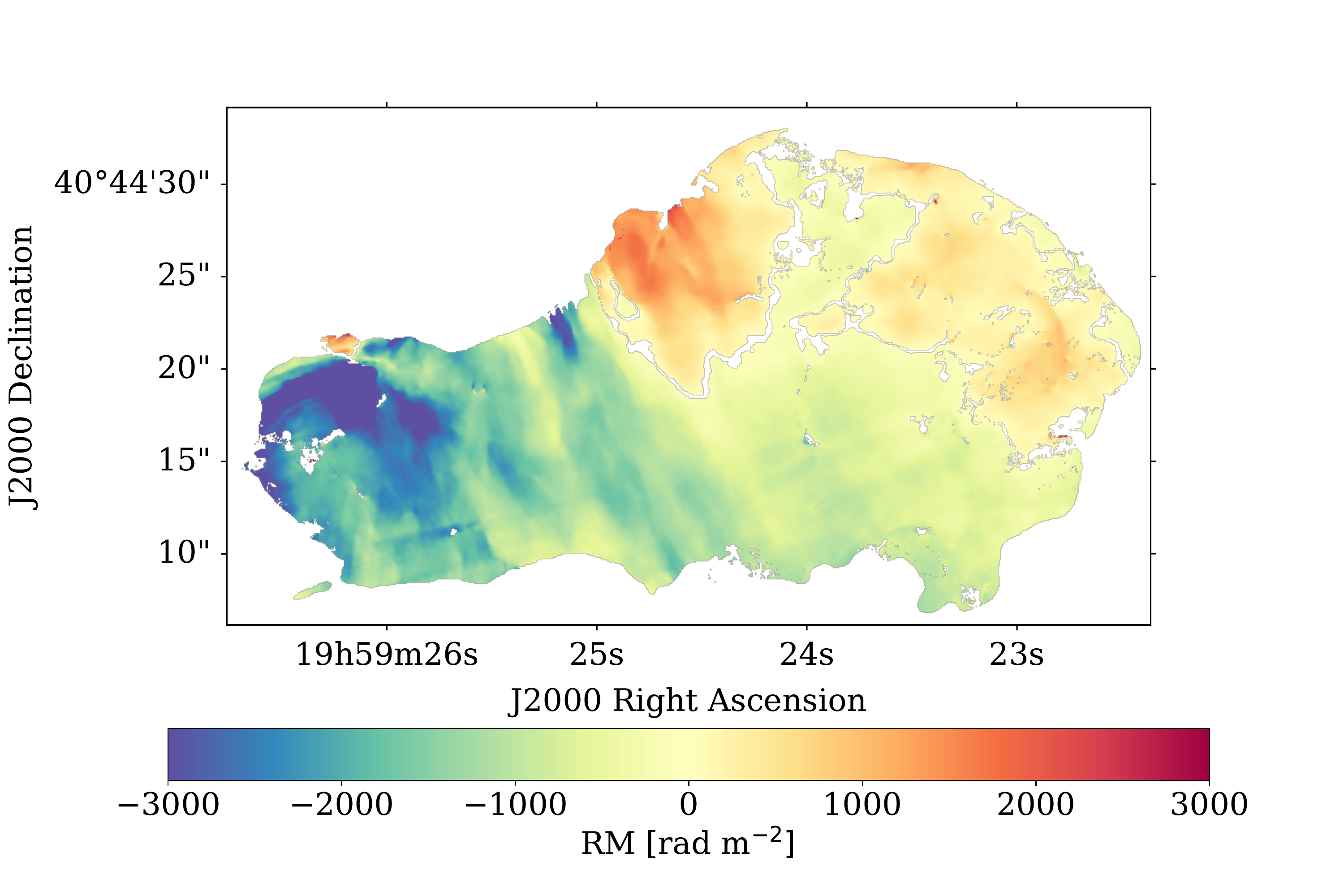}    
    \end{minipage}
   \begin{minipage}[b]{0.45\linewidth}
   \includegraphics[width=1.1\linewidth]{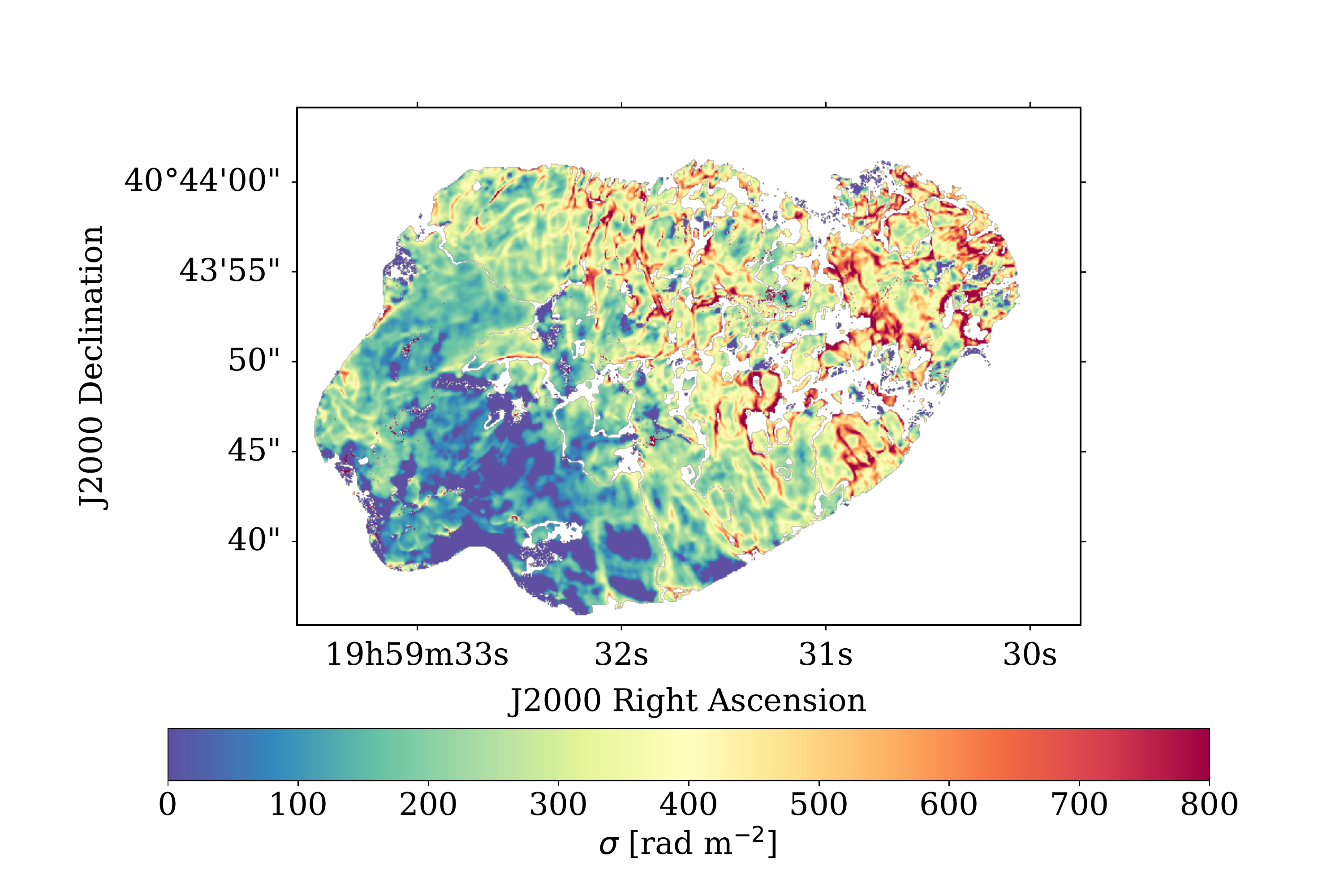}
    \end{minipage}
    \begin{minipage}[b]{0.45\linewidth}
   \includegraphics[width=1.1\linewidth]{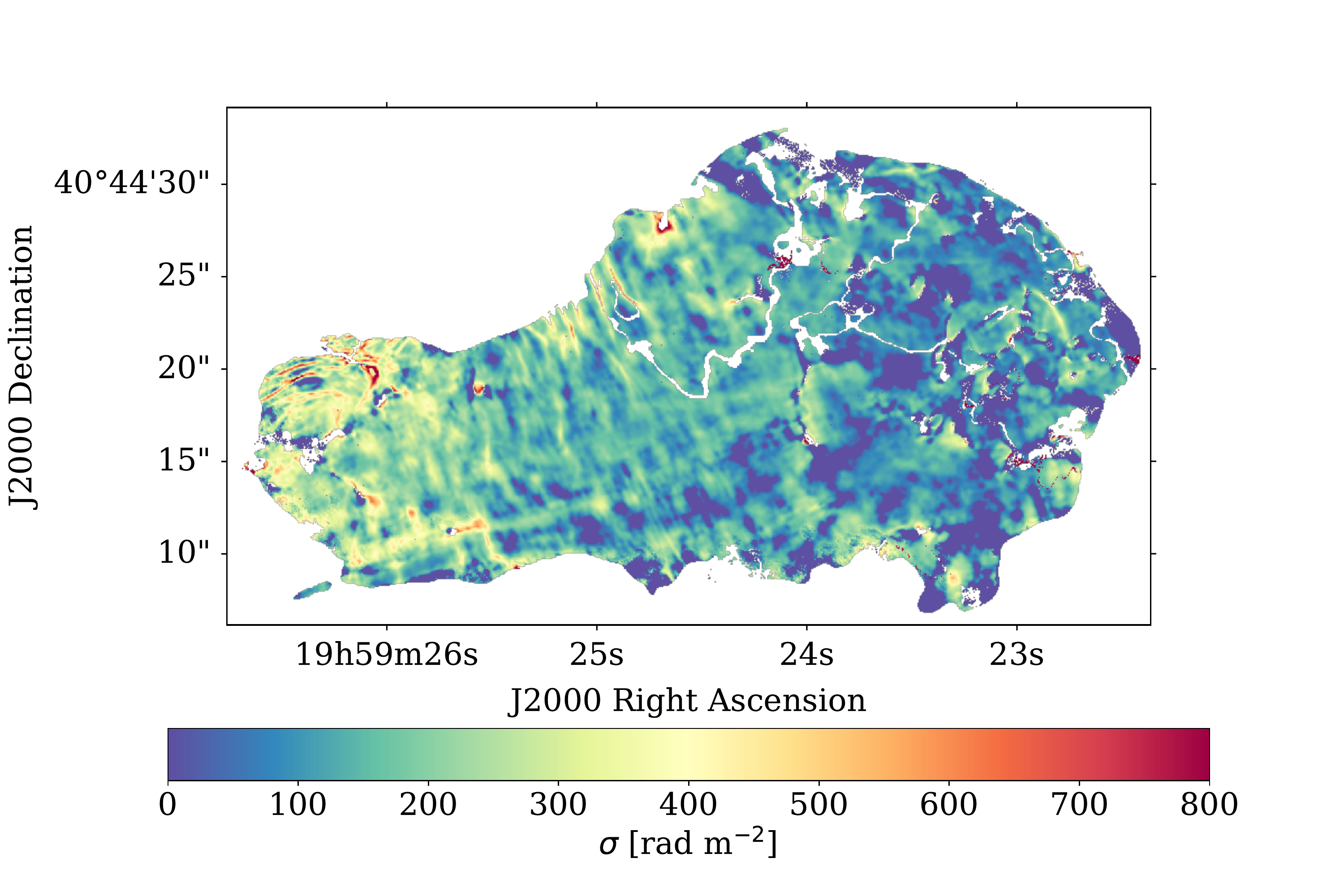}    
    \end{minipage}
   \caption{$QU$-fitting parameter maps at $0.30\arcsec$ ($6-18$
     GHz). The top row shows the intrinsic fractional polarization,
     the middle row shows the rotation measures, and the bottom row
     the $RM$ dispersions.  The intrinsic fractional polarization is
     consistent with the maps at high frequencies and high resolution,
     see Fig. \ref{fig:polarizationmaps} and
     \ref{fig:resolutionmaps1}.  Rotation measures range between
     $-4500$ to $+6400$ \radm in the eastern lobe and $-5000$ and
     $+3000$ \radm in the western lobe.  The RM changes sign on $3-20$
     kpc scales, with a banded structure orthogonal to the source
     axis, particularly evident in the western lobe. The lobes
     depolarize much more rapidly in the inner regions towards the AGN
     as seen in the $\sigma$ map. Pixels shown have relative error in
     $RM$ and $p_0$ of less than 0.2.
 \label{fig:fittedparams} }
 \end{figure*}

The bottom row of Fig. \ref{fig:fittedparams} shows the dispersion
images from the $0.3\arcsec$ modeling.  This parameter essentially
describes the rate of depolarization between $18$ and $6$ GHz, as seen
in Fig. \ref{fig:fitexample}.  The error in these measurements is
$\leq 10$ \radm for the brighter, outer regions in the lobe, rising to
$\leq 50$ \radm at the tails of the lobe. 

Typical dispersion values across the western lobe are $ \lesssim 300$
\radm with a few regions having dispersions of up to $\sim 400$ \radm. Regions with very small dispersions close to zero may be a result of a lack of significant depolarization, or due to repolarization. The latter can be real or due to the noise -- the form of Eq. \ref{eqn:tofit} does not account for this effect.
The dispersion values in the eastern lobe are similar to those in the
western lobe, with typical $RM$ dispersions of $\lesssim 200$ \radm in
the extreme parts of the lobe close to the hotspots, and regions
closest to the AGN having dispersions of $\sim 400$ \radm, with a few
spatially narrow regions having dispersion of $\sim 800$
\radm. However, these latter regions are associated with very large
errors -- 100 -- 300 \radm, and are mostly coincident with regions of
large $RM$ gradients and low polarization, similar to 3C 31; see
Appendix C in \citet{2008LAING}. The discussion of the actual origin
of these narrow filaments will be addressed in paper 2.  Note that the
dispersions are rather chaotic in the innermost regions of the eastern
lobe, likely indicating considerable structure remains on scales less
than $0.3\arcsec$. The similarity in $\sigma$ across both lobes is
also suggestive of a common medium encompassing both lobes, with the
more complex structures in the eastern lobe likely a result of the
Laing-Garrington effect
\citep{1988GARRINGTON,1991GARRINGTON,1988LAING} -- with the eastern lobe further from us than the western lobe. However, it should
be noted that this effect cannot not explain the banded $RM$
distribution across the lobes. Instead, the bands indicate the ordering of the large-scale magnetic field component, with a relatively large field ordering across the western lobe. 

\subsection{Intrinsic Magnetic Field Orientations}
\begin{figure*}
 \centering
    \begin{minipage}[b]{1\linewidth}
    \centering
   \includegraphics[width=0.95\linewidth]{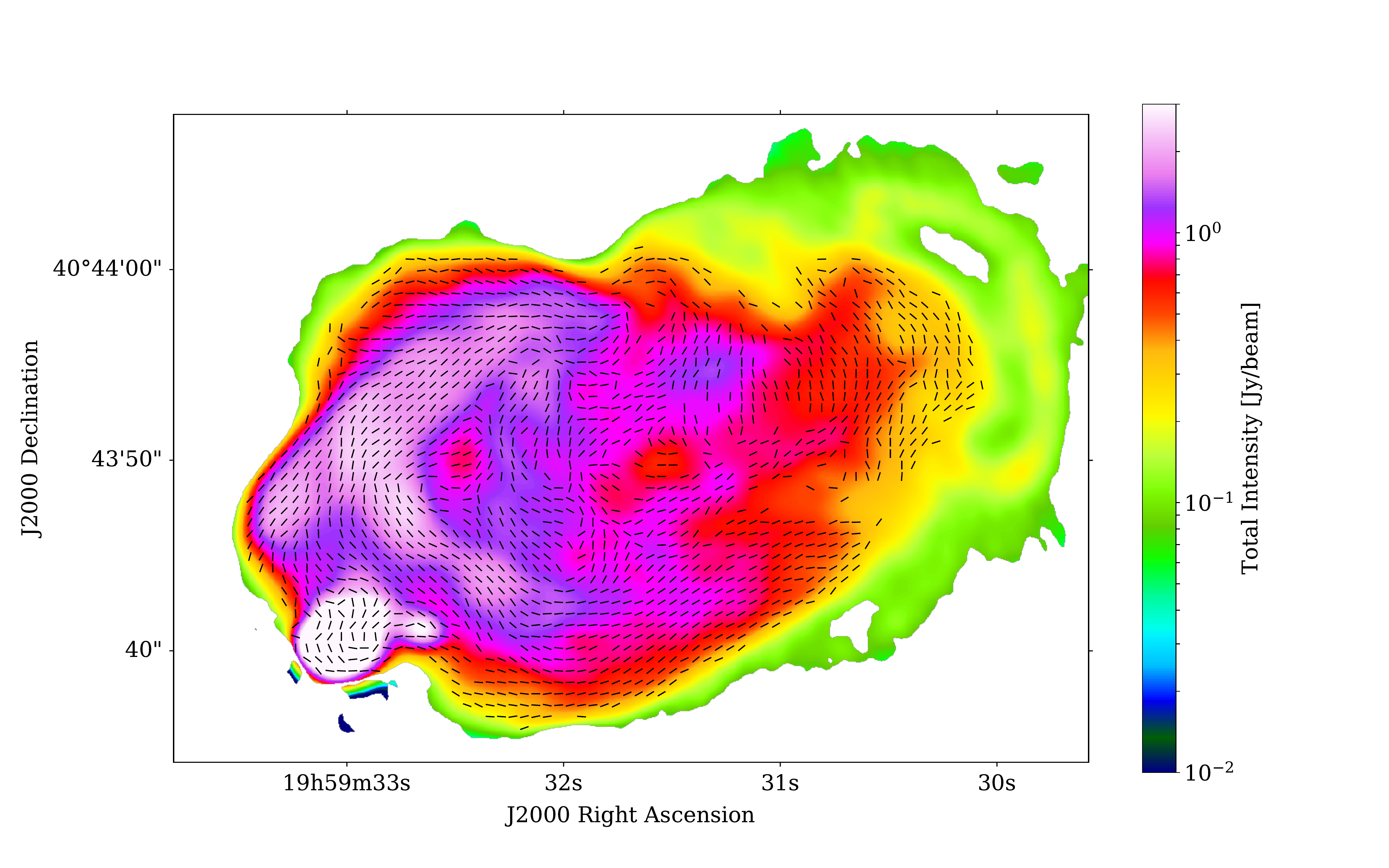}
    \end{minipage}
    \begin{minipage}[b]{1\linewidth}
    \centering
   \includegraphics[width=0.95\linewidth]{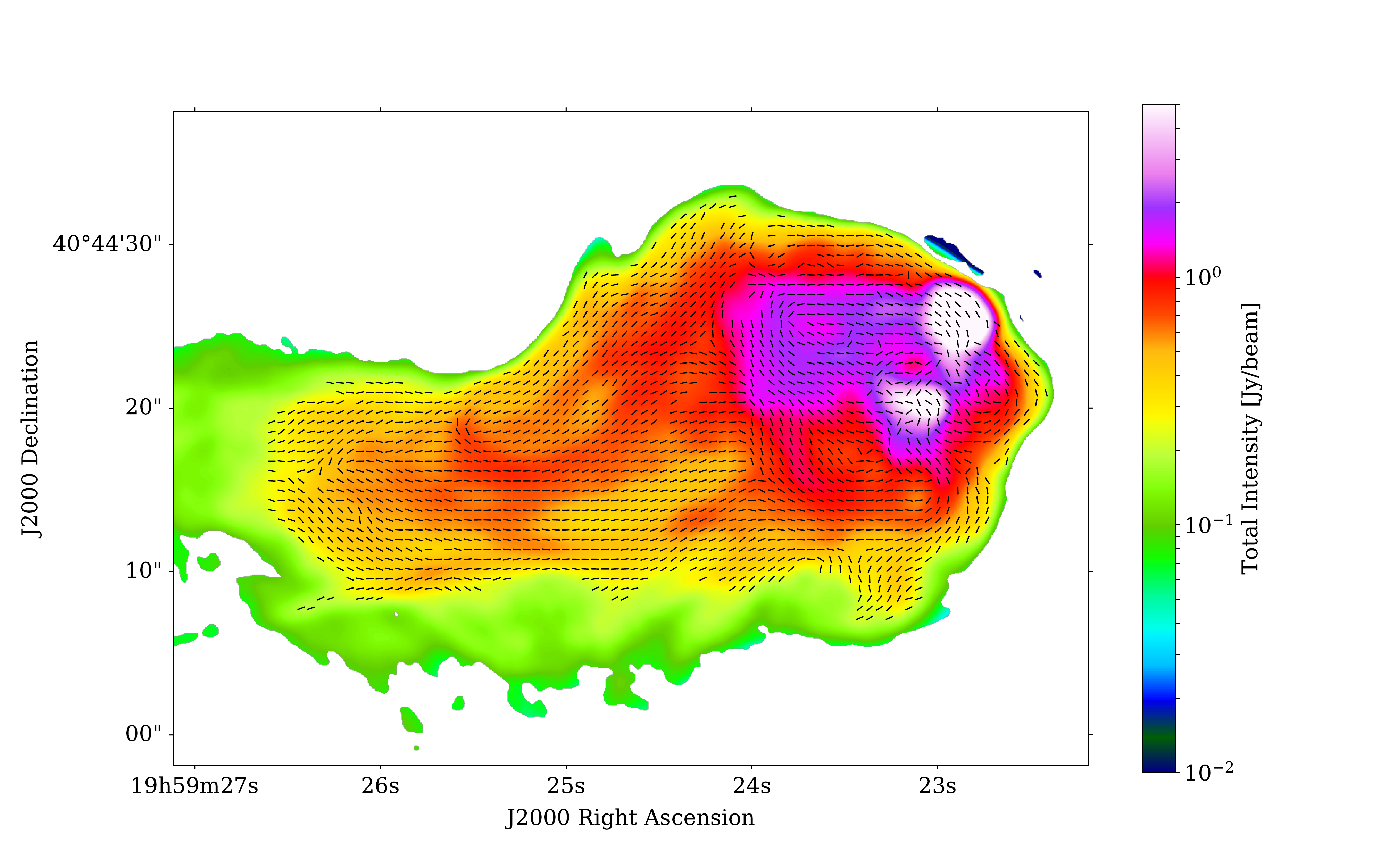}    
    \end{minipage}
      \caption{Intrinsic magnetic field orientation at
        $0.30\arcsec$ resolution across the lobes. These are
        superimposed on a color image of the total intensity at
        $3.979$ GHz and $0.54\arcsec \times 0.44\arcsec$ resolution.
        The field vectors follow the boundaries of the source as well
        as the filamentary structures.  \label{fig:bvectors}}
 \end{figure*}

The fitting described above determines the intrinsic polarization
(`zero-wavelength') angle values, from which we can directly determine
the intrinsic projected magnetic field ($\chi_0 +\pi/2$) of the
source. Figure \ref{fig:bvectors} shows the magnetic field
orientations superposed on a color-coded image of the total intensity
at 4 GHz. The fields follow the boundaries and filamentary structures
of the lobe emission.  This behavior is quite common in radio
galaxies, seen for example in 3C 465 \citep{2002EILEK}, Hydra A
\citep{1990TAYLOR}, and Pictor A \citep{1997PERLEY}, and is generally
understood as an effect resulting from shearing (and compression at
outer parts of the lobes) of the tangled lobe magnetic field at the
lobe boundary, resulting in suppression of field components normal to
the lobe boundaries \citep{1980LAING}.
The field vectors are
generally smooth across the western lobe and brighter parts of the
eastern lobe, while becoming slightly chaotic in the inner region of
the eastern lobe.  As with the other fitted parameters, this is likely
due to significant structures on scales less than the $0.3\arcsec$
resolution utilized here.
 
 \subsection{$RM$ vs. Random $RM$ Fluctuations}
\begin{figure}
\centering
 \includegraphics[width=0.8\linewidth]{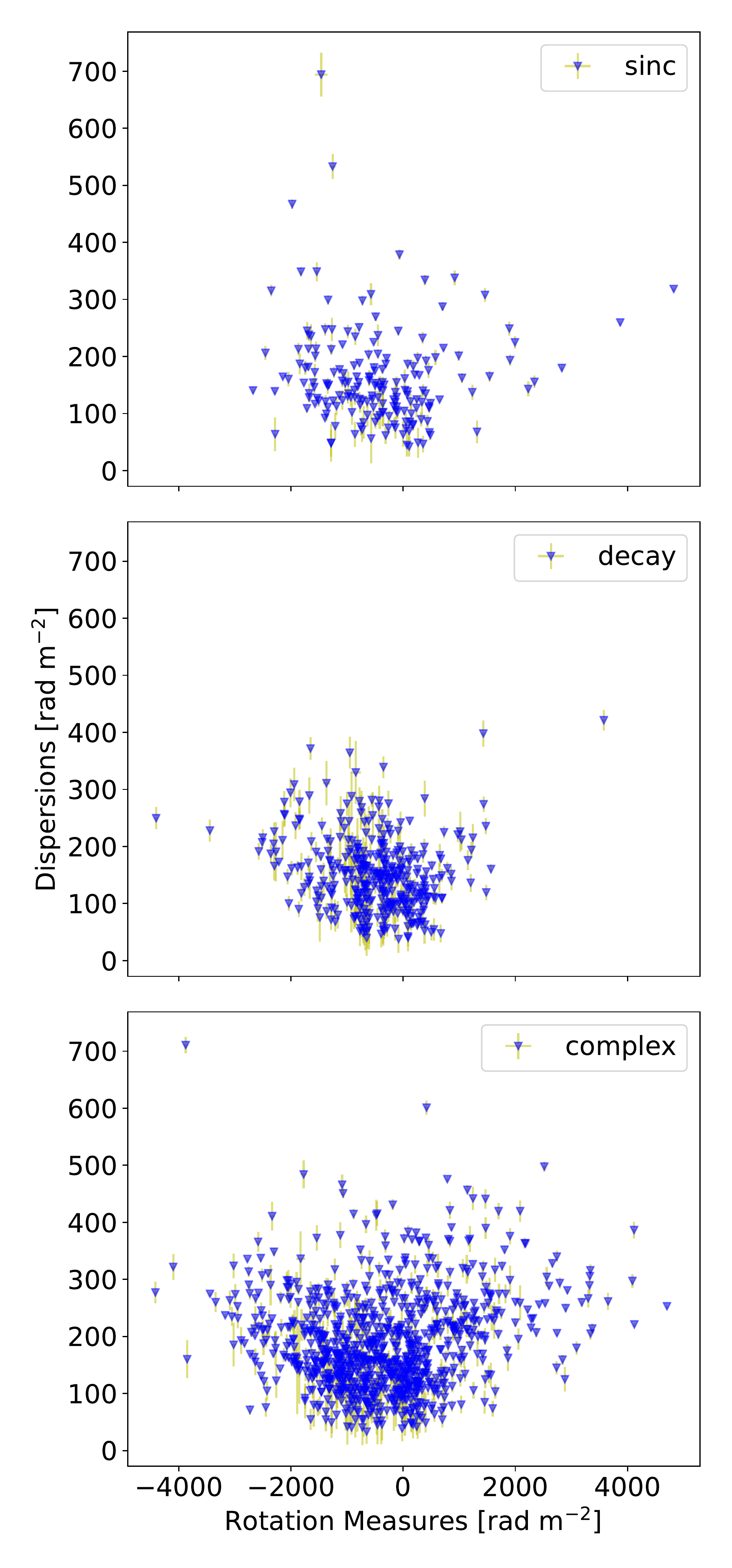}
 \caption{Faraday dispersions, $\sigma$, as a function of rotation measures, $RM$. These are separated based on their classes (see section
   \ref{polfreq}): sinc-like, smooth decay and complex decay.  There
   is no clear correlation between the $RM$ and $\sigma$. Suggesting
   that the observed $RM$s are not directly related to the observed
   depolarizations (decrease of fractional polarization with
   increasing $\lambda^2$). \label{fig:correlation}}
\end{figure}

One question that arises is whether the observed $RM$ dispersions are
correlated with the rotation measures. This might occur if the major
$RM$ structures are associated with a mixing layer surrounding the
source. The higher $RM$s might then be associated with higher field
values within the turbulent cells. To address this question, we plot
in Fig. \ref{fig:correlation} the $RM$ vs $\sigma$ for lines of sight
classified as described in section \ref{polfreq}. Shown are only those
lines of sight with relative errors in $RM$ dispersion less than 80\%.
There is no clear correlation between these two quantities, from which
we conclude that the $RM$ dispersions are not directly a result of the
observed $RM$, suggesting that the observed large-scale $RM$ are not
coming from a mixed region.  However, there is a clear tendency for
regions of very high $RM$ to have high dispersions. A likely
explanation is that a high $RM$ gradient across the resolution beam
will cause a larger apparent depolarization, leading to an apparent
increase in the dispersion.  The regions of the high gradients are the
inner lobes, where the highest $RM$s are accompanied by high
dispersions.

\subsection{Jet Rotation Measures and Magnetic Field Orientation}
\begin{figure}
   \begin{minipage}[b]{1\linewidth}
   \centering
     \includegraphics[width=1.2\linewidth]{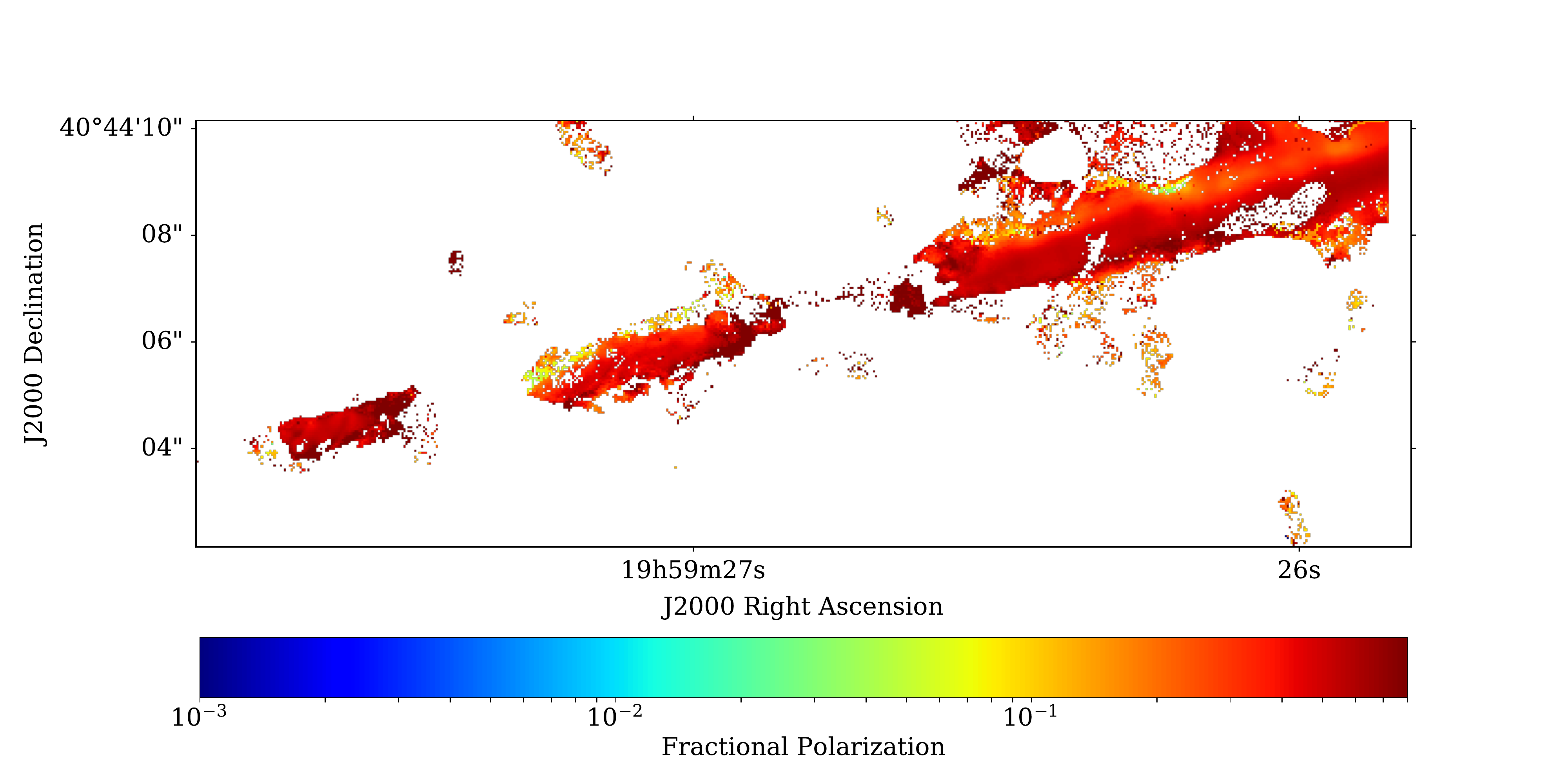}
   \end{minipage}\\
   \begin{minipage}[b]{1\linewidth}
     \includegraphics[width=1.2\linewidth]{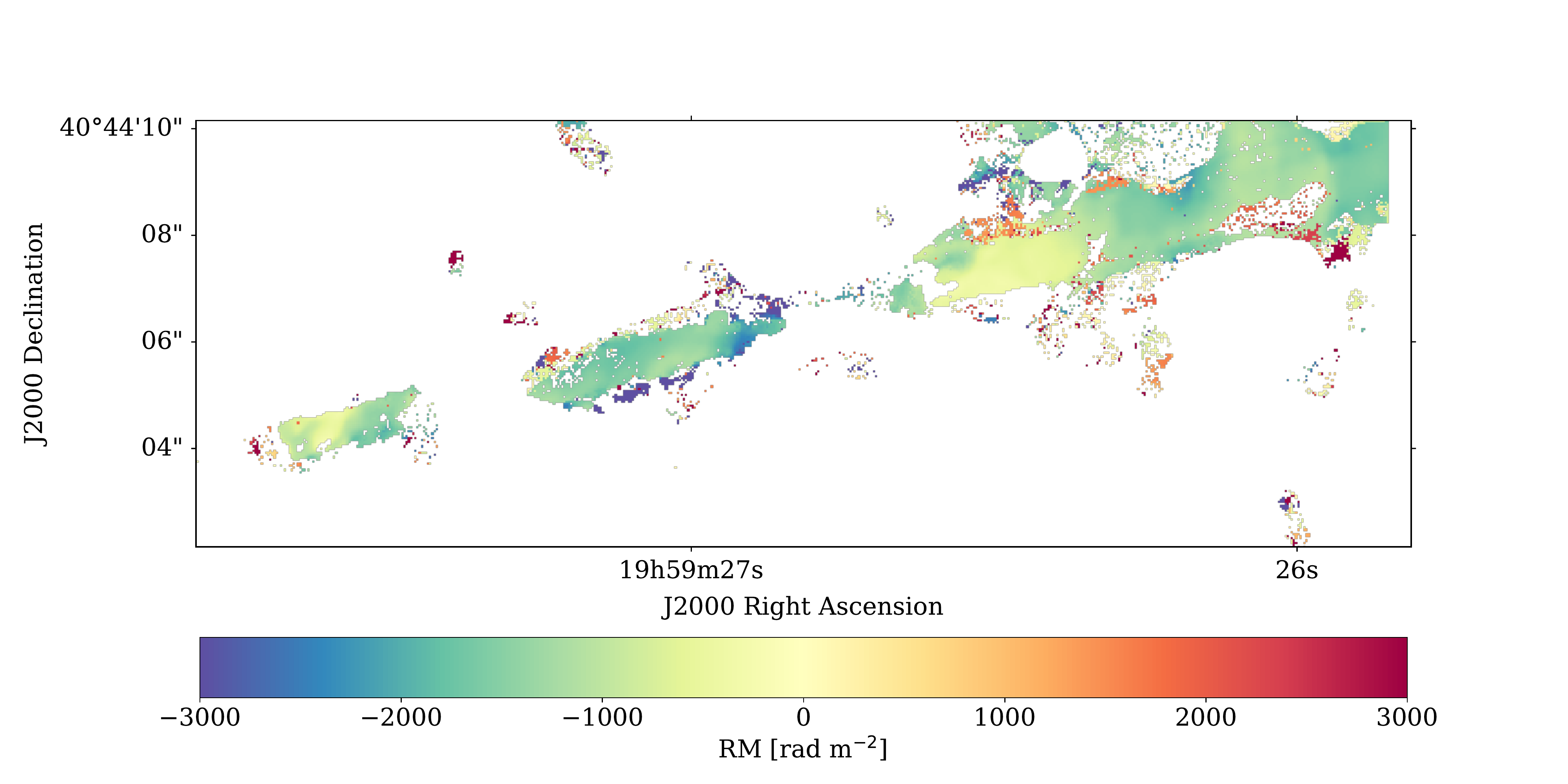}
   \end{minipage}\\
       \begin{minipage}[b]{1\linewidth}
      \includegraphics[width=1.2\linewidth]{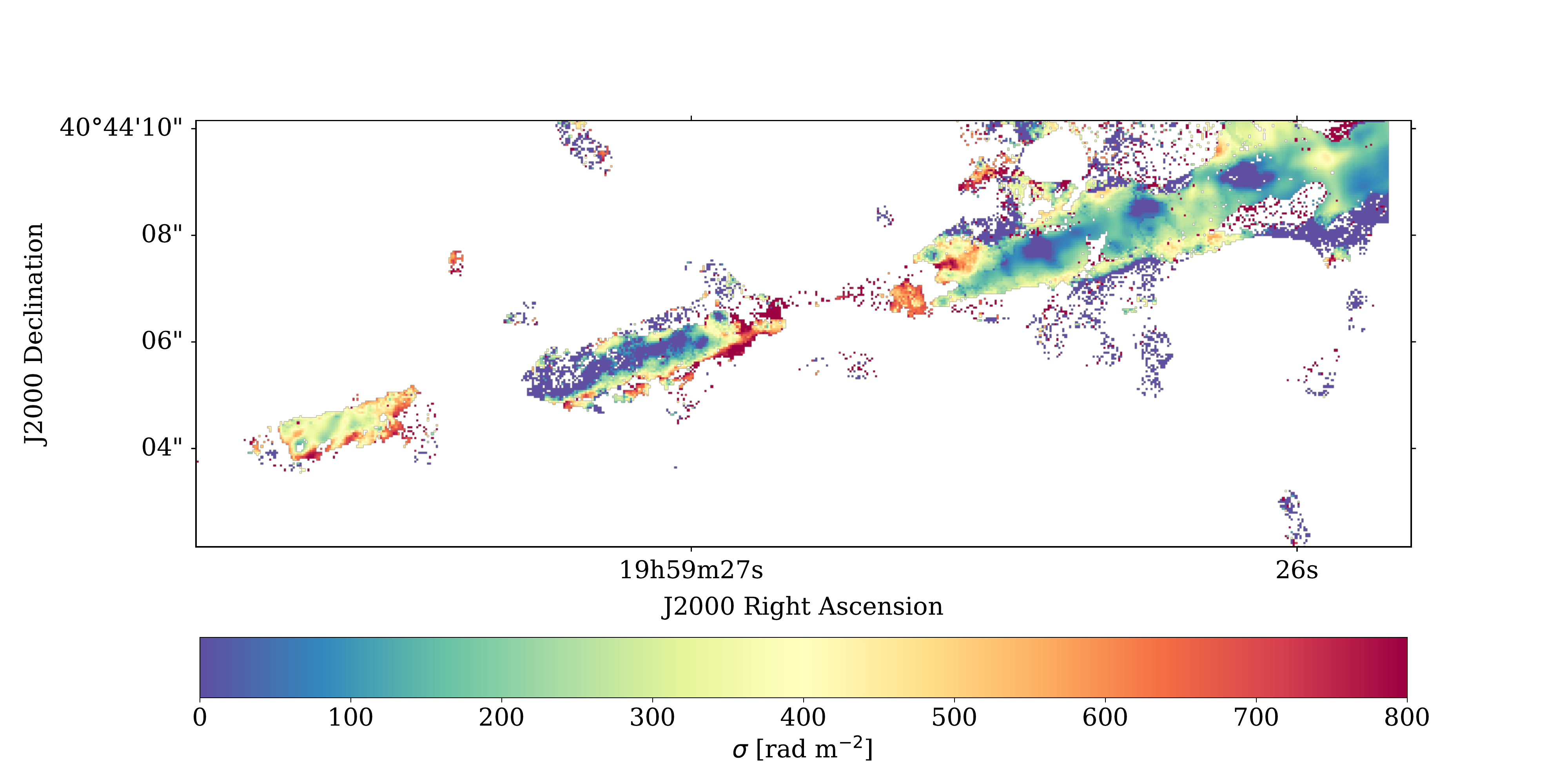}
    \end{minipage}
    \begin{minipage}[b]{1\linewidth}
      \includegraphics[width=1.2\linewidth]{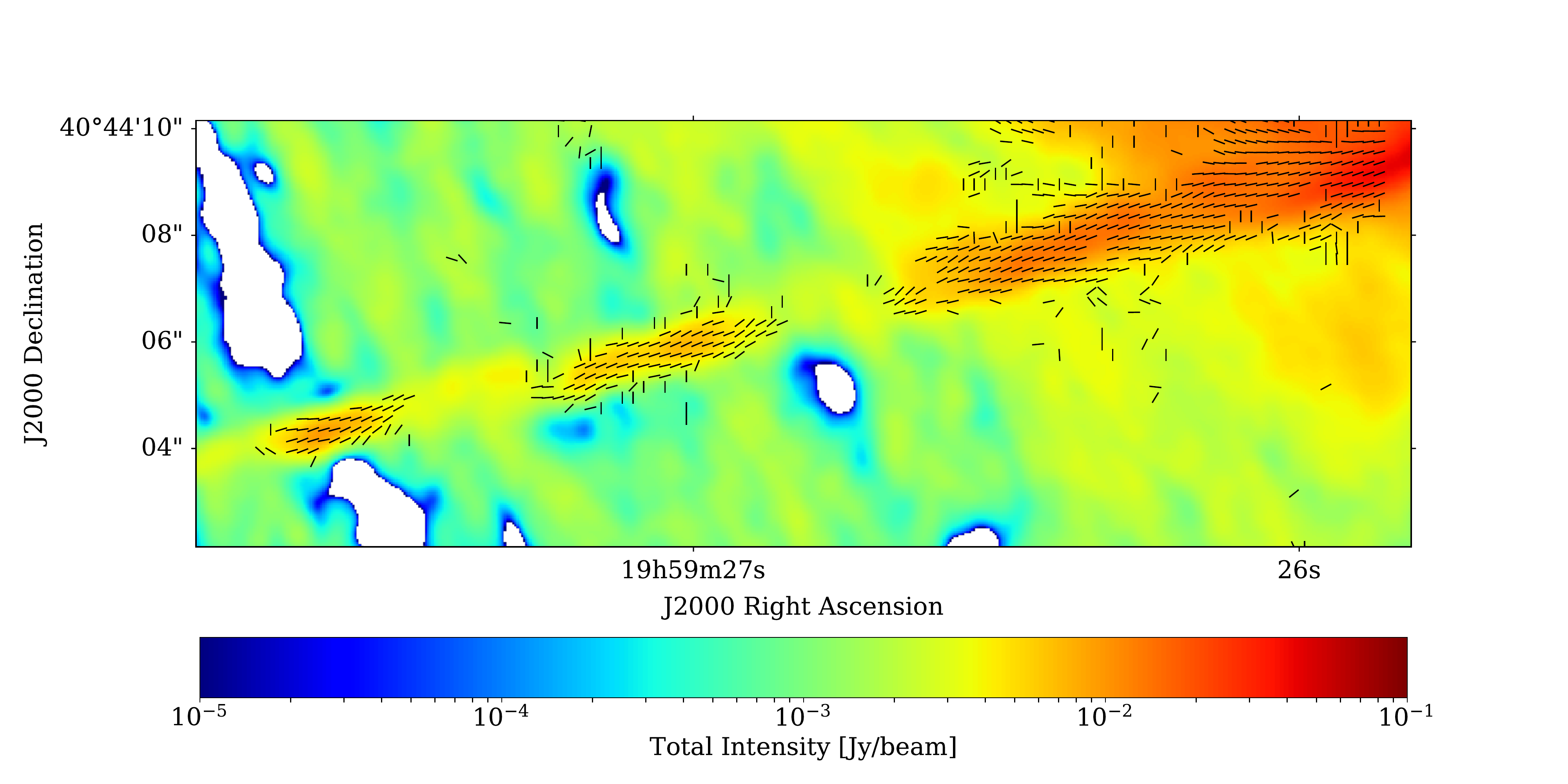}
    \end{minipage}
   \caption{Fitted parameters across the western jet. Top to bottom are the intrinsic fractional polarization, rotation measure, Faraday dispersion and magnetic field vectors along the jet, respectively. The jet is intrinsically highly polarized $\sim$ 20\% - 50\%. The $RM$s and $\sigma$ within the jet are similar to the $RM$s of the  adjacent lobe -- indicating a similar origin for both. The field vectors lie parallel to the jet axis. \label{fig:rmjet}}
 \end{figure}

To obtain detailed solutions for the jets, we performed the above fitting separately for regions enclosing the jet emission.
 Figure \ref{fig:rmjet} shows fractional polarization, rotation measure, Faraday dispersion and magnetic field
 orientation across the western jet. The pixels shown have $\sigma_{RM}/RM <0.2$. No reliable solutions were found on or close to the counter jet – even with RM-synthesis. The analysis of this region (and nearby structures such as the nucleus and the ring structure in the eastern lobe) awaits more sensitive observations at higher frequencies. 
 
 The intrinsic field orientations lie
 parallel to the jet axis across most parts of the jet where the jet polarization is visible. This polarization behavior along the jet is
 common in strong FR II sources \citep{1984BRIDLE}. The rotation measures of the jet are very similar to those of the lobe (see Fig. \ref{fig:fittedparams}) both in
 magnitude $\sim$ -300 to -2000 rad m$^{-2}$, and angular scale $\sim$ 2 - 5 kpc, again supporting the argument that the
 origin of the $RM$ is exterior to the source. The jet has similar Faraday dispersions as those of the lobe emission in the vicinity of the jet: $\sim$ a few 100 rad m$^{-2}$ -- with slight (not conclusive) indication of larger dispersions towards the nucleus. The jet is intrinsically highly polarized -- with fractional polarizations $\sim$ 20\% to 50\%.

The similarity of the jet $RM$ to that of the lobes has been used to argue
against mixing of external gas into the synchrotron-emitting regions
-- see \citet{1993TAYLOR} in the case of Hydra A.

\section{Predicting Low Frequency Depolarization}\label{sec:beamtest}
Section \ref{polresol} showed that beam depolarization --
depolarization from transverse fluctuations unresolved by the
resolution beam -- is an important factor in the depolarization of
Cygnus A. Can we state that the observed depolarizations are entirely
explained by this effect? This is an important question, which can
only be fully answered by observations with much higher resolution --
likely better than $0.20\arcsec$ at $2$ GHz -- which are not possible
with current instruments. However, there is a hint in
Fig. \ref{fig:polarizationmaps} and \ref{fig:resolutionmaps1}, showing
the polarization as a function of resolution, and frequency, that a
significant fraction of the foreground depolarization is resolved out
by a resolution of $0.45\arcsec$, which raises the question of whether
the high-resolution, high frequency foreground screen images shown in
Fig. \ref{fig:fittedparams} can be used to predict the
lower-frequency, lower-resolution images where the depolarization
effects are very strong. An accurate prediction would be strong
evidence for a screen origin of all, or nearly all, the observed
depolarization.

We assume that the fractional polarization, $p_0$ and polarization
angles, $\chi_0$, derived in section \ref{sec:faradayrotation}
represent the intrinsic polarization properties of the
source. Further, we assume that the rotation measure map is an
accurate representation of the foreground Faraday rotating screen on a
$0.3\arcsec$ scale -- essentially assuming that this resolution has
fully resolved the $RM$ screen.  With these, we can calculate the low
frequency polarization ($<$6 GHz) at $0.3\arcsec$ resolution, by simply
rotating the observed polarization angle by $RM\lambda^2$ as follows
\begin{equation}\label{eqn:model}
  P(\lambda^2) = p_0 I e^{2 \text{i} \chi_0} e^{2\text{i}RM\lambda^2}.
 \end{equation}
We estimated $I$ at $0.30\arcsec$ across 2-18 GHz first by determining
 the spectral index \footnote{Using task {\tt SPIXR} in AIPS.} at this resolution between 6-18 GHz, 
 and then using this spectral index
 to extrapolate the total intensity at low frequencies. 
We then convolved (using AIPS task 'CONVL') the resulting estimates of the $Q$, $U$ and $I$
to the desired resolution ($0.75\arcsec$) providing the model
images. These were then compared to the observed polarizations.
 
\begin{figure*}
 \center
     \begin{minipage}[b]{1\linewidth}
     \centering
     \includegraphics[width=0.6\linewidth]{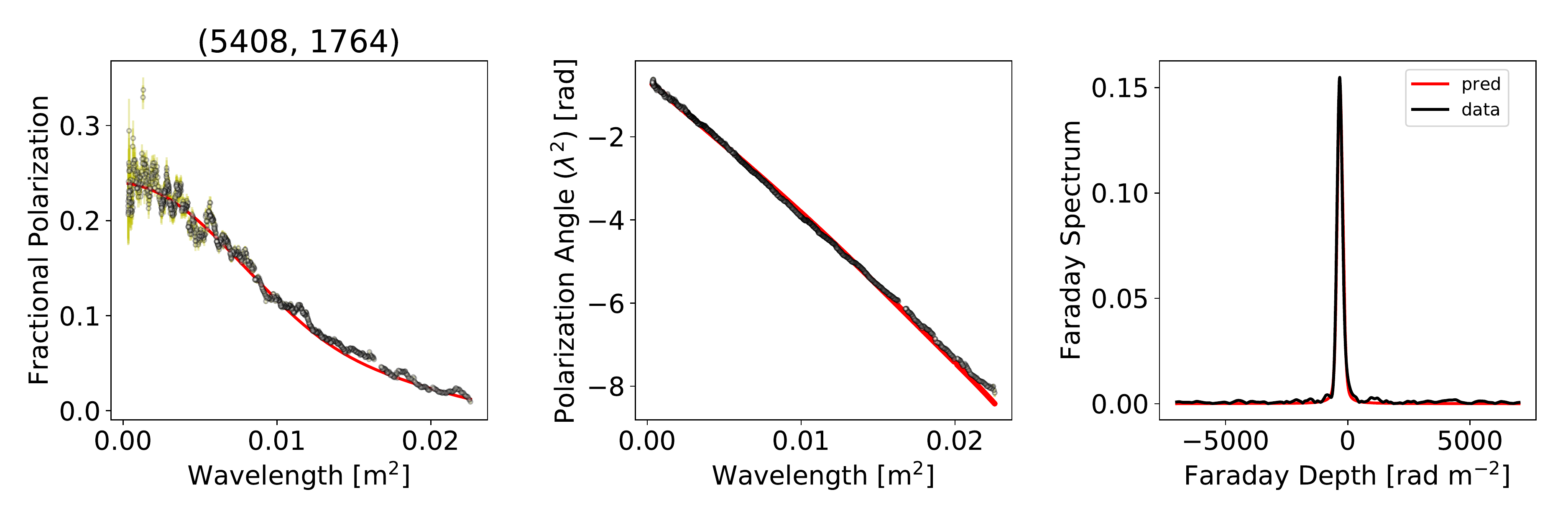} 
   \end{minipage}
       \begin{minipage}[b]{1\linewidth}
    \centering
    \includegraphics[width=0.6\linewidth]{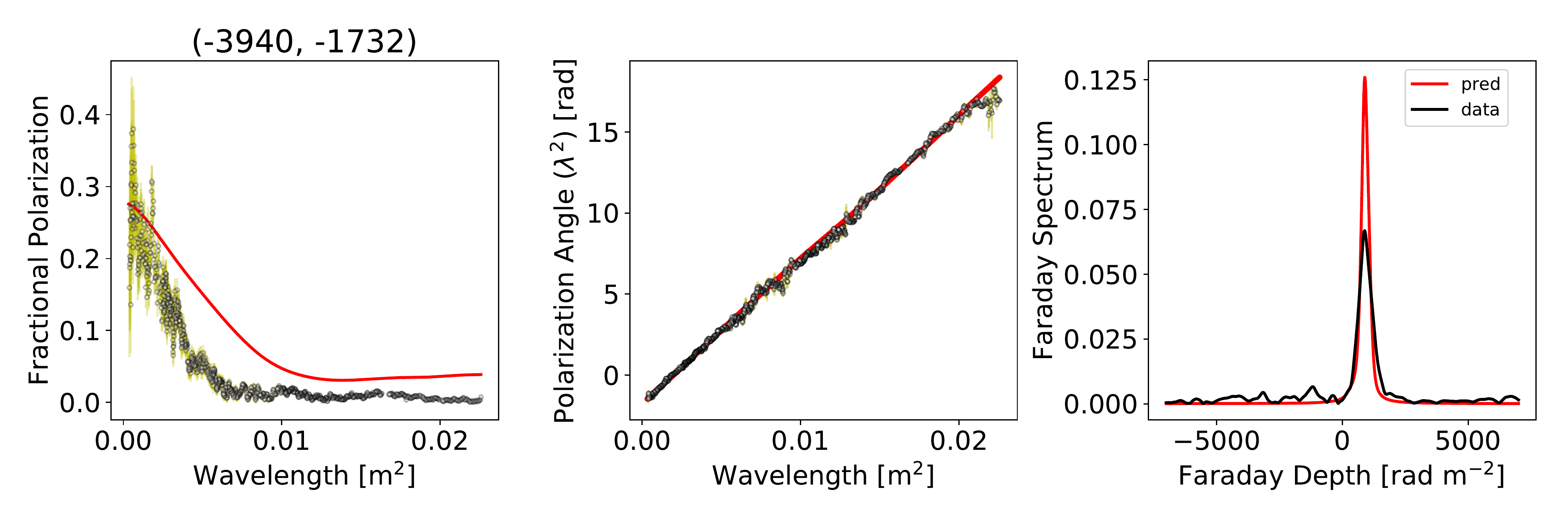}  
  \end{minipage}

    \begin{minipage}[b]{1\linewidth}
     \centering
     \includegraphics[width=0.6\linewidth]{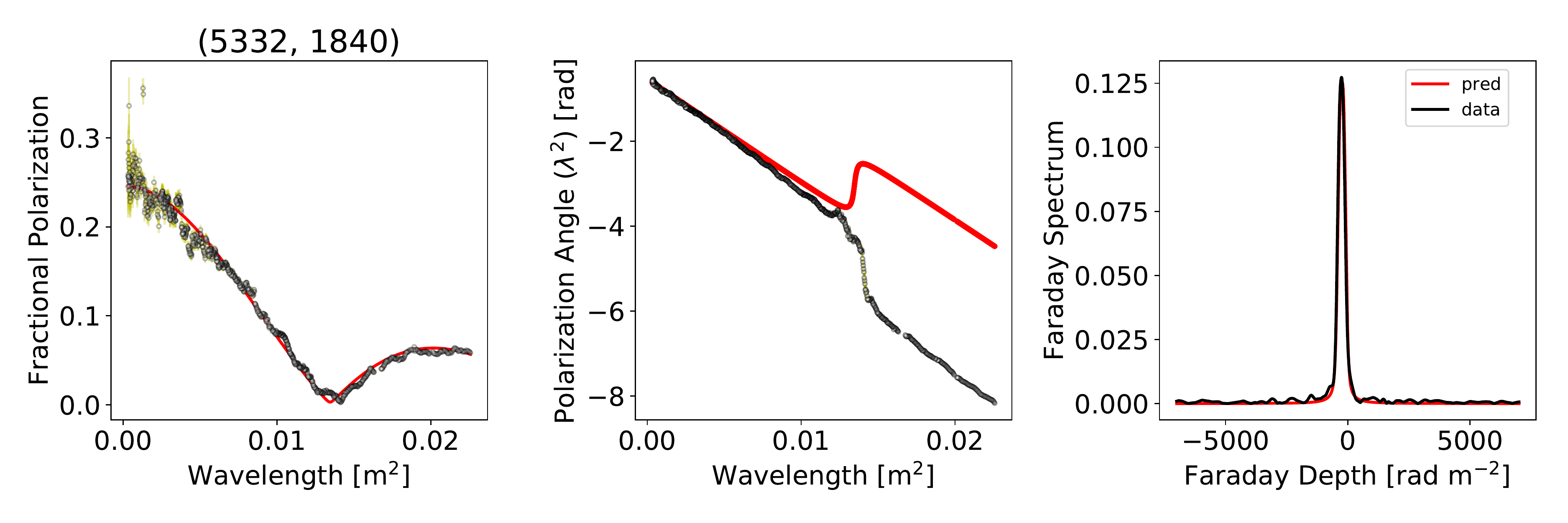}
   \end{minipage}
          \begin{minipage}[b]{1\linewidth}
    \centering
    \includegraphics[width=0.6\linewidth]{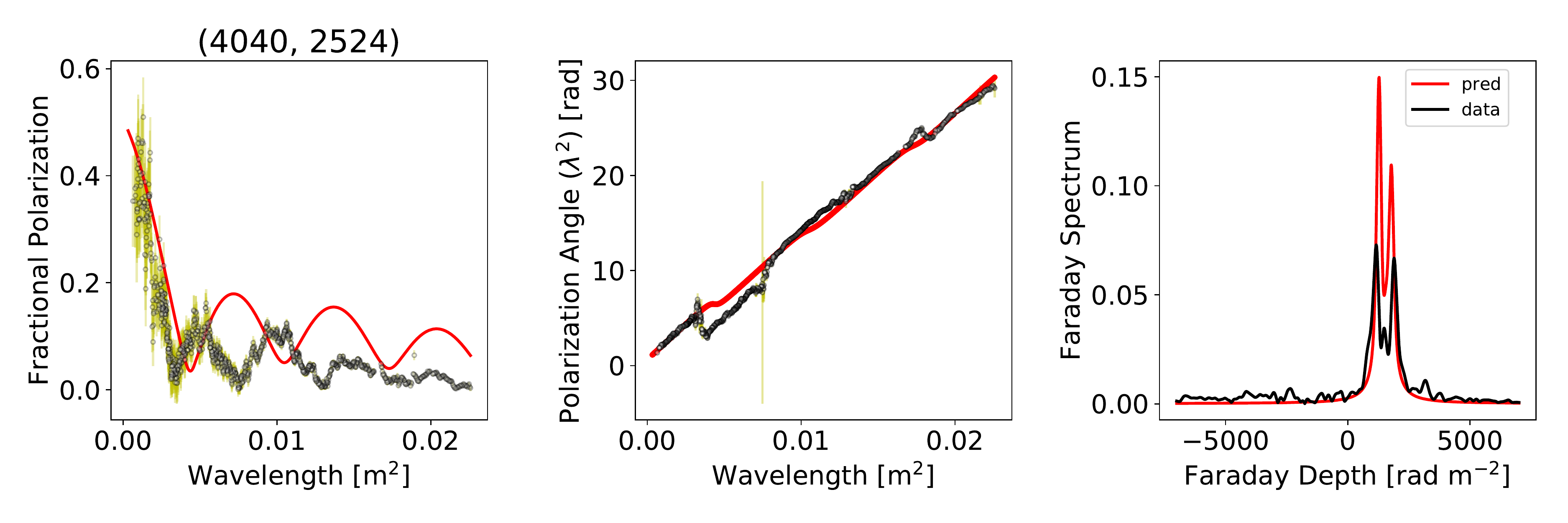} 
  \end{minipage}
    \begin{minipage}[b]{1\linewidth}
     \centering
     \includegraphics[width=0.6\linewidth]{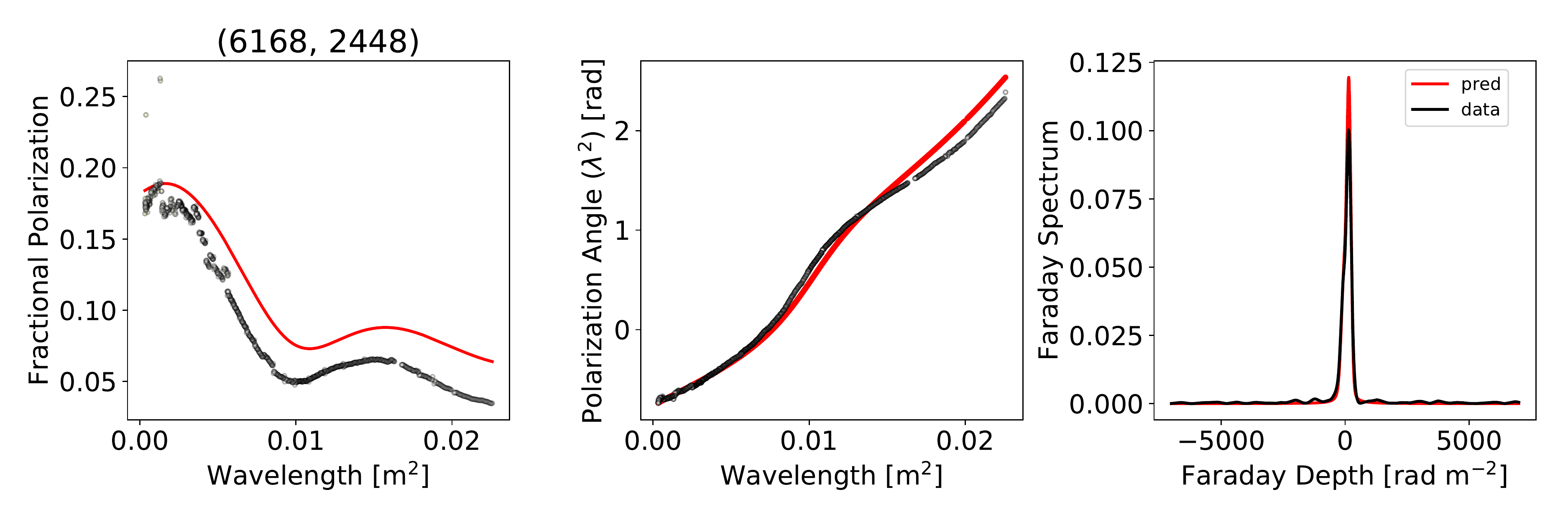} 
   \end{minipage}
   \begin{minipage}[b]{1\linewidth}
      \centering
      \includegraphics[width=0.6\linewidth]{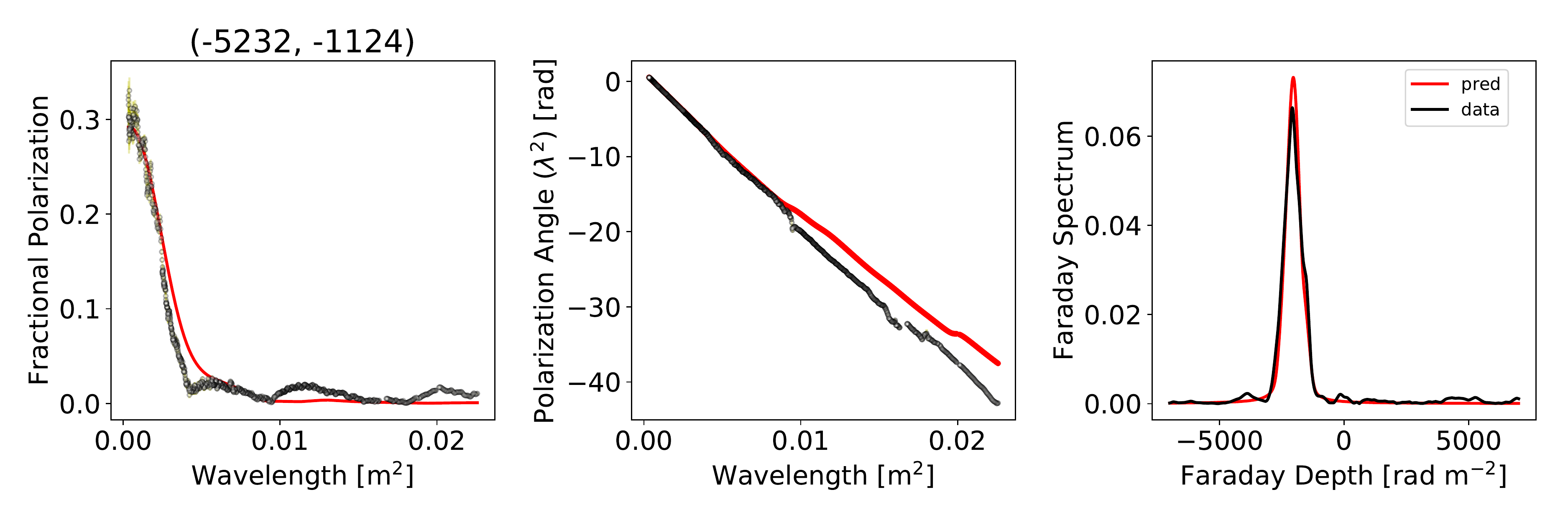}  
   \end{minipage} 
  \caption{Model predictions of low resolution, $0.75\arcsec$, data
    using the high resolution, high frequency data. The left column
    shows the fractional polarization, the middle column the
    polarization position angle, and the right column the Faraday
    spectra. Black shows the original data, and red shows the
    predictions. The simple model fits the data remarkably well for
    about 2/3 of the LoS. \label{fig:predictions}}
\end{figure*}

Figure \ref{fig:predictions} shows a few illustrative examples showing
our original data at $0.75\arcsec$ (in black) and the model predictions
(in red).  The left column shows the fractional polarization as a
function of $\lambda^2$, the middle column shows the polarization angle as a function of $\lambda^2$, and the right column shows
 the amplitude of the 
Faraday depths.

Table \ref{tab2} shows the result of the predictions.
The top two rows show examples with smoothly decaying fractional polarization as a function of $\lambda^2$. The top row shows an example where the predictions match the data very well -- $\sim 13\%$ of the smooth decaying lines of sight fit as well as this. The second row shows an example in which the prediction underestimates the rate
of depolarization -- denoted in Table \ref{tab2} as 'approx'. About $\sim 72\%$ of the lines of sight fall within this category.  The remaining lines of sight ($\sim 15\%$) are not
accurately predicted by our simple model  (denoted in Table \ref{tab2} as 'N/A').  

The bottom four rows show examples of lines of sight with oscillatory
behavior. In general, the presence of an oscillatory depolarization is
well predicted by our model. This result suggests that although the
dominant depolarization mechanism is associated with unresolved
fluctuations, there must be present in these cases larger-scale components
responsible for the oscillatory depolarization characteristics, likely
operating on scales only a little less than our standard $0.75\arcsec$
resolution.

The extraordinarily good fit shown in the third row is not common,
however, and in general, the predicted nulls occur at longer
$\lambda^2$ values than the data, so that the predicted peaks in the
Faraday spectrum are too close together, as can be seen in
the second column. See results of the predictions for sinc-like lines of sight in first row, column 12 to 14. Results for complex lines of sight are given in row 3 to 5 and same columns.

Thus,
overall, on average (weighted by the total number of lines of sight in
each class) we find that $\sim 14\%$ of the lines of sight are
accurately predicted (within measurement errors), and $\sim 72\%$ have
the correct functional form, but with a scaling error,
and $ \sim 14\%$ of the predictions are incorrect.

What could be causing the shifting of the predicted nulls to higher
$\lambda^2$ as well as underestimating the depolarizations?  The
simplest explanation for the latter is that the $0.3 \arcsec$
resolution image from which these estimates are made is missing some
smaller-scale structure. This is a reasonable conclusion given that
we are still seeing depolarizations at $0.30\arcsec$ which we
characterized using $\sigma$. The shifting of the nulls to higher
$\lambda^2$, on the other hand, is caused by an underestimation of the
$RM$ difference between the `patches' of polarized emission as resolved
on the $0.3 \arcsec$ $RM$ image.  It is not clear to us why this small
underestimation occurs.

\section{Summary}\label{sec:discussion}

In this paper we have presented the results from our wideband ($2-18$
GHz), high spectral resolution polarimetry data on Cygnus A. We have
shown how the fractional polarization varies as a function of
frequency and resolution. We also conducted a Faraday rotation
analysis using high frequency data, $>6$ GHz at $0.30\arcsec$
resolution. In this Faraday rotation study, we derived maps of
rotation measures ($RM$), intrinsic fractional polarization ($p_0$),
intrinsic polarization angles ($\chi_0$), and $RM$ dispersions
($\sigma$), from which we predict the depolarization behavior at lower
frequencies and resolutions, with remarkable success. A complete
analysis (physical interpretation) of our full band data will be presented in paper 2. The data are to be made publicly available immediately once our second paper -- the analysis paper-- gets accepted.

A summary of our results shows:
\begin{enumerate}
 \item All lines of sight through the lobes depolarize significantly
   with decreasing frequency. There is no relation between
   depolarization behavior and the various structural features (hot
   spots, filaments, jet, etc.).  The eastern lobe, particularly those
   regions nearer the nucleus, depolarize at higher frequencies than
   the western lobe.
 \item The distribution of the fractional polarization across the
   lobes is smooth at high frequencies, but becomes increasingly clumpy
   on $\sim 1$ kpc scales, at low frequencies.
\item The fractional polarization as a function of $\lambda^2$ for the
  different lines of sight reveal complex depolarization behaviors:
  some lines of sight smoothly depolarize, while others show
  oscillations, which in extreme cases have `sinc-like' behavior. By
  classifying these polarization behaviors into three simple
  categories: i) smooth decaying, i) sinc-like, and iii) complex, we
  find that $25\%$ fall into the first class, $11\%$ the second and
  $50\%$ are complex. There is a tendency for more complex decay in
  the eastern lobes, with relatively fewer smooth decay and sinc-like
  behaviors.  The distribution across the the western lobe, on the
  other hand, consists mostly of smooth decay, particularly at the
  extremes of the lobes, and complex behavior. There is no clear
  spatial correlation between these decay patterns across the lobes,
  except that they are not random but are clumped on few-kpc scales.
  
\item Lines of sight with oscillations in their depolarization behavior also show strong deviations of the polarization angle from linearity.
\item The lobe emission also depolarizes significantly with decreasing
  resolution. The inner region of the eastern lobe (closest to the
  AGN), depolarizes more rapidly than other regions of this lobe.
\item Plots of fractional polarization as a function of resolution and
  frequency for different lines of sight reveal that the beam-related
  effects are complicated, and that our observations do not have
  sufficient resolution to completely resolve the depolarizing
  structures present in or around Cygnus at frequencies below $6$ GHz.
\item The $RM$ values derived are consistent with those found by
  \citet{1987DREHER,1996PERLEY}, and extend those results to the
  weaker emission nearer the nucleus. Our new observations reveal very
  high $RM$ values at the tails of the lobes -- ranging between $
  -4500$ and $+6400$ \radm in the eastern lobe, and $-5000$ and $+3000$
  \radm in the western lobe. We find that the $RM$ values in the eastern lobe
  occurs in bands of low and high $RM$ along the source-axis. The $RM$ values
  are ordered on $3-20$ kpc scales.  There is no relation between the
  RM values and the presence of structural features in the total
  intensity (hotspots, jet, filaments, etc.).
\item The intrinsic magnetic fields of the source follow the boundary
  and filamentary structures of the lobes -- consistent with other
  radio galaxies. The only exception is the inner region of the
  eastern lobe where the fields appear to be slightly chaotic.
\item The derived intrinsic fractional polarization of the source
  found from analysis of the high frequency data alone at $0.3
  \arcsec$ resolution shows that Cygnus A is intrinsically highly
  polarized, with typical fractional polarizations between 15\% to 45\%, and some
  regions as high as $70\%$. The intrinsic fractional polarization of
  the inner region of the eastern lobe is very similar to that of the
  rest of the lobe, indicating that this region is not intrinsically
  unique, and that the observed depolarizations, and rapid depolarizations have an
  external origin.
\item We find that the $RM$ dispersions across both lobes decrease
  with increasing distance from the nucleus.  The outermost regions in
  both lobes have small RM dispersions $\sim 0$ rad m$^{-2}$.  These
  increase towards the nucleus to typical values of $\sim 200$ \radm
  to $400$ \radm. The innermost regions of the eastern lobe show the
  highest dispersions -- up to $\sim 800$ \radm.  The $RM$ dispersions
  are weakly correlated with the observed $RM$, again suggesting that
  the $RM$ are external to the lobes.
\item Assuming that the derived $p_0$, $\chi_0$, and $RM$ generated
  from the high frequency, high resolution data represent the
  intrinsic polarization properties of the source, and that there is
  no synchrotron emission in the external medium, we generated
  predictions of the low frequency polarization emission at $0.3
  \arcsec$ resolution, which were then convolved to our resolution
  of $0.75\arcsec$.  Comparison of the predictions with the
  observations show remarkable agreement, with about $14\%$ of the
  lines of sight predicted accurately, and a further $72\%$ with the
  correct shape, but with small scale errors.  This result supports
  the interpretation that the majority of the observed depolarization
  is due to unresolved fluctuations, on the $0.3$ -- $0.7$ arcsecond
  scale, in the external medium.
 
\end{enumerate}

\acknowledgments
\section{acknowledgments}
Support for this work was provided in part by the National Aeronautics
and Space Administration through Chandra Award Number GO5-16117B
issued by the Chandra X-ray Center, which is operated by the
Smithsonian Astrophysical Observatory for and on behalf of the
National Aeronautics Space Administration under contract NAS8-03060.
The financial assistance of the South African Radio Astronomy
Observatory (SARAO) towards this research is hereby acknowledged
(www.ska.ac.za). Additional financial assistance of the National Radio Astronomy Observatory (NRAO) a
facility of the National Science Foundation operated under cooperative
agreement by Associated Universities, Inc is acknowledged. This work is based upon
research supported by the South African Research Chairs Initiative of
the Department of Science and Technology and National Research
Foundation. We thank Larry Rudnick for his enthusiastic assistance in
the interpretation of complex polarimetric data. We also thank the reviewer for providing insightful suggestions for improving our findings.

\end{document}